\documentclass[%
 aip,
 amsmath,amssymb,
preprint,%
]{revtex4-1}

\usepackage{graphicx}
\usepackage{dcolumn}
\usepackage{bm}

\usepackage[utf8]{inputenc}
\usepackage[T1]{fontenc}
\usepackage{mathptmx}
\usepackage{etoolbox}

\usepackage{hyperref}
\usepackage{graphicx}
\usepackage{wrapfig}
\usepackage{caption}

\makeatletter
\def\@email#1#2{%
 \endgroup
 \patchcmd{\titleblock@produce}
  {\frontmatter@RRAPformat}
  {\frontmatter@RRAPformat{\produce@RRAP{*#1\href{mailto:#2}{#2}}}\frontmatter@RRAPformat}
  {}{}
}%
\makeatother

\begin{document}

\title[]{Lie Algebra Contractions and Interbasis Expansions on Two-Dimensional Hyperboloid IIA. Subgroup Basis.}
\author{\framebox{G.S.~Pogosyan}}
 \affiliation{Yerevan State University, Yerevan, Armenia}
\author{A.~Yakhno}%
 \email{alexander.yakhno@academicos.udg.mx}
\affiliation{%
Departamento de Matematicas, CUCEI, Universidad de Guadalajara, Mexico
}%

\def\vphi{\varphi}
\def\rphi{\vphi\in[0,2\pi)}
\def\vtheta{\vartheta}
\def\SCPM{{\SCPZ\over\SCPN}}
\def\SCPZ{{\xi^2+\eta^2}}
\def\SCPN{{\xi^2\eta^2}}

\def\ralpha{\alpha\in(\i K',\i K'+2K)}
\def\rPalpha{\alpha\in(\i K',\i K'+K)}
\def\rbeta{\beta\in[0,4K')}
\def\rPbeta{\beta\in(0,K')}
\def\rmu{\mu\in(\i K',\i K'+2K)}
\def\rPmu{\mu\in(\i K',\i K'+K)}
\def\reta{\eta\in[0,4K')}
\def\rPeta{\eta\in(0,K')}

\newdimen\theight
\newcommand{\ds}{\displaystyle}
\newcommand{\be}{\begin{equation}}
\newcommand{\ee}{\end{equation}}
\newcommand{\bi}{\begin{itemize}}
\newcommand{\ei}{\end{itemize}}
\newcommand{\x}{{\ensuremath{\times}}}
\newcommand{\bb}[1]{\makebox[16pt]{{\bf#1}}}
\newtheorem{theorem}{Theorem}
\newtheorem{definition}{Definition}

\newtheorem{lemma}{Lemma}
\newtheorem{comment}{Comment}
\newtheorem{corollary}{Corollary}
\newtheorem{example}{Example}
\newtheorem{examples}{Examples}
\newcommand{\sn}{\mbox{sn}}
\newcommand{\cn}{\mbox{cn}}
\newcommand{\dn}{\mbox{dn}}
\newcommand{\ba}{\begin{array}}
\newcommand{\ea}{\end{array}}
\newcommand{\bea}{\begin{eqnarray}}
\newcommand{\eea}{\end{eqnarray}}
\newcommand{\Res}{\mbox{Res}}
\newcommand{\arcsinh}{\mbox{arsinh}}
\newcommand{\arccosh}{\mbox{arcosh}}
\newcommand{\sign}{\mbox{sign}}
\newcommand{\diag}{\mathop{\mathrm{diag}}}

\def \Column{%
             \vadjust{\setbox0=\hbox{\sevenrm\quad\quad tcol}%
             \theight=\ht0
             \advance\theight by \dp0    \advance\theight by \lineskip
             \kern -\theight \vbox to \theight{\rightline{\rlap{\box0}}%
             \vss}%
             }}%

\catcode`\@=11
\def\qed{\ifhmode\unskip\nobreak\fi\ifmmode\ifinner\else\hskip5\p@\fi\fi
 \hbox{\hskip5\p@\vrule width4\p@ height6\p@ depth1.5\p@\hskip\p@}}
\catcode`@=12 

\def\cents{\hbox{\rm\rlap/c}}
\def\miss{\hbox{\vrule height2pt width 2pt depth0pt}}

\def\vvert{\Vert}                

\def\tcol#1{{\baselineskip=6pt \vcenter{#1}} \Column}

\def\dB{\hbox{{}}}                 
\def\mB#1{\hbox{$#1$}}             
\def\nB#1{\hbox{#1}}               

\date{\today}

\begin{abstract}
Three subgroup type eigenfunctions of the Laplace-Beltrami operator on a two-dimensional two-sheeted hyperboloid are considered and all interbasis expansions between them are calculated. It is shown how the coefficients determining the expansions and the expansions themselves between subgroup basis contract from the hyperboloid to the Euclidean plane.
\end{abstract}

\maketitle

\tableofcontents

\newpage

\section{Introduction}
\label{sec:Introduction}

This article is the second step after the  Ref. \onlinecite{POG-YAKH4} (see also
 Refs. \onlinecite{IPSW1, IPSW2, KMP-CONT, PW1, IPSW4, POG-YAKH2, POG-YAKH3, POG-YAKH0}), where we continue studies on Lie algebra contractions and separation of variables for 
Laplace-Beltrami (LB) equation on the upper sheet of two-dimensional two-sheeted hyperboloid (real Lobachevsky space) of radius $R > 0$, namely 
$H_2^+:$ $u_0^2 - u_1^2 - u_2^2 = R^2$, \, $u_0 \ge R$, embedded into 3D pseudo-Euclidean space $E_{2,1}$ with Cartesian coordinates $(u_0, u_1, u_2)$.
In particular\cite{POG-YAKH4}, we examined in detail the contraction limit $R \to \infty$
from separated systems of coordinates on one- and two-sheeted hyperboloids to
2D pseudo-Euclidean $E_{1,1}$ and Euclidean $E_2$ spaces  respectively. 

Another important problem that has not yet been sufficiently studied is the contraction limit of 
eigenfunctions of the LB equation and interbasis expansions from  $H_2^+$ to $E_2$.  In Refs. \onlinecite{POG-YAKH4,  IPSW2, KMP-CONT, IPSW4, PW1, POG-YAKH2} a connection was established between separable coordinate systems on two-dimensional curved and flat spaces, associated by the contraction of their isometry groups $SO(2,1)$ to $E(2)$. It was shown \cite{POG-YAKH4, PW1, IPSW4, POG-YAKH2} that all nine coordinate systems on $H_2^+$ can be contracted to at least one of the four systems on $E_2$. The $n$-dimensional case, including the contraction limit of eigenfunctions and their interbasis expansions, was considered only for $S_n$ spheres\cite{POG-YAKH1,IPSW5, IPSW6}. The approach presented here uses specific implementations of  In\"{o}n\"{u}-Wigner
contractions, so-called {\it analytical contractions}\cite{IPSW1}. This means that in separable coordinate systems, in Lie algebra operators, in eigenvalues and eigenfunctions of the Laplace-Beltrami operator $\Delta_{LB}$ there appears a contraction parameter $\varepsilon = 1/R$. Thus, it becomes possible to trace the contraction limit $\varepsilon \sim 0$ at different levels, including separated (ordinary) differential equations and interbasis expansions.



The free quantum motion on  $H_2^+$ is described by the Schr\"odinger equation 
\begin{eqnarray}
\label{HE1}
{\cal H} \Psi = - \frac{\Delta_{LB}}{2}  \Psi =  {\cal E}  \Psi, \qquad {\cal E} = \mathrm{const}.
\end{eqnarray}
The ambient space $E_{2,1}$ has the metric $G_{\mu \nu}=\diag (-1, 1, 1)$, $\mu, \nu = 0,1,2$ and the line element satisfies $dL^2 = - du_0^2 + du_1^2 + du_2^2$. Let $(\xi^1, \xi^2)$ be the curvilinear coordinates on $H_2^+$. The relation of the line element $dl$ on this manifold with components of the metric tensor $g_{ik}(\xi^1, \xi^2)$ is $dl^2 = g_{ik} d\xi^i d\xi^k$, and the operator $\Delta_{LB}$ has the form
\begin{equation}
\label{ALGEBRA4}
\Delta_{LB} = \frac{1}{\sqrt{g}}\frac{\partial}{\partial \xi^i}
\sqrt{g} g^{ik}\frac{\partial}{\partial \xi^k},
\qquad
g=|\det (g_{ik})|,
\
g_{ik}g^{k j} = \delta_{i j}, \ i, k, j = 1, 2,
\end{equation}
where $\delta_{i j}$ is Kronecker delta. The following relations between components $g_{ik}(\xi^1, \xi^2)$ of the metric tensor and Cartesian coordinates of the ambient space  are valid:
\begin{equation}
\label{METRIC}
g_{ik} =  G_{\mu \nu} \frac{\partial u_\mu}{\partial \xi^i}
\frac{\partial u_\nu}{\partial \xi^k}.
\end{equation}

The isometry group of   $H_2^+$ is the pseudo-orthogonal group $SO(2,1)$.
This group corresponds to Lie algebra $so(2,1)$, containing three linearly independent elements:
\begin{equation}
\label{GEN}
K_1 =   - u_0 \frac{\partial}{\partial u_2} - u_2 \frac{\partial}{\partial u_0},
\qquad
K_2 =   - u_0 \frac{\partial}{\partial u_1} - u_1 \frac{\partial}{\partial u_0},
\qquad
M =  u_1 \frac{\partial}{\partial u_2} - u_2 \frac{\partial}{\partial u_1},
\end{equation}
with non-zero commutation relations:
\begin{equation}
\label{GEN-1}
[K_1, K_2] =  - M,
\qquad
[K_2, M] =  K_1,
\qquad
[M, K_1] =  K_2.
\end{equation}
Two operators $K_1$ and $K_2$ are the Lorentz transformations with respect to $u_1$ and $u_2$ axes, respectively, and $M$ is the rotation in the $u_1u_2$ plane. The Casimir invariant of the Lie algebra $so(2,1)$ is a quadratic operator ${\cal C}$
defining by ${\cal C}= K_1^2 + K_2^2 - M^2$, and is connected with the Laplace-Beltrami
operator ${\cal C} = R^2 \Delta_{LB}$. All operators (\ref{GEN}) are hermitian with respect to the scalar product of the wave functions on the manifold
\begin{eqnarray}
\label{HERMIT1}
(\Psi_1,  \Psi_2)_{H_2^{+}} = \int_{H_2^{+}}\, \Psi_1^{*} \, \Psi_2  d{\rm s},
\end{eqnarray}
where $d{\rm s} = \sqrt{g}d\xi^1d\xi^2$ is the area element. 

Olevskii \cite{OLEV} have shown that Eq. (\ref{HE1}) on two-dimensional  
hyperboloid admits separation of variables in nine orthogonal coordinate systems.
Each solution of Eq. (\ref{HE1}), relating  with the separated system of coordinates, 
is an eigenfunction of the pair of commuting operators, namely Casimir operator  ${\cal C}$ and one of the nine second order operators $L^\beta$  in the enveloping algebra of $so(2,1)$:
\begin{eqnarray}
\label{COM-HE1}
R^2 \Delta_{LB} \Psi = {\cal C} \Psi =   \sigma(\sigma+1) \, \Psi,
\qquad
L^\beta \Psi = \lambda \Psi, \qquad
\Psi_{\sigma\lambda} (\xi^1, \xi^2) = \Omega_{\sigma\lambda}(\xi^1)\Phi_{\sigma\lambda}(\xi^2),
\end{eqnarray}
where $\sigma(\sigma + 1)$  is eigenvalue of Casimir operator, and  $\lambda$ represents the separation constant\footnote{For certain systems of coordinates the 
situation is not so straightforward. Examples are the EQ and EP coordinates, when operators (\ref{COM-HE1}) do not constitute a complete set of commuting self-adjoint operators. In this cases, to obtain a complete set, one needs to add one operator 
to them.  The same fact  occurs in parabolic coordinates on $E_2$ space. 
For details about this phenomena see Ref. \onlinecite{MUKUNDA}.}. Following Ref. \onlinecite{WLS}, we will denote these nine operators $L^\beta$ as follows:
{\it 1.} Pseudo-Spherical system $L^{S}= M^2$,
{\it 2.} Equidistant system $L^{EQ} = K_2^2$,
{\it 3.} Horocyclic system  $L^{HO} = (K_1+M)^2$,
{\it 4.} Semi-Circular-Parabolic  system $L^{SCP} = K_1 K_2 + K_2 K_1 + K_2 M + M K_2$,
{\it 5.} Elliptic Parabolic system $L^{EP} = (K_1 + M)^2 + \gamma K_2^2$,  $\gamma > 0$,
{\it 6.} Hyperbolic Parabolic system $L^{HP} = (K_1 + M)^2 - \gamma K_2^2$, $\gamma > 0$,
{\it 7.} Elliptic system $L^{E} = M^2 + \alpha K_2^2$,  $\alpha \in R$,
{\it 8.} Hyperbolic system $L^{H} = K_2^2- \alpha M^2$, $0 < \alpha < 1$,
{\it 9.} Semi-Hyperbolic system $L^{SH} = M K_1 + K_1 M + \alpha K_2^2$,  $0 < \alpha < \infty$.


Conventionally, these nine coordinate systems can be divided into three classes.
Pseudo-spherical, equidistant and horocyclic systems correspond to the group
reductions $SO(2,1)\supset SO(2)$,  $SO(2,1) \supset SO(1,1)$ and 
$SO(2,1)\supset E(1)$ respectively and belong to the first class.
The procedure for separating variables leads  to one-dimensional Schrödinger equation, in which  the quantity $-\frac12 (\sigma+\frac12)^2$ plays the role of energy, and $\lambda$ acts as a coupling constant. Solutions of the LB equation related to $L^{S}$, $L^{EQ}$ and $L^{HO}$ are studied in many works, see for example Refs. \onlinecite{PW1}, \onlinecite{MUKUNDA},  \onlinecite{VERDIYEV}.

The second class of coordinates includes three systems of non-subgroup type: semi-circular parabolic,
elliptic parabolic and hyperbolic parabolic. Two of  them,  namely EP and HP coordinate systems contain a dimensionless parameter $\gamma$.  In the limiting case $\gamma \to 0$ and $\gamma \to \infty$  they recover the horocyclic and equidistant systems of coordinates  \cite{POG-YAKH4}.  
 It is interesting to note that  parameter 
$\gamma $ {is not included} in the Laplace-Beltrami operator and, consequently,  in separated equations. Typically $\gamma=1$ is chosen for simplicity, with the exception of cases of the limit transitions of the eigenfunctions of the operator $L^{EP}$ and its contractions.
Some results concerning non-normalized eigenfunctions of the operators $L^{SCP}$, $L^{EP}$ and $L^{HP}$, as well as some examples of interbasis expansions, were presented in Ref. \onlinecite{KAL-MIL1} (see also \onlinecite{GROab} and \onlinecite{GROSCHE-BOOK1}).

The last three systems: elliptic, hyperbolic and semi-hyperbolic contain dimensional parameter $\alpha$ and form the third class. The LB equation is separated into ordinary differential equations {with parameter} $\alpha$. Unlike the previous classes, solutions in these coordinate systems cannot be represented in terms of usual special functions and are expressed through  three-term recurrent relations.  The corresponding eigenfunctions of $L^{E}$, $L^{H}$ and 
$L^{SH}$ are investigated in articles  \onlinecite{KAL-MIL1, MACWI, PAWI}.

In the presented paper (IIA) (and in the next one, which we call (IIB)) instead of study of (\ref{HE1})
through the standard one-dimensional model \cite{KAL-MIL1} or within path integral approach \cite{GROab}, we directly use the LB equation and the interbasis expansions to calculate an orthonormal and complete set of eigenfunctions corresponding to the first two classes of coordinates.  We calculate the coefficients of interbasis expansions between various eigenfunctions of the Laplace-Beltrami operator  (\ref{ALGEBRA4}). Many of them are expressed in simple form and are represented in terms of gamma functions or classical Wilson-Racah polynomials, or in the form of an exponential and  Bessel functions. 

We analyze the eigenfunctions of Eq. (\ref{HE1}) by calculating the coefficients of interbasis expansions for subgroup-type coordinates. The cases of non-subgroup coordinates will be discussed in detail in the paper (IIB). Special attention is paid to the contractions from $H_2^{0}$ to the corresponding basis on the Euclidean plane $E_2$. 

Sec. \ref{sec: Subgroup bases} is devoted to the description of coordinate systems and normalized wave functions. Interbasis expansions between subgroup bases are calculated in Sec. \ref{INTER-BASIS-1}. Some of their properties are discussed. The contraction limit is realized in Sec. \ref{sec:CONT} for both wave functions and overlap coefficients. 


Finally we note that the study of solutions to the LB equation on a two-dimensional hyperboloid, in addition to its explicit application in the theory of special functions (for example, establishing new identities between special functions) can serve as the basis for constructing a unitary irreducible representation of the group $SO(2,1)$, different from the eigenvalues $m$ of the operator $i M$. The obtained results can be used in modern physics.

\section{Subgroup basis on $H_2^{+}$ hyperboloid}
\label{sec: Subgroup bases}

\subsection{Horocyclic wave functions}
\label{sec:Horocyclic Wave Functions}

The { horocyclic} coordinate system has the form \cite{POG-YAKH4} 
\begin{equation}
\label{sys_horicyclic}
u_0=R{\tilde x^2+ \tilde y^2+1\over 2\tilde y},\qquad
u_1=R{\tilde x^2+ \tilde y^2-1\over 2\tilde y},\qquad
u_2=R{\tilde x\over \tilde y},
\end{equation}
where $\xi^{1} = \tilde y>0$, $\xi^{2} = \tilde x\in\mathbb{R}$. In this case $g_{11} = g_{22} = R^2/ \tilde y^2$ and the area element is $d {\rm s}  = 
\frac{R^2}{{\tilde{y}^2}}\, d\tilde x d\tilde y$.  The couple of equations (\ref{COM-HE1}) is given by
\begin{equation}
\label{0-MACDONALD-1}
R^2 \Delta_{LB} \Psi^{HO}   =
\tilde y^2\left(\frac{\partial^2 \Psi^{HO} }{\partial\tilde x^2} +
\frac{\partial^2 \Psi^{HO} }{\partial\tilde y^2}\right) =  \sigma(\sigma + 1) \Psi,
\qquad
L^{HO} \Psi^{HO}   =  \frac{\partial^2 \Psi^{HO}}{\partial \tilde{x}^2} = - s^2  \Psi^{HO} ,
\end{equation}
where $s^2$ is a separation constant.  Substituting  $\Psi^{HO}_{\rho s}(\tilde{y},\tilde{x})$ of the following form 
\begin{equation}
\label{01-MACDONALD-1}
\Psi^{HO}_{\rho s}(\tilde{y},\tilde{x}) = N_{\rho s}\, \psi_{\rho s}(\tilde{y}) \,
\frac{e^{i s \tilde{x}}}{\sqrt{2\pi}},
\qquad
s \in  {\mathbb R}\setminus\{0\},
\end{equation}
where $N_{\rho s}$  denotes the normalization constant,  into the left equation 
(\ref{0-MACDONALD-1})  and applying the change of variable $\tilde{y} = 
e^{-a}$,  $a\in\mathbb{R}$, $\psi_{\rho s} = \psi(a) \exp(-a/2)$, we obtain the following equation
\begin{equation}
\label{02-MACDONALD-1}
\frac{d^2 \psi}{d a^2}   +  \left[- \left(\sigma+\frac12\right)^2  - \frac{s^2}{e^{2a}}\right] \psi = 0.
\end{equation}
The above equation describes the quantum motion in repulsive potential 
$V^{HO}(a) =  \frac{s^2}{2}  e^{-2a}$ with the separation constant $s^2$ as a coupling 
constant  and energy $E := - \frac12 (\sigma+\frac12)^2$. 

The eigenfunction $\psi(a)$ is square-integrable as $a \to \infty$ and oscillates as $a \to - \infty$.    
The spectrum of energy $E$ { is pure continuous, covering the positive axis}.  
It follows that  $\sigma$  is  a  complex quantity.  For further research it is more convenient to introduce { real parameter} $\rho$  according to
\begin{equation}
\label{102-MACDONALD-1}
\sigma = -\frac12 + i\rho,
\qquad
\rho = \sqrt{2E} \geq 0,
\quad\qquad
E \geq 0 .
\end{equation}
The eigenvalue of  energy ${\cal E}$ in (\ref{HE1}) is determined by formula  
${\cal E} = \frac{1}{2R^2} \left(\rho^2 + \frac14\right)$ which coincides with the formula  obtained in the book \onlinecite{GROab}. 
We restricted  $\rho$ to be nonnegative since the both signs are equivalent for Eq. (\ref{02-MACDONALD-1}). In what follows we will use the parameter $\rho$ everywhere instead of $\sigma$.

Applying now the transformation  $\psi_{\rho s}(\tilde{y}) = \sqrt{\tilde{y}}\, w_{\rho s} (\tilde{y})$ in  (\ref{01-MACDONALD-1}) we arrive to the Bessel type equation \cite{BE2}
\begin{equation}
\label{MACDONALD-1}
\tilde{y}^2 w_{\rho s}^{\prime\prime} +  \tilde{y} w_{\rho s}^{\prime} +
\left(\rho^2 - s^2 \tilde{y}^2 \right) w_{\rho s} =0.
\end{equation}
The general solution is a linear combination of the modified Bessel
function of the first kind $I_{i\rho}(|s|\tilde{y})$ and of the second kind (MacDonald function) $K_{i\rho}(|s|\tilde{y})$
\begin{equation}
\label{MACDONALD-2}
w_{\rho s} (\tilde{y})  = A K_{i\rho}(|s|\tilde{y}) + B I_{i\rho}(|s|\tilde{y}).
\end{equation}
Function $I_{i\rho}(|s|\tilde{y})$ is given by (12) 7.2.2.\cite{BE2}
\bea
\label{000-MACDONALD-4}
I_{i\rho}(|s|\tilde{y})
=
\frac{(|s|\tilde{y}/2)^{i\rho}}
{\Gamma(1+i\rho)}\, {_0F_1} \left(1+i\rho; \frac{|s|^2 \tilde{y}^2}{4} \right)
=
\frac{(|s|\tilde{y}/2)^{i\rho} e^{-|s|\tilde{y}}}
{\Gamma(1+i\rho)}\, {_1F_1} \left(\frac12+i\rho, 1 + 2i\rho;  2|s|\tilde{y} \right).
\eea
The MacDonald function $K_{i\rho}(|s|\tilde{y})$ (for $\rho\not=0$)  can be written as an
hypergeometric series in increasing degrees of  variable $|s|\tilde{y}$  ((13) 7.2.2.\cite{BE2})
\begin{eqnarray}
K_{i\rho}(|s|\tilde{y}) &=& \frac{\pi \left[I_{-i\rho}(|s|\tilde{y}) - I_{i\rho}(|s|\tilde{y})\right]}{2i \sinh \pi\rho}
= \frac{\pi e^{-|s|\tilde{y}}}{2i \sinh \pi\rho}
\left[
\frac{\left(|s|\tilde{y}/2\right)^{-i\rho}}{\Gamma(1-i\rho)}\,
{_1}F_1\left(\frac{1}{2} - i\rho, 1 - 2 i\rho; 2|s|\tilde{y}\right)
\right. - 
\nonumber \\[2mm]
&-&
\left.
\frac{\left(|s|\tilde{y}/2\right)^{i\rho}}{\Gamma(1+i\rho)}\,
{_1}F_1\left(\frac{1}{2} + i\rho, 1+2 i\rho;  2|s|\tilde{y}\right)
\right],
\label{00-MACDONALD-4}
\end{eqnarray}
or in decreasing degrees  (see 6.9.1.(14) and 6.9.(5) from Ref. \onlinecite{BE1}, $s \ne 0$)
\bea
\label{01-MACDONALD-4}
K_{i\rho}(|s|\tilde{y})
=
\sqrt{\frac{\pi}{2|s|\tilde{y}}} \,
e^{-|s|\tilde{y}}\,
{_2F_0} \left(\frac12+i\rho, \frac12-i\rho; \,  - \frac{1}{2|s|\tilde{y}}
\right).
\eea

The asymptotics at $\tilde{y} \sim \infty$ for functions $K_{i\rho}(|s|\tilde{y})$  and   $I_{i\rho}(|s|\tilde{y})$ follow from (\ref{01-MACDONALD-4}), (\ref{000-MACDONALD-4}) and formula (3) from 6.13.1\cite{BE1} :
\begin{equation}
\label{MACDONALD-3}
K_{i\rho}(|s|\tilde{y}) \sim \sqrt{\frac{\pi}{2|s|\tilde{y}}}\, e^{-|s|\tilde{y}},
\qquad
I_{\nu}(|s|\tilde{y})  \sim \frac{1}{\sqrt{2\pi |s|\tilde{y}}}\, e^{|s|\tilde{y}}.
\end{equation}
Hence, the requirement of quadratic integrability of wave function (\ref{MACDONALD-2})
at $\tilde{y}\sim \infty$ leads to $B=0$. Therefore the regular solution of equation (\ref{MACDONALD-1})
is given by 
\begin{equation}
\label{00-MACDONALD-3}
\psi_{\rho s}(\tilde{y}) = \sqrt{|s| \tilde{y}} \, K_{i\rho}(|s|\tilde{y}).
\end{equation}
From (\ref{00-MACDONALD-4}) it follows immediately that $K_{i\rho}(|s|\tilde{y})  =  K_{- i\rho}(|s|\tilde{y})$, i.e. the solution (\ref{00-MACDONALD-3}) is real. 
The behavior of function $K_{i\rho}(|s|\tilde{y})$ at  $\tilde{y} \sim 0$,  
can be easily obtained with the help of (\ref{00-MACDONALD-4}):   
\begin{equation}
\label{AS-MACDONALD-3}
K_{i\rho}(|s|\tilde{y}) \sim  \, 
\sqrt{\frac{\pi}{\rho \sinh \pi\rho}}
\,\,
\cos\left(\rho\ln\frac{|s| \tilde{y}}{2} +  \mathrm{arg}\left\{ \Gamma(-i\rho)\right\}\right).
\end{equation}
Thus, the wave function  $\psi_{\rho s} (\tilde{y}) \sim 0$ in the both limits $\tilde{y} \sim 0$ and $\tilde{y} \sim \infty$ (see Fig. \ref{fig:yK_HO}).

\begin{center}
\begin{minipage}{0.5\textwidth}
\includegraphics[width=\textwidth]{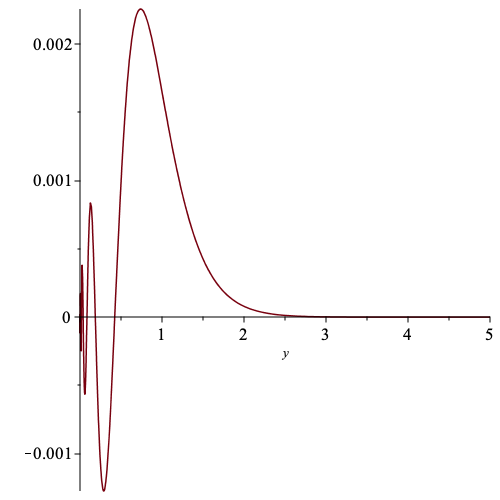}
\captionof{figure}{\small Wave function $\sqrt{\tilde{y}} K_{i\rho}(|s|\tilde{y})$ for $\rho = s = 4$.}
\label{fig:yK_HO}
\end{minipage}
\end{center}

Mutual orthogonality of wave functions (\ref{01-MACDONALD-1}), for different $s$,  
is provided by exponential functions $e^{i s\tilde{x}}$. The eigenfunctions $\psi_{\rho s}(\tilde{y})$ satisfy orthogonality and completeness conditions given in Appendix \ref{bsection: Completeness of HO}.  Therefore, 
horocyclic wave functions $\Psi^{HO}_{\rho s}(\tilde{y},\tilde{x})$
\begin{equation}
\label{Psi_HO}
\Psi^{HO}_{\rho s}(\tilde{y},\tilde{x}) = N_{\rho s}
\, \sqrt{|s| \tilde{y}} \,
K_{i\rho}(|s|\tilde{y}) \, \frac{e^{i s \tilde{x}}}{\sqrt{2\pi}},
\qquad
N_{\rho s} = \frac{1}{R \pi} \,
\sqrt{\frac{2 \rho \sinh \pi \rho}{|s|}}
\end{equation}
form an orthonormal and complete set with relations:
\bea
R^2 \int\limits^{\infty}_{-\infty} d \tilde{x} \int\limits^{\infty}_{0}
\Psi^{HO\ast }_{\rho^\prime s^\prime}(\tilde{y},\tilde{x})
\Psi^{HO}_{\rho s}(\tilde{y},\tilde{x}) \frac{d \tilde{y}}{\tilde{y}^2}
= \delta(\rho - \rho^\prime)\delta(s - s^\prime),
\label{normHO}
\eea
\bea
R^2 \int\limits^{\infty}_{-\infty} d s \int\limits^{\infty}_{0}
\Psi^{ HO \ast}_{\rho s}(\tilde{y}',\tilde{x}')
\Psi^{HO}_{\rho s}(\tilde{y},\tilde{x}) d \rho =
{\tilde{y}^2} \,
\delta(\tilde{y} - \tilde{y}^\prime)
\delta(\tilde{x} - \tilde{x}^\prime).
\label{COMPLET-01}
\eea

\subsection{Pseudo-spherical basis}
\label{sec: Pseudo-spherical basis}

The {pseudo-spherical} coordinates  \cite{POG-YAKH4,WLS} are determine as
\begin{equation}
\label{spherical}
 u_0=R\cosh\tau,\qquad
 u_1=R\sinh\tau\cos\vphi,\qquad
 u_2=R\sinh\tau\sin\vphi,
\end{equation}
where $\xi^{1} = \tau \ge 0$, $\xi^{2} = \rphi$. From (\ref{spherical}) and (\ref{METRIC})
for the metric tensor and area element we have: $g_{i k} = R^2 \diag (1, \, \sinh^2\tau)$,  $d{\rm s} = R^2 \sinh \tau  d \tau d \vphi$.
The pair of equations (\ref{COM-HE1}) in  spherical system of coordinates is given by
\begin{eqnarray}
\label{1-METRIC}
\frac{\partial ^2 \Psi}{\partial\tau^2} + \coth\tau\frac{\partial \Psi}{\partial\tau} + \frac{1}{\sinh^2\tau}
\frac{\partial^2 \Psi}{\partial\varphi^2} = - \left(\rho^2 + \frac14 \right) \Psi,
\quad
L^{S} \Psi = \frac{\partial^2 \Psi}{\partial\varphi^2} = - m^2 \Psi.
\end{eqnarray}
The substitution $\Psi (\tau, \vphi) = (\sinh\tau)^{-\frac12}\, f(\tau) \, e^{i m \vphi}$ with
$m = 0, \pm 1, \pm 2, ...$ reduces  the left Eq.  (\ref{1-METRIC}) to the one-dimensional
Schr\"odinger equation
\begin{eqnarray}
\label{00-1-METRIC}
\frac{d^2 f(\tau)}{d \tau^2} +  \left(\rho^2  - \frac{m^2-\frac14}{\sinh^2\tau} \right) f(\tau) = 0,
\end{eqnarray}
which describes a quantum motion in repulsive hyperbolic centrifugal potential 
$V_m^{S}(\tau) = \frac{ m^2 -  \frac14}{2\sinh^2\tau}$, and the energy $E=  \rho^2/2$.
It  is obvious that $E \geq 0$ and the spectrum of energy is pure continuous
(see Fig. \ref{fig:v_PS}.) Note that for $m=0$  the potential
$V_m^{S}(\tau) = - {1}/{8\sinh^2\tau}$ is an attractive singular, however the wave functions form an  orthogonal and complete set (see below), so the energy spectrum is again continuous. 

To solve Eq. (\ref{1-METRIC}) we apply the anzatz
\bea
\label{0-SCHROD-SPHERIC-2}
\Psi^S_{\rho m}(\tau,\vphi) =  N_{\rho m} \, Y_{\rho m} (\tau)\, \frac{e^{i m \vphi}}{\sqrt{2\pi}},
\eea
with the consequent notation $z = \cosh \tau$. It  leads to the following differential equation
\begin{equation}
\label{SCHROD-SPHERIC-2}
(1-z^2) \frac{d^2 Y_{\rho m} }{dz^2} - 2z \frac{d Y_{\rho m} }{dz} + \left[- \left(\rho^2 + \frac{1}{4}\right)
- \frac{m^2}{1-z^2}\right] Y_{\rho m}  = 0
\end{equation}
 for the associated Legendre functions \cite{BE1}. 
Two linearly independent solutions of this equation are the first and second kind 
associated Legendre functions  $P^{\mu}_{\nu}(z)$ and  $Q^{\mu}_{\nu}(z)$ respectively ($\nu = -1/2+i\rho$, $\mu = |m|$). 
They are defined in the form of  hypergeometric functions as follows (see (3) and (5) 3.2.\cite{BE1}):
\bea
\label{SOLUTION-HP3}
P_{\nu}^{\mu} (z)
&=& \frac{1}{\Gamma(1-\mu)}
\left(\frac{z + 1}{z-1}\right)^{\frac{\mu}{2}}
{_2F_1}\left(- \nu, 1 + \nu;  1-\mu;  \frac{1-z}{2}\right),
\\[2mm]
Q_{\nu}^{\mu} (z)
&=& e^{i\mu\pi} 2^{-\nu-1}\sqrt{\pi}\, \frac{\Gamma(\nu + \mu + 1)}{\Gamma\left(\nu + \frac{3}{2}\right)}
\,  z^{-\nu-\mu-1}
(z^2 - 1)^{\frac{\mu}{2}} \times
\nonumber\\[2mm]
&\times&
{_2F_1}\left(\frac{\nu}{2} + \frac{\mu}{2}+1, \frac{\nu}{2} + \frac{\mu}{2}+\frac{1}{2}; \nu + \frac{3}{2};
 z^{-2}\right).
\eea
The Legendre functions are regular single-valued and uniquely defined in the 
region $|1-z| < 2$ and $|z| >1$ respectively 
\footnote{We follow the book \onlinecite{BE1} and use the notation $P^\mu_\nu(z)$, $Q^\mu_\nu(z)$ for the Legendre functions and {P}$^\mu_\nu(z)$, Q$^\mu_\nu(z)$ for analytic continuation in the region $z \in (-1, 1)$.}.

In case of $\mu = m\in {\mathbb Z}$ there are the following relations (see  3.3.1.\cite{BE1} (2) and (7)):
\bea
\label{LEGENDRE-1}
P_{\nu}^{m} (z) =  \frac{\Gamma(1+\nu+m)}{\Gamma(1+\nu-m)}\, P_{\nu}^{-m} (z),
\qquad
Q_{\nu}^{m} (z) = (-1)^{m} \, \frac{\Gamma(1+\nu+m)}{\Gamma(1+\nu-m)}\,
Q_{\nu}^{-m} (z).
\eea
Therefore we can choose the general solution of equation (\ref{SCHROD-SPHERIC-2})
in the form
\bea
\label{SOL.LEGENDRE-1}
Y_{\rho m} (\cosh\tau) =  A  P^{|m|}_{-1/2 + i\rho}(\cosh\tau) + B 
Q^{|m|}_{-1/2 + i\rho}(\cosh\tau), \quad A,\, B = \mathrm{const}.
\eea
From the asymptotic formulas for $z \sim 1$
\bea
\label{SCHROD-SPHERIC-3}
P^{|m|}_{-1/2 + i\rho}(z) \sim (z-1)^{\frac{|m|}{2}},
\qquad
{\mbox{and}}
\qquad
Q^{|m|}_{-1/2 + i\rho}(z) \sim
\left\{
\begin{array}{cc}
(z-1)^{-\frac{|m|}{2}}, & \quad |m| > 0,
\\
\ln (z-1), & \quad m = 0,
\end{array}
\right.
\eea
it follows that for integer $m$ only the function $P^{|m|}_{-1/2 + i\rho}(z)$ 
is square integrable at  the  point $z \sim 1$ and can be used as the regular solution at 
$\tau \sim 0$, so $B=0$. The function $P^{|m|}_{-1/2 + i\rho}(\cosh\tau)$ is real because of the special property 
that  comes from the general relation $P_{\nu}^{|m|} (z) = P_{-\nu - 1}^{|m|} (z)$.  
The orthogonality and completeness conditions of functions $P^{|m|}_{-1/2 + i\rho}(\cosh\tau)$
are special cases of a more general formulas for functions $P_{-1/2 + i\rho}^{\mu} (z)$:
\bea
\label{LEGENDRE-4}
\left|\frac{\Gamma(1/2-\mu + i\rho)}{\Gamma(i\rho)}\right|^2 \,\,
\int\limits^{\infty}_{1} P_{-1/2 + i\rho}^{\mu} (z)
P_{-1/2 + i\rho'}^{\mu} (z) \, d z = \delta(\rho-\rho'),
\eea
and
\bea
\label{LEGENDRE-5}
\int\limits^{\infty}_{0} \left|\frac{\Gamma(1/2-\mu + i\rho)}
{\Gamma(i\rho)}\right|^2 \,P_{-1/2 + i\rho}^{\mu} (z)
P_{-1/2 + i\rho}^{\mu} (y) \, d\rho = \delta(z-y).
\eea
These two relations follow from the generalized Mehler transformations (see Ref. \onlinecite{MAGNUS}, page 202, and Refs. \onlinecite{BANDER, GRO-STEINER}). Using the properties of orthogonality exponential function $e^{i m \vphi}$ on $\vphi \in [0, 2\pi)$, we get that the 
pseudo-spherical wave functions $\Psi^S_{\rho m}(\tau,\vphi)$ form a complete orthonormal  
basis 
\bea
\label{normS}
R^2 \int\limits^{\infty}_{0} \sinh\tau d\tau \int\limits^{2\pi}_{0}
\Psi^{S \ast}_{\rho^\prime m^\prime}(\tau,\vphi) \Psi^S_{\rho m}(\tau,\vphi)
d\vphi = \delta(\rho - \rho^\prime)\delta_{m m^\prime},
\\[2mm]
\label{LEGENDRE-8}
R^2 \sum\limits_{m= - \infty}^{\infty} \, \int\limits^{\infty}_{0}
\Psi^{S \ast}_{\rho m}(\tau^\prime,\vphi^\prime) \Psi^S_{\rho m}(\tau,\vphi)
d\rho
=
(\sinh\tau)^{-1}\, \delta(\tau - \tau^\prime) \delta(\vphi-\vphi^\prime),
\eea
if they are determined by (see Fig. \ref{fig:SOL_PS})
\begin{equation}
\label{sol_S}
\Psi^S_{\rho m}(\tau,\vphi) = N_{\rho m} \,
P^{|m|}_{i\rho-1/2}(\cosh\tau) \frac{e^{i m \vphi}}{\sqrt{2\pi}},
\qquad
N_{\rho m} =  \sqrt{\frac{\rho\sinh \pi\rho}{ \pi \, R^2}}
\left| \Gamma\left(\frac{1}{2}-|m| + i\rho \right)\right|,
\end{equation}
where ((1), (3) from 3.6.1\cite{BE1})
\bea
\label{00-sol_S}
P^{|m|}_{i\rho-\frac12}(\cosh\tau) 
&=&
\frac{\Gamma(\frac12 + i\rho + |m|)}{\Gamma(\frac12 + i\rho - |m|)}
 \frac{(\sinh{\tau})^{|m|}}{2^{|m|} |m|!}
\nonumber
\\[2mm]
&\times&
{_2F_1}\left(\frac{1}{2} + |m| + i\rho,
\frac{1}{2} + |m| - i\rho; 1 + |m|; - \sinh^2\frac{\tau}{2}\right).
\eea

Alternatively, relations (\ref{normS}) and (\ref{LEGENDRE-8}) can be proven
using expansions between pseudo-spherical and equidistant wave 
functions with orthogonality and completeness conditions for equidistant basis
(see details in Appendix \ref{bsection:Completeness_of_SH}).

\begin{minipage}{0.45\textwidth}
\includegraphics[width=\textwidth]{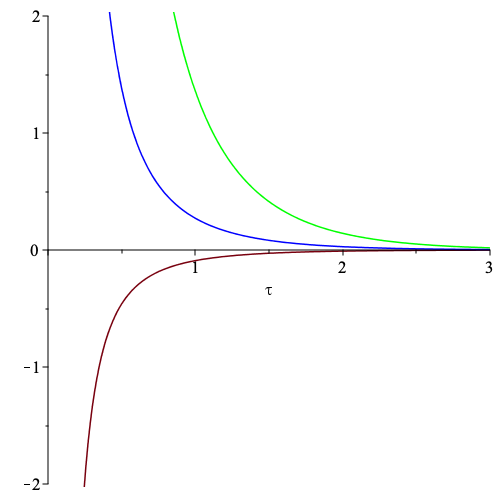}
\captionof{figure}{\small Graphics of potential $V_m^S(\tau)$ for $m = 0$ (red line), $m = 1$ (blue line) and $m = 2$ (green line).}
\label{fig:v_PS}
\end{minipage}
\hfill
\begin{minipage}{0.45\textwidth}
\includegraphics[width=\textwidth]{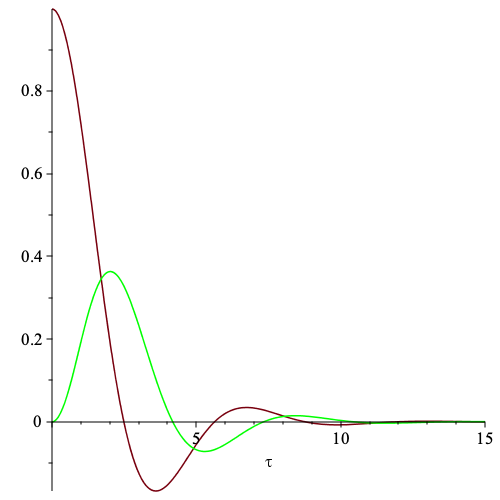}
\captionof{figure}{\small Garphics of function $N_{\rho m} P^{|m|}_{i\rho-1/2}(\cosh\tau)$ for $\rho = 1$, $R = 1$, $m = 0$ (red line) and $m = 2$ (green line).}
\label{fig:SOL_PS}
\end{minipage}

Let us note that, unlike the Legendre polynomials, the functions $P_{-1/2 + i\rho}^{|m|}(\cosh\tau)$ are not orthogonal with the measure ${dz}/{(z^2-1)}$ for index $|m|$ and satisfy the following relation for $|m'| \not= |m|$:
\bea
\label{LEGENDRE-ORTH}
&&\int\limits^{\infty}_{1} \,P_{-1/2 + i\rho}^{|m|} (z)
P_{-1/2 + i\rho}^{|m'|}(z) \, \frac{dz}{z^2-1} = 
\nonumber
\\[2mm]
&=&
\frac{i}{\pi}\, \frac{\coth\pi\rho}{|m|^2 - |m'|^2}\,
\left[ (-1)^{m}\, \frac{\Gamma\left(\frac{1}{2} - i \rho + |m|\right)}{\Gamma\left(\frac{1}{2} 
- i \rho - |m'|\right)} \right.  -  \, \left.(-1)^{m'}\,
\frac{\Gamma\left(\frac{1}{2} - i \rho + |m'|\right)}{\Gamma\left(\frac{1}{2} - i \rho - |m|\right)}
\right],
\eea
and for $|m'| = |m| \not= 0$
\bea
\label{LEGENDRE-ORTH-A}
\int\limits^{\infty}_{1} \left[P_{-1/2 + i\rho}^{|m|} (z) \right]^2 \frac{dz}{z^2-1}
=
\frac{\pi}{|m| \cosh\pi\rho} \left| \Gamma\left(\frac{1}{2} + i \rho - |m|\right) \right|^{-2}.
\eea

\subsection{Equidistant basis}
\label{sec:Equidistant Basis}

\begin{figure}[htbp]
\begin{center}
\includegraphics[scale=0.4]{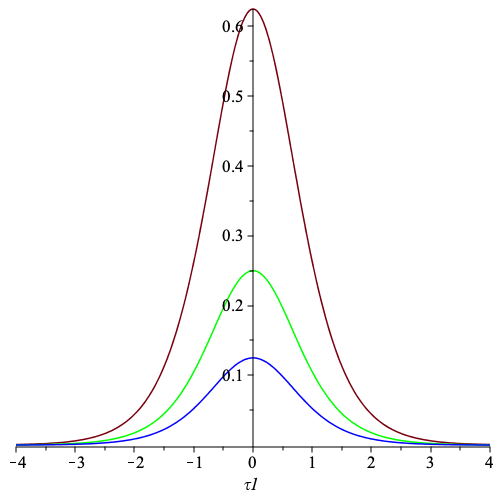}
\captionof{figure}{\small Graphics of potential $V^{EQ}_\nu(\tau_1)$ for $\nu = 0$ (blue line), $\nu = 1/2$ (green line) and $\nu = 1$ (red line).}
\label{fig:v_EQ}
\end{center}
\end{figure}

The coordinate system is the following:
\begin{eqnarray}
\label{sys_equi}
 u_0=R\cosh\tau_1\cosh\tau_2,\,\
 u_1=R\cosh\tau_1\sinh\tau_2,\,\
 u_2=R\sinh\tau_1,
\end{eqnarray}
 $\xi^1=\tau_1\in\mathbb{R}$, $\xi^2 = \tau_2\in\mathbb{R}$, $g_{ik} = R^2 {\mbox{diag}}(1, \cosh^2\tau_1)$,
$d{\rm s} = R^2 \cosh\tau_1 d\tau_1 d\tau_2$. The Laplace--Beltrami operator in equidistant coordinates is given by
\begin{equation}
\label{0-EQUID-EQ1}
\Delta_{LB} = \frac{1}{R^2} \, \left(\frac{\partial^2}{\partial\tau_1^2} +
\tanh\tau_1\frac{\partial}{\partial\tau_1} + \frac{1}{\cosh^2\tau_1}
\frac{\partial^2}{\partial\tau_2^2}\right).
\end{equation}
Equation (\ref{HE1}) is equivalent to
\begin{equation}
\label{EQUID-EQ1}
\left(\frac{\partial^2}{\partial\tau_1^2} +
\tanh\tau_1\frac{\partial}{\partial\tau_1} + \frac{1}{\cosh^2\tau_1}
\frac{\partial^2}{\partial\tau_2^2}\right) \Psi^{EQ}(\tau_1,\tau_2)
= - (\rho^2 + 1/4) \Psi^{EQ}(\tau_1,\tau_2).
\end{equation}
After the substitution $ \Psi^{EQ}(\tau_1,\tau_2)=(\cosh\tau_1)^{-\frac12}  u(\tau_1) e^{i\nu \tau_2}$,  $\nu \in  {\mathbb R}$, $L^{EQ} \Psi^{EQ}   =  {\partial^2 \Psi^{EQ}}/{\partial \tau_2^2} = - \nu^2 \Psi^{EQ}$, equation (\ref{EQUID-EQ1}) transforms to one dimensional Schr\"odinger  equation
\begin{equation}
\label{00-EQUID-EQ1}
\frac{d^2 u}{d\tau_1^2}
+ \left(\rho^2 - \frac{\nu^2 + \frac14}{\cosh^2\tau_1}\right)
u =  0,
\end{equation}
describing the movement in a field of repulsive potential $V^{EQ}_\nu(\tau_1) = \frac{\nu^2+1/4}{2 \cosh\tau_1^2}$
(see Fig. \ref{fig:v_EQ}). The spectrum of energy $E = \rho^2/2 > 0$ is purely continuous.

Applying the substitution
\begin{equation}
\label{0-EQUID-EQ1}
\Psi^{EQ}_{\rho \nu} (\tau_1,\tau_2) = N_{\rho \nu} \, \psi_{\rho \nu} (\tau_1)\, \frac{e^{i\nu\tau_2}}{\sqrt{2\pi}}
\end{equation}
in (\ref{EQUID-EQ1}) we arrive to the equation
\begin{eqnarray}
\label{EQUID-EQ3}
\frac{d^2 \psi_{\rho \nu} }{d \tau_1^2}  + \tanh\tau_1 \frac{d \psi_{\rho \nu} }{d \tau_1} +
\left(\rho^2 + \frac{1}{4} - \frac{\nu^2}{\cosh^2\tau_1}\right)
\psi_{\rho \nu} = 0.
\end{eqnarray}
The above equation is invariant under the transformation $\tau_1 \to - \tau_1$. Hence, one can choose the solutions of equation (\ref{EQUID-EQ3})  in form of even $\psi_{\rho\nu}^{(+)}$
and odd $\psi_{\rho\nu}^{(-)}$ functions  with respect to variable 
$\tau_1$.  The bases $\left\{\psi_{\rho\nu}^{(+)}, \psi_{\rho\nu}^{(-)}\right\}$
correspond to the complete set of commuting operators $\left\{\Delta_{LB}, L^{EQ}, {\cal P}\right\}$, where ${\cal P}$  is operator of parity.  The action of the  operator  ${\cal P}$ on the wave function $\psi_{\rho\nu}^{(\pm)}(\tau_1)$ is the change of the sign of variable $\tau_1 \to - \tau_1$ and ${\cal P}\psi_{\rho\nu}^{(\pm)}  = \pm \psi_{\rho\nu}^{(\pm)}$.

Application of the change $x = \tanh^2 \tau_1$ and further substitution of
$\psi (x) = (1-x)^{\frac{1}{4}+\frac{i\rho}{2}} \, w(x)$ transform Eq. (\ref{EQUID-EQ3})
to the one of hypergeometric type.  Thus,  in equidistant system of coordinates the space of solutions splits into two sets of regular wave functions at the point $x=0$ ($\tau_1 = 0$). Namely
\begin{eqnarray}
\label{H2}
\psi_{\rho\nu}^{(+)}(\tau_1) &=&
(\cosh \tau_1)^{-\frac{1}{2} - i\rho}
\,
_2F_1 \left(\frac{1}{4}+ i\frac{\rho-\nu}{2},
\frac{1}{4}+ i\frac{\rho + \nu}{2};\frac{1}{2}; \tanh^2 \tau_1 \right) = 
\nonumber\\[2mm]
&=&
(\cosh \tau_1)^{ i\nu}
\,
_2F_1 \left(\frac{1}{4} - i\frac{\rho-\nu}{2},
\frac{1}{4}+ i\frac{\rho + \nu}{2};\frac{1}{2}; - \sinh^2 \tau_1 \right),
\end{eqnarray}
\begin{eqnarray}
\label{H3}
\psi_{\rho\nu}^{(-)}(\tau_1)
&=&
\tanh \tau_1 (\cosh \tau_1)^{-\frac12 - i\rho}\,
_2F_1 \left(\frac{3}{4} + i\frac{\rho-\nu}{2},
\frac{3}{4} + i\frac{\rho+\nu}{2};\frac{3}{2}; \tanh^2 \tau_1 \right) = 
\nonumber\\[2mm]
&=&
\sinh \tau_1 (\cosh \tau_1)^{i\nu}\,
_2F_1 \left(\frac{3}{4} - i\frac{\rho-\nu}{2},
\frac{3}{4} + i\frac{\rho+\nu}{2};\frac{3}{2};  -\sinh^2 \tau_1 \right).
\end{eqnarray}
Since functions  $\psi_{\rho\nu}^{(+)}(\tau_1)$ and $\psi_{\rho\nu}^{(-)}(\tau_1)$ have 
different parities relative to variable $\tau_1$, then 
\begin{equation}
\label{EQUIDIS-NORM-3}
\int\limits^{\infty}_{-\infty}
\psi^{(\pm)}_{\rho \nu}(\tau_1) 
\psi^{(\mp)\ast}_{\rho^\prime \nu}(\tau_1)
 \cosh \tau_1 d\tau_1 = 0,
\end{equation}
and, therefore, none of the sets is complete. Wave functions (\ref{H2})
and (\ref{H3}) satisfy the following orthonormal conditions
\bea
\label{EQUIDIS-NORM-00}
\left| N_{\rho \nu}^{(\pm)}\right|^2 
R^2 
\int\limits^{\infty}_{-\infty} 
\psi^{(\pm)}_{\rho \nu}(\tau_1)
\psi^{(\pm)\ast}_{\rho^\prime \nu}(\tau_1)
\cosh \tau_1 d\tau_1
= \delta(\rho - \rho^\prime).
\eea
The normalization constants $N_{\rho \nu}^{(\pm)}$, ensuring these conditions are given by (see Appendix \ref{bsection:Completeness_of_EQ})
\begin{eqnarray}
\label{EQUIDIS-NORM-01}
N_{\rho \nu}^{(+)} =
\frac{\left|\Gamma\left(\frac{1}{4}
+  i\frac{\rho+\nu}{2}\right) \Gamma\left(\frac{1}{4}+
i\frac{\rho-\nu}{2}\right)\right|}{2\sqrt{\pi^3} R
(\rho \sinh\pi \rho)^{-1/2}},
\quad
N_{\rho \nu}^{(-)} =
\frac{\left|\Gamma\left(\frac{3}{4}
+ i\frac{\rho+\nu}{2}\right)\Gamma\left(\frac{3}{4}+
i\frac{\rho-\nu}{2}\right)\right|}{\sqrt{\pi^3} R (\rho \sinh\pi \rho)^{-1/2}}.
\end{eqnarray}
Thus, the wave functions  $\Psi^{EQ(\pm)}_{\rho \nu}(\tau_1,\tau_2)$ are orthonormal  
\bea
\label{COMPLET-00-005}
R^2 
\int\limits^{\infty}_{-\infty}\cosh \tau_1  d \tau_1
\int\limits^{\infty}_{-\infty} \Psi^{EQ(\pm)}_{\rho \nu}(\tau_1,\tau_2)
\Psi^{EQ(\pm)\ast}_{\rho' \nu'}(\tau_1, \tau_2) d \tau_2
= \delta(\rho-\rho')  \delta(\nu-\nu'),
\eea
and form the complete set with condition
\bea
R^2
\int\limits^{\infty}_{-\infty}d\nu
\int\limits^{\infty}_{0} \left[  \Psi^{EQ(+)}_{\rho \nu}(\tau_1,\tau_2)
\Psi^{EQ(+)\ast}_{\rho \nu}(\tau_1^\prime,\tau_2^\prime) \right.
&+&
\left. \Psi^{EQ(-)}_{\rho \nu}(\tau_1,\tau_2)\Psi^{EQ(-)\ast}_{\rho \nu}(\tau_1^\prime,\tau_2^\prime) \right] d\rho = 
\nonumber
\\[2mm]
&=&
\frac{1}{\cosh \tau_1} \delta(\tau_1 -  \tau_1^\prime) \delta(\tau_2 - \tau_2^\prime).
\label{COMPLET-005}
\eea
From the above condition it follows that 
\bea
R^2
\int\limits^{\infty}_{0} \left[ \left|N_{\rho \nu}^{(+)}\right|^2 \psi^{(+)}_{\rho \nu}(\tau_1)
\psi^{(+)\ast}_{\rho \nu}(\tau_1^\prime)
+
\left|N_{\rho \nu}^{(-)}\right|^2 \psi^{(-)}_{\rho \nu}(\tau_1)
\psi^{(-)\ast}_{\rho \nu}(\tau_1^\prime) \right] d\rho
=
\frac{1}{\cosh \tau_1} \delta(\tau_1 - \tau_1^\prime).
\label{COMPLET-005-1}
\eea

There is another complete set of wave functions which we label as
$\left\{\psi_{\rho\nu}^{(1)}(\tau_1), \psi_{\rho\nu}^{(2)}(\tau_1)\right\}$\footnote{This phenomena was  detailed considered in the series of articles \onlinecite{MUKUNDA}.}. To construct the explicit form of this basis we use relations of  the Legendre function on the cut 
with hypergeometric function 2.4.3\cite{MAGNUS}, p. 53-54, and obtain:
\begin{eqnarray}
\label{0-H2-1}
\psi_{\rho\nu}^{(+)}(\tau_1)
&=&
\frac{1}{2 C^{(+)}_{\rho \nu} \sqrt{\cosh \tau_1}} \,
\left({\rm P}^{- i\rho}_{-\frac12 + i\nu} (- |\tanh \tau_1|) +
{\rm P}^{- i\rho}_{-\frac12 + i\nu} (|\tanh \tau_1|)
\right),
\nonumber
\\[2mm]
\psi_{\rho\nu}^{(-)}(\tau_1)
&=&
\frac{\mathrm{sign}(\tau_1)}{2 C^{(-)}_{\rho \nu} \sqrt{\cosh \tau_1}} \,
\left({\rm P}^{- i\rho}_{-\frac12 + i\nu} (- |\tanh \tau_1|)  -
{\rm P}^{- i\rho}_{-\frac12 + i\nu} (|\tanh \tau_1|)
\right),
\nonumber
\label{1-H2-1}
\end{eqnarray}
where constants $C^{(\pm)}_{\rho \nu}$ are
\begin{eqnarray}
\label{H2-1-0}
C^{(+)}_{\rho \nu}  =  \frac{2^{- i\rho} \sqrt{\pi}}
{\Gamma\left(\frac{3}{4} + i\frac{\rho + \nu}{2}\right)
\Gamma\left(\frac{3}{4} + i\frac{\rho - \nu}{2}\right)},
\quad
C^{(-)}_{\rho \nu}  = \frac{2^{1-i\rho} \sqrt{\pi}}
{\Gamma\left(\frac{1}{4} + i\frac{\rho + \nu}{2}\right)
\Gamma\left(\frac{1}{4} + i\frac{\rho - \nu}{2}\right)},
\end{eqnarray}
and we use the signum function 
$\mathrm{sign}(\tau_1) = \{1,\, {\rm  if}\ \tau_1 > 0; -  1, \, {\rm  if}\ \tau_1 < 0;  0, \, {\rm  if}\ \tau_1 = 0\}$, 
due to the fact that $\lim\limits_{\tau_1 \to 0}\psi_{\rho\nu}^{(-)} = 0$.

Therefore the functions
\begin{eqnarray}
\label{H2-1}
\psi_{\rho \nu}^{(1,2)} (\tau_1) := C^{(+)}_{\rho \nu}
\psi_{\rho\nu}^{(+)}(\tau_1)
\pm  C^{(-)}_ {\rho \nu} \,  \psi_{\rho\nu}^{(-)}(\tau_1)
= \frac{{\rm P}^{- i\rho}_{-1/2 + i\nu} (\mp \tanh \tau_1)}{\sqrt{\cosh \tau_1}},
\end{eqnarray}
are the eigenfunctions of equation (\ref{EQUID-EQ3}). 
From the above formula it follows that $\psi_{\rho \nu}^{(1)} (\tau_1) = \psi_{\rho \nu}^{(2)} (-\tau_1)$ 
and $\psi_{\rho \nu}^{(1)} (0) = \psi_{\rho \nu}^{(2)} (0) =  {\rm P}^{- i\rho}_{-1/2 + i\nu} (0) 
=  C^{(+)}_{\rho \nu}$. 
The quality drawing of functions $\psi_{\rho\nu}^{(\pm)}(\tau_1)$ and  $\psi_{\rho\nu}^{(1,2)}(\tau_1)$ is shown in Figures \ref{fig:1a} and \ref{fig:1b} for $\rho = 1$, $\nu = 2$.
Due to relation (\ref{EQUIDIS-NORM-3}), the functions $\psi^{(1)}_{\rho\nu}$ and
$\psi^{(2)}_{\rho\nu}$ are mutually orthogonal:
\begin{eqnarray}
\label{normH-B100}
\int\limits^{\infty}_{-\infty}
{\rm P}^{-i\rho}_{-\frac12 + i\nu} (-\tanh \tau_1)\,
{\rm P}^{i{\rho^\prime}}_{-\frac12 - i\nu} (\tanh \tau_1)\,
\, d \tau_1 =  0.
\end{eqnarray}

\begin{minipage}{0.45\textwidth}
\includegraphics[width=\textwidth]{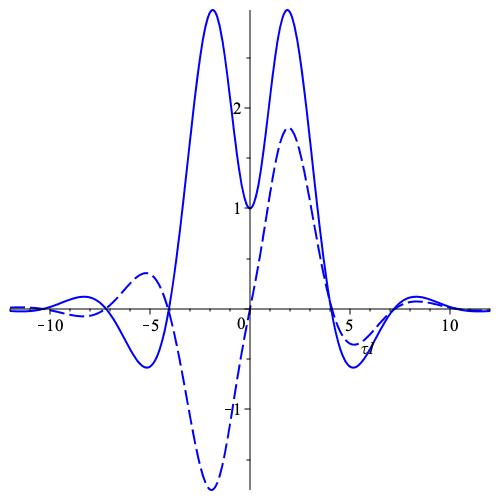}
\captionof{figure}{\small Graphics of wave functions $\psi_{\rho\nu}^{(+)}$ (solid) and $\psi_{\rho\nu}^{(-)}$ (dashed).}
\label{fig:1a}
\end{minipage}
\hfill
\begin{minipage}{0.45\textwidth}
\includegraphics[width=\textwidth]{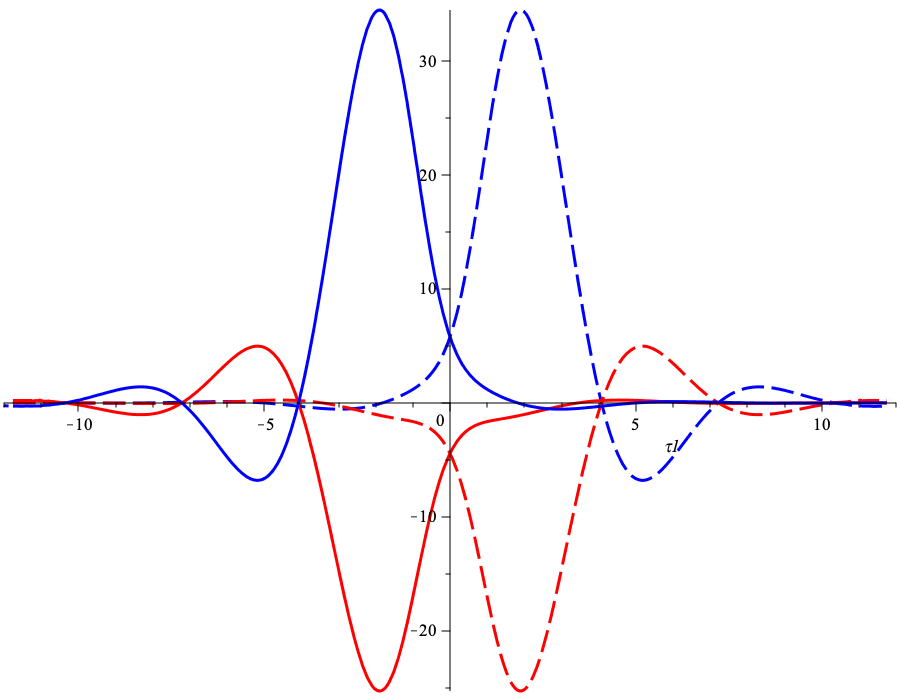}
\captionof{figure}{\small  Graphics of real parts (blue lines) of wave functions $\psi_{\rho\nu}^{(1)}$ (dashed) and $\psi_{\rho\nu}^{(2)}$  (solid). Imaginary parts of $\psi_{\rho\nu}^{(1, 2)}$ are represented by red lines.}
\label{fig:1b}
\end{minipage}

Therefore, each of the $\psi^{(1)}_{\rho\nu}$ and $\psi^{(2)}_{\rho\nu}$ functions separately does not form a complete system of functions.  
To obtain the completeness and orthonormalization of $\psi_{\rho\nu}^{(1,2)}(\tau_1)$ we can use (\ref{EQUIDIS-NORM-3}), (\ref{EQUIDIS-NORM-00}), and take into account that $|C_{\rho \nu}^{(+)}|/|N_{\rho \nu}^{(+)}| = |C_{\rho \nu}^{(-)}|/|N_{\rho \nu}^{(-)}| =: R/|N_{\rho \nu}|$.
Thus, we obtain:
\bea
 \frac{1}{2}
\int\limits^{\infty}_{0} |N_{\rho \nu}|^2 \left[ {\rm P}^{-i\rho}_{-\frac12 + i\nu}(\tanh \tau_1)
{\rm P}^{i\rho}_{-\frac12 - i\nu} (\tanh \tau_1') \right.
&+& \left.
{\rm P}^{-i\rho}_{-\frac12 + i\nu} (- \tanh \tau_1)
{\rm P}^{i\rho}_{-\frac12 - i\nu} (- \tanh \tau_1')
\right] 
d\rho = 
\nonumber
\\[2mm]
&=&
\delta(\tau_1 - \tau_1^\prime),
\label{COMPLET-005-2}
\eea
and
\be
\frac12
\int\limits^{\infty}_{-\infty}
{\rm P}^{-i\rho}_{-\frac12 + i\nu} (\pm \tanh \tau_1)
{\rm P}^{i{\rho^\prime}}_{-\frac12- i\nu} (\pm \tanh \tau_1) d\tau_1
= \frac{\delta(\rho - \rho^\prime)}{\left| N_{\rho \nu}\right|^2 },
\quad
|N_{\rho \nu}|^2 = 
\frac{\rho \sinh \pi \rho}{\sinh^2\pi\rho + \cosh^2\pi \nu}.
\label{int_N_psi1_psi2}
\ee
Finally,
\begin{eqnarray}
\label{H2-0001}
\Psi^{EQ(1,2)}_{\rho\nu}(\tau_1, \tau_2) = \frac{1}{R \sqrt{2}} 
\sqrt{\frac{\rho \sinh \pi \rho}{\sinh^2\pi\rho + \cosh^2\pi \nu}}  \,
\frac{{\rm P}^{- i\rho}_{-\frac12 + i\nu}(\mp \tanh \tau_1)}{\sqrt{\cosh \tau_1}}
\frac{e^{i\nu \tau_2}}{ \sqrt{2\pi}}
\end{eqnarray}
is an alternative form of the eigenfunctions of LB operator in equidistant coordinates.

\section{Interbasis expansions between subgroup bases}
\label{INTER-BASIS-1}

In this section we consider interbasis expansions between three sets of the subgroup wave functions. Each of the basis has the form $\Psi_{\rho \lambda}(\xi^1, \xi^2) = \psi_{\rho \lambda}(\xi^1) e^{i\lambda \xi^2}$ (where the quantum number $\lambda$ takes a discrete or continuous range of values) on $H_2^+$  and forms an orthogonal and complete set of wave functions. The functions $\Psi_{\rho \lambda}$ in an arbitrary state with a given value $\rho$  are connected to each other by unitary transformations. This fact allows us to use the orthogonality property of the exponential functions $e^{i\lambda \xi^2}$ when calculating the overlap coefficients. The method we will follow is based on the simple behavior of the eigenfunctions as $u_0 \sim R$.

{\bf 1.}  Expansion  between   horocyclic and equidistant bases is as follows 
\footnote{Below, when calculating interbasis expansions, where necessary, we will use only the functions $\Psi_{\rho \nu}^{EQ (\pm)} (\tau_1,\tau_2)$. Using functions $\Psi_{\rho \nu}^{EQ (1,2)}$ instead of $\Psi_{\rho \nu}^{EQ (\pm)}$ leads to a different form of coefficients of the interbasis expansions.}
\begin{equation}
\label{EXPANSION-02}
\Psi_{\rho s}^{HO} (\tilde{x},\tilde{y}) =
\int\limits_{-\infty}^{\infty} {\cal W}_{\rho s}^{\nu (+)} \Psi_{\rho \nu}^{EQ (+)}
(\tau_1,\tau_2) d\nu +
\int\limits_{-\infty}^{\infty} {\cal W}_{\rho s}^{\nu (-)} \Psi_{\rho \nu}^{EQ (-)}
(\tau_1,\tau_2) d\nu,
\end{equation}
and vicse versa
\begin{equation}
\label{HORIC-EQUIDIST-EXPAN-01}
\Psi_{\rho \nu}^{EQ (\pm)}(\tau_1,\tau_2) =
\int\limits_{-\infty}^{\infty} {{\cal W}_{\rho \nu}^{s (\pm)}} \Psi_{\rho s}^{HO} (\tilde{x},\tilde{y}) \, d s,
\end{equation}
where ${\cal W}_{\rho \nu}^{s (\pm)} =  {\cal W}_{\rho s}^{\nu (\pm)\ast}$. The connection between
horocyclic (\ref{sys_horicyclic}) and equidistant (\ref{sys_equi}) coordinates is given by the following relations:
\begin{eqnarray}
\label{COOR-01}
\tilde{x} =  e^{\tau_2} \tanh\tau_1,
\qquad
\tilde{y} = \frac{e^{\tau_2}}{\cosh\tau_1},
\end{eqnarray}
and
\begin{eqnarray}
\label{COOR-02}
\sinh \tau_1 = \frac{\tilde{x}}{\tilde{y}},
\qquad
\sinh \tau_2 = \frac12
\left(\sqrt{\tilde{x}^2  + \tilde{y}^2} - \frac{1}{\sqrt{\tilde{x}^2  + \tilde{y}^2} }\right).
\end{eqnarray}

{\bf 2.}  Expansion  between  equidistant and pseudo-spherical bases has the form 
\bea
\label{EQUIDIST-SPHERIC- EXPANSION-02}
\Psi_{\rho \nu}^{^{EQ}(\pm)}(\tau_1,\tau_2) =
\sum_{m = -\infty}^{\infty}  {\cal U}_{\rho \nu}^{m (\pm)}  \Psi_{\rho m}^{S} (\tau, \varphi),
\eea
and vice versa
\begin{equation}
\label{HORIC-EQUIDIST-EXPAN-03}
\Psi_{\rho m}^{S} (\tau, \varphi) =
\int\limits_{-\infty}^{\infty} {{\cal U}_{\rho m}^{\nu (+)}} \, \Psi_{\rho \nu}^{EQ (+)}
(\tau_1,\tau_2) d\nu +
\int\limits_{-\infty}^{\infty} {{\cal U}_{\rho m }^{\nu (-)}} \, \Psi_{\rho \nu}^{EQ (-)}
(\tau_1,\tau_2) d\nu,
\end{equation}
where ${\cal U}_{\rho m }^{\nu (\pm)} = {\cal U}_{\rho \nu}^{m (\pm)\ast}$.
The connection between coordinates is given by 
\be
\label{EQUIDIST-SPHERIC-03}
\sinh\tau_1 = \sinh \tau \sin\vphi,
\qquad
\tanh\tau_2 = \tanh\tau \cos\vphi, 
\ee
and
\be
\label{EQUIDIST-SPHERIC-1-03}
\cosh \tau  = \cosh \tau_1 \cosh \tau_2,
\qquad
\tan\vphi  = \frac{\tanh\tau_1}{\sinh \tau_2}.
\ee

{\bf 3.}  Expansions  between  horocyclic and pseudo-spherical bases have the form:
\bea
\label{EQUIDIST-SPHERIC- EXPANSION-04}
\Psi_{\rho s}^{HO}  (\tilde{x},\tilde{y}) =
\sum_{m = -\infty}^{\infty}  {\cal V}_{\rho s}^{m}  \Psi_{\rho m}^{S}  (\tau, \varphi),
\qquad\quad
\Psi_{\rho m}^{S}  (\tau, \varphi) =
\int\limits_{-\infty}^{\infty} {\cal V}_{\rho  m}^{s} 
\Psi_{\rho s} ^{HO} (\tilde{x},\tilde{y}) d s,
\eea
where ${\cal V}_{\rho m}^{s} = {\cal V}_{\rho s}^{m \ast}$ are the overlap coefficients.
The connection between horocyclic and spherical coordinates is
\begin{eqnarray}
\label{HORIC-SPHERICAL-01A}
\cosh\tau =  {\tilde x^2+ \tilde y^2+1\over 2\tilde y},
\qquad
\cot\vphi =  \frac{\tilde x^2+ \tilde y^2-1}{2\tilde{x}},
\end{eqnarray}
and
\begin{eqnarray}
\label{HORIC-SPHERICAL-001A}
\tilde{x} = \frac{\sinh\tau \sin\vphi}
{\cosh\tau-\sinh\tau\cos\vphi} ,
\qquad
\tilde{y} = \frac{1}{\cosh\tau-\sinh\tau\cos\vphi}.
\end{eqnarray}

\subsection{Connection between equidistant and horocyclic bases}
\label{section:EQ_HO}

Let us construct the decomposition (\ref{EXPANSION-02}) of  the horocyclic wave function $\Psi^{HO}_{\rho s}$ over the equidistant one, at the fixed values of quantum numbers $\rho$ and $s$.
First note that the transformation $\tau_1 \to - \tau_1$ is equivalent to transformation $\tilde{x} \to - \tilde{x}$.
Next, taking into account that $ \Psi_{\rho \nu}^{EQ(\pm)}(- \tau_1,\tau_2) =
\pm \Psi_{\rho \nu}^{EQ(\pm)}(\tau_1,\tau_2)$ and $\Psi^{HO}_{\rho, s} (- \tilde{x}, \tilde{y}) =
\Psi^{HO}_{\rho, \,-s} (\tilde{x}, \tilde{y})$ we can rewrite the interbasis expansion 
in form of two expansions over even and odd functions
\bea
\label{HORIC-EQUIDIST-02}
\frac{1}{2}\,
\left[\Psi_{\rho, s}^{HO} (\tilde{x}, \tilde{y}) + \Psi_{\rho, -s}^{HO}  (\tilde{x},\tilde{y})\right]
&=&
\int\limits_{-\infty}^{\infty} {\cal W}_{\rho  s}^{\nu(+)} \Psi_{\rho \nu}^{EQ (+)}
(\tau_1,\tau_2) d\nu,
\\[2mm]
\label{HORIC-EQUIDIST-03}
\frac{1}{2}\,
\left[\Psi_{\rho, s}^{HO}  (\tilde{x}, \tilde{y})  - \Psi_{\rho, -s}^{HO}  (\tilde{x},\tilde{y})\right]
&=&
\int\limits_{-\infty}^{\infty}  {\cal W}_{\rho s}^{\nu(-)} \Psi_{\rho \nu}^{EQ(-)}
(\tau_1,\tau_2) d\nu.
\eea
Substitution of the horocyclic $\Psi_{\rho, \pm s}^{HO}$ and equidistant $ \Psi_{\rho \nu}^{EQ(+)}$ wave functions into (\ref{HORIC-EQUIDIST-02}) and  the change  (\ref{HORIC-SPHERICAL-001A}), give in the limit $\tau_1 \sim 0$ ($\tilde{x} \sim 0$, $\tilde{y}\sim e^{\tau_2}$)
\be
\label{HORIC-EQUIDIST-04}
{\cal W}_{\rho s}^{\nu(+)} =
\frac{\sqrt{2/\pi}}{\left|\Gamma\left(\frac{1}{4} +  i\frac{\rho+\nu}{2}\right)
\Gamma\left(\frac{1}{4} + i\frac{\rho-\nu}{2}\right)\right|} \,
\int\limits_{-\infty}^{\infty} e^{\left(\frac{1}{2} - i\nu \right)\tau_2}K_{i\rho}(|s|
e^{\tau_2}) d\tau_2,
\ee
where we use the orthogonality of  the functions $e^{i\nu\tau_2}$ in the region $\nu \in (-\infty, \infty)$. The integral in Eq. (\ref{HORIC-EQUIDIST-04}) can be easily calculated by passing to the new variable $z = e^{\tau_2}$ and taking into account the formula (27) from 7.7.3 \cite{BE2}:
\be
\label{HORIC-EQUIDIST-05}
\int\limits_{0}^{\infty} K_{\alpha}(\beta z)  z^{\mu-1}d z = \frac{2^{\mu-2}}{\beta^{\mu}}
\, \Gamma\left(\frac{\mu+\alpha}{2}\right) \Gamma\left(\frac{\mu-\alpha}{2}\right),
\qquad \Re(\mu \pm \alpha) > 0, \quad \Re(\beta) > 0.
\ee
Thus, for interbasis coefficients we have
\be
\label{HORIC-EQUIDIST-106}
{\cal W}_{\rho s}^{\nu (+)} =
\frac{(|s|/2)^{i\nu}}{2\sqrt{\pi |s|}}\,
F^{(+)}(\rho, \nu),
\qquad\qquad
s \not= 0.
\ee
where we denote
\be
\label{HORIC-EQUIDIST-006}
F^{(+)}(\rho, \nu) :=
\sqrt{\frac{\Gamma\left(\frac{1}{4} + i \frac{\rho - \nu}{2}\right)
\Gamma\left(\frac{1}{4} - i \frac{\rho + \nu}{2}\right)}
{\Gamma\left(\frac{1}{4} - i \frac{\rho - \nu}{2}\right)
\Gamma\left(\frac{1}{4} + i \frac{\rho + \nu}{2}\right)}}.
\ee
Similarly, we can calculate the expansion coefficients $W_{\rho s}^{\nu (-)}$. The difference is that before taking $\tau_1 \sim 0$ we divide the both side of expansion (\ref{HORIC-EQUIDIST-03}) by $\tanh\tau_1$ and use $\lim\limits_{\tau_1 \to 0} {\sin(s e^{\tau_2} \tanh\tau_1)}/{\tanh\tau_1} = s e^{\tau_2}$. Thus, we get
\be
\label{HORIC-EQUIDIST-107}
{\cal W}_{\rho s}^{\nu(-)} =
\frac{i s/\sqrt{2\pi}}{\left|\Gamma\left(\frac{3}{4} +  i\frac{\rho+\nu}{2}\right)
\Gamma\left(\frac{3}{4} + i\frac{\rho-\nu}{2}\right)\right|} 
\int\limits_{-\infty}^{\infty} e^{\left(\frac{3}{2} - i\nu \right)\tau_2}K_{i\rho}(|s|
e^{\tau_2}) d\tau_2,
\ee
and after a short calculation we obtain
\bea
\label{HORIC-EQUIDIST-08}
{\cal W}_{\rho s}^{\nu(-)}
=
\frac{is}{|s|} \,
\frac{(|s|/2)^{i\nu}}{2 \sqrt{ \pi |s|}}
F^{(-)}(\rho, \nu),
\qquad\qquad
s \not= 0,
\eea
where now we denote
\be
\label{HORIC-EQUIDIST-0006}
F^{(-)}(\rho, \nu) :=
\sqrt{\frac{\Gamma\left(\frac{3}{4} +  i \frac{\rho - \nu}{2}\right)
\Gamma\left(\frac{3}{4} - i\frac{\rho + \nu}{2}\right)}
{\Gamma\left(\frac{3}{4} - i\frac{\rho - \nu}{2}\right)
\Gamma\left(\frac{3}{4} + i\frac{\rho + \nu}{2}\right)}}.
\ee
It is easy to see that $\left|F^{(\pm)}(\rho, \nu)\right|^2 = 1$,
$F^{(\pm)}(\rho,  - \nu) = F^{(\pm)*}(\rho, \nu)$  and
\bea
\label{HORIC-EQUIDIST-0-8}
F^{(-)}(\rho, \nu) F^{(+)*}(\rho, \nu) =
\left(\frac{\cosh\pi\rho - i\sinh\pi\nu}{\cosh\pi\rho + i\sinh\pi\nu}\right)^{\frac{1}{2}},
\quad
F^{(\pm)}(\rho, 0)=1.
\eea

\subsubsection{Properties of interbasis coefficients ${\cal W}_{\rho s}^{\nu (\pm)}$}
The following properties for interbasis coefficients come from equations
(\ref{HORIC-EQUIDIST-106}), (\ref{HORIC-EQUIDIST-006}), (\ref{HORIC-EQUIDIST-08}),
(\ref{HORIC-EQUIDIST-0006}) and (\ref{HORIC-EQUIDIST-0-8}):
\bea
\label{HORIC-INTER-00}
{\cal W}_{\rho,  - s}^{\nu (\pm)} = \pm {\cal W}_{\rho, s}^{\nu(\pm)},
\qquad
{\cal W}_{\rho s}^{- \nu(\pm)} = \pm {\cal W}_{\rho s}^{\nu (\pm)\ast},
\qquad
\left|{\cal W}_{\rho s}^{\nu (\pm)}\right|^2  =
\frac{1}{4 \pi |s|},
\eea
and
\bea
\label{HORIC-INTER-10}
{\cal W}_{\rho s}^{0 (+)} = \frac{1}{2\sqrt{\pi |s|}},
\qquad
{\cal W}_{\rho s}^{0 (-)} =
\frac{is}{|s|} \,
\frac{1}{2\sqrt{\pi |s|}}.
\eea
Additionally one can prove the orthogonality relations of coefficients
${\cal W}_{\rho, s}^{\nu (\pm)}$ for quantum number $\nu$ and $s$:
\be
\label{HORIC-EQUIDIST-10}
\int\limits_{-\infty}^{\infty} {\cal W}_{\rho s}^{\nu (\pm)}
{{\cal W}_{\rho  s}^{\nu^\prime (\mp)\ast}}  ds = 0,
\qquad
\int\limits_{-\infty}^{\infty} {\cal W}_{\rho s}^{\nu (\pm)}
{{\cal W}_{\rho s^\prime}^{\nu (\mp)\ast}}  d\nu = 0,
\ee
\begin{equation}
\label{HORIC-EQUIDIST-1-01}
\int\limits_{-\infty}^{\infty} {\cal W}_{\rho s}^{\nu (\pm)}
{{\cal W}_{\rho s}^{\nu^\prime (\pm)\ast }} ds = \delta(\nu - \nu^\prime),
\quad
\int\limits_{-\infty}^{\infty} \left[{\cal W}_{\rho s}^{\nu (+)}
{{\cal W}_{\rho s^\prime}^{\nu (+)\ast}}
+
{\cal W}_{\rho s}^{\nu (-)}
{{\cal W}_{\rho s^\prime}^{\nu(-)\ast}}
\right]
d\nu
= \delta\left(s - s^\prime \right).
\end{equation}
Indeed
\be
\label{HORIC-EQUIDIST-11}
\int\limits_{-\infty}^{\infty} {\cal W}_{\rho s}^{\nu (\pm)}
{{\cal W}_{\rho  s}^{\nu^\prime (\pm)}}^\ast  ds
=
F^{(\pm)}(\rho, \nu) F^{(\pm)*}(\rho, \nu^\prime) 
\frac{2^{i( \nu^\prime - \nu)}}{4\pi}
\int\limits_{-\infty}^{\infty} e^{i(\nu-\nu^\prime)\ln |s|} \frac{d s}{|s|}
= \delta(\nu - \nu\,^\prime),
\ee
and
\bea
\label{HORIC-EQUIDIST-1-ORT}
\int\limits_{-\infty}^{\infty} {\cal W}_{\rho s}^{\nu (+)}
{{\cal W}_{\rho s^\prime}^{\nu (+)}}^\ast d\nu
&=&
\frac{1}{4\pi \sqrt{|s s'|}}
\int\limits_{-\infty}^{\infty} e^{i\nu(\ln |s|- \ln |s'|)} d \nu
=
\frac{\delta(\ln |s|- \ln |s'|)}{2|s|},
\nonumber\\[2mm]
\int\limits_{-\infty}^{\infty} {\cal W}_{\rho s}^{\nu (-)}
{{\cal W}_{\rho s^\prime}^{\nu (-)}}^\ast d\nu
&=&
\frac{1}{4\pi \sqrt{|s s'|}} \frac{s s'}{|s s'|}
\int\limits_{-\infty}^{\infty} e^{i\nu(\ln |s|- \ln |s'|)} d \nu
=
\frac{s s'}{|s s'|} \frac{\delta(\ln |s|- \ln |s'|)}{2|s|},
\eea
therefore 
\bea
\int\limits_{-\infty}^{\infty} {\cal W}_{\rho s}^{\nu(\pm)}
{{\cal W}_{\rho s^\prime}^{\nu (\pm)}}^\ast d\nu = 
\frac{1}{2} \left[\delta\left(s - s^\prime \right)
\pm  \delta\left(s + s^\prime \right) \right].
\eea

Multiplying the both sides of expansion (\ref{EXPANSION-02}) on ${{\cal W}_{\rho s}^{\nu' (\pm)\ast}}$, integrating over $s$ and using the orthogonality relations (\ref{HORIC-EQUIDIST-1-01}), we get the inverse expansion (\ref{HORIC-EQUIDIST-EXPAN-01}). 
And vice versa, the formula (\ref{HORIC-EQUIDIST-1-01}) allows to get the ''direct'' expansion (\ref{EXPANSION-02}) from  (\ref{HORIC-EQUIDIST-EXPAN-01}).

\subsubsection{Particular cases}
The knowledge of  interbasis coefficients  ${\cal W}_{\rho s}^{\nu (\pm)}$  permits  us  to write 
out some interesting integral representations of MacDonald and hypergeometric function 
${_2F_1}(a,b;c; x)$.   From (\ref{HORIC-EQUIDIST-02}) and  (\ref{HORIC-EQUIDIST-03}) 
we get                                                                                                                                                                                                                                                                                                                                                                                                                                                                                                                                                                                                                                                                                                                                                                                                                                                                                                                                                                                                                                                                                                                                                                                                                                                                                                                                                                                                                                             
\bea
\label{HORIC-EQUIDIST-EXPLIC-1}
 && 4\pi \sqrt{2|s| \tilde{y}}\,   K_{i\rho}(|s| \tilde{y}) 
\cos s \tilde{x}
=
\int\limits_{-\infty}^{\infty}
\Gamma\left(\frac14 +  i\frac{\rho-\nu}{2}\right)
\Gamma\left(\frac14 -  i\frac{\rho+\nu}{2}\right)\times
\nonumber\\[2mm]
&\times&
{_2F_1}\left(\frac14 - i \frac{\rho-\nu}{2}, \frac14 +  i \frac{\rho+\nu}{2}; 
\frac12;  -\frac{\tilde{x}^2}{\tilde{y}^2} \right) 
\left[\frac{|s| (\tilde{x}^2+\tilde{y}^2)}{2\tilde{y}}\right]^{i\nu}
d \nu,
\eea
\bea
\label{HORIC-EQUIDIST-EXPLIC-2}
 && 2\pi \frac{|s|}{s} \sqrt{2|s| \tilde{y}}\,  K_{i\rho}(|s| \tilde{y}) 
\sin s \tilde{x}
=
\int\limits_{-\infty}^{\infty}
\Gamma\left(\frac34 +  i\frac{\rho-\nu}{2}\right)
\Gamma\left(\frac34 -  i\frac{\rho+\nu}{2}\right) \times
\nonumber\\[2mm]
&\times&
{_2F_1} \left(\frac34 - i \frac{\rho-\nu}{2}, \frac34 +  i \frac{\rho+\nu}{2}; 
\frac32;  -\frac{\tilde{x}^2}{\tilde{y}^2} \right) \frac{\tilde{x}}{\tilde{y}}
\left[\frac{|s| (\tilde{x}^2+\tilde{y}^2)}{2\tilde{y}}\right]^{i\nu}
d \nu.
\eea
We can also write down the inverse transformation 
\bea
\label{HORIC-EQUIDIST-EXPLIC-3}
&&
2\frac{ (2\tilde{y})^{\frac12 + i\nu}}{(\tilde{x}^2+\tilde{y}^2)^{i\nu}}
\int\limits_{0}^{\infty} \frac{\cos s \tilde{x}}{s^{\frac12 + i\nu}} K_{i\rho}(s \tilde{y})
d s = 
\nonumber\\[2mm]
&=&
\Gamma\left(\frac14 +  i\frac{\rho-\nu}{2}\right)
\Gamma\left(\frac14 -  i\frac{\rho+\nu}{2}\right)
\,
{_2F_1} \left(\frac14 - i \frac{\rho-\nu}{2}, \frac14 +  i \frac{\rho+\nu}{2}; 
\frac12;  -\frac{\tilde{x}^2}{\tilde{y}^2} \right),
\eea
which generalizes integral (\ref{HORIC-EQUIDIST-05}) and coincides with it for $\tilde{x} = 0$. In the same way one can obtain another generalization
\bea
\label{HORIC-EQUIDIST-EXPLIC-3}
&&
\frac{ (2\tilde{y})^{\frac12 + i\nu}}{(\tilde{x}^2+\tilde{y}^2)^{i\nu}}
\int\limits_{0}^{\infty} \frac{\sin s \tilde{x}}{s^{\frac12 + i\nu}} K_{i\rho}(s \tilde{y}) d s =
\nonumber\\[2mm]
&=&
\frac{\tilde{x}}{\tilde{y}}\, \Gamma\left(\frac34 +  i\frac{\rho-\nu}{2}\right)
\Gamma\left(\frac34 -  i\frac{\rho+\nu}{2}\right)
\,
{_2F_1} \left(\frac34 - i \frac{\rho-\nu}{2}, \frac34 +  i \frac{\rho+\nu}{2};
\frac32;  -\frac{\tilde{x}^2}{\tilde{y}^2} \right).
\eea

As $\nu \sim 0$ in both parts of the expansion (\ref{HORIC-EQUIDIST-EXPAN-01})
we obtain
\bea
\label{HORIC-EQUIDIST-PART-1}
\sqrt{8 \tilde{y}} \int\limits_{0}^{\infty}  K_{i\rho} (s \tilde{y}) 
\cos s\tilde{x}\, \frac{d s}{\sqrt{s}}
=
\left|\Gamma\left(\frac14 + \frac{i\rho}{2}\right)\right|^2
{_2F_1} \left(\frac14 - \frac{i\rho}{2}, \frac14 + \frac{i\rho}{2};  \frac12;  -\frac{\tilde{x}^2}{\tilde{y}^2} \right),
\eea
and 
\bea
\label{HORIC-EQUIDIST-PART-2}
\sqrt{2 \tilde{y}^3} 
\int\limits_{0}^{\infty}  K_{i\rho}\left(s \tilde{y}\right) \,
\frac{\sin s \tilde{x}}{\tilde{x}} \, \frac{d s}{\sqrt{s}}
=
\left|\Gamma\left(\frac34 + \frac{i\rho}{2}\right)\right|^2
{_2F_1} \left(\frac34 - \frac{i\rho}{2}, \frac34 + \frac{i\rho}{2};   \frac32; -\frac{\tilde{x}^2}{\tilde{y}^2}
 \right).
\eea
At a further limit $\tilde{x} \sim 0$  formulas  (\ref{HORIC-EQUIDIST-PART-1}) and  (\ref{HORIC-EQUIDIST-PART-2}) give particular cases of (\ref{HORIC-EQUIDIST-05}).

\subsection{Connection between equidistant and pseudo-spherical bases}
\label{subsection:P-EQ_SPH}
Let us now compute the coefficients of interbasis expansion ${\cal U}_{\rho \nu}^{m (\pm)}$
between equidistant and  pseudo-spherical bases (\ref{EQUIDIST-SPHERIC- EXPANSION-02}).
Firstly we multiply  the both sides of this expansions by $e^{-i m \vphi}$ and then integrate over the interval $\vphi\in [0, 2\pi)$.  Thus, the calculation of  ${\cal U}_{\rho \nu}^{m (\pm)}$ is reduced to the following expression
\be
\label{EQUIDIST-SPHERIC-07}
{\cal U}_{\rho \nu}^{m (\pm)} 
P^{|m|}_{i\rho-1/2}(\cosh\tau)
=
\frac{N_{\rho \nu}^{(\pm)}}{N_{\rho m}}
\frac{1}{2\pi}\,
\int\limits_{0}^{2\pi}
\psi_{\rho \nu}^{(\pm)}(\tau_1) e^{i\nu \tau_2}
e^{- i m \vphi} d \vphi.
\ee
Taking into account that
\bea
\label{EQUIDIST-SPHERIC-09}
(e^{\tau_2} \cosh\tau_1 )^{i\nu}  = (\cosh\tau + \sinh\tau \cos\vphi)^{i\nu}
= \sum_{k=0}^{\infty} \frac{\Gamma(-i\nu+k)}{\Gamma(-i\nu) k!}
\, \frac{(- \sinh\tau)^{k}}{(\cosh\tau)^{k-i\nu}} \, (\cos\vphi)^{k},
\eea
we obtain
\bea
\label{EQUIDIST-SPHERIC-10}
{\cal U}_{\rho \nu}^{m (+)} 
P^{|m|}_{i\rho-1/2}(\cosh\tau)
&=&
\frac{N_{\rho \nu}^{(+)}}{N_{\rho m}}
\sum_{n, k = 0}^{\infty}
(-1)^{n+k}
\frac{(-i\nu)_k (v)_n (w)_n}{(1/2)_n \, n!\, k!} \times
\nonumber
\\[2mm]
&\times&
(\sinh\tau)^{2n+k} (\cosh\tau)^{i\nu-k} A_{n m}^{k},
\eea
\bea
\label{EQUIDIST-SPHERIC-11}
{\cal U}_{\rho \nu}^{m (-)} P^{|m|}_{i\rho-1/2}(\cosh\tau)
&=&
\frac{N_{\rho \nu}^{(-)}}{N_{\rho m}}
\sum_{n, k = 0}^{\infty}
(-1)^{n+k}
\frac{(-i\nu)_k  (1/2+ v)_n (1/2+ w)_n}
{(3/2)_n\,  n!\,  k!} \times
\nonumber
\\[2mm]
&\times&
(\sinh\tau)^{2n+1+k} 
(\cosh\tau)^{i\nu-k}
B_{n m}^{k},
\eea
where we have made the following notations: $v := 1/4+ i(\nu-\rho)/2$, $w := 1/4+ i(\nu+\rho)/2$ and
\bea
\label{EQUIDIST-SPHERIC-12}
A_{n m}^{k}
&:=& \frac{1}{2\pi} \int_{0}^{2\pi} (\cos\vphi)^{k} 
(\sin\vphi)^{2n}  e^{-i m \vphi}  d \vphi,
\\[2mm]
\label{EQUIDIST-SPHERIC-13}
B_{n m}^{k}
&:=& \frac{1}{2\pi} \int_{0}^{2\pi} (\cos\vphi)^{k} 
(\sin\vphi)^{2n+1}  e^{ - i m\vphi}  d \vphi.
\eea
It is obvious that equalities (\ref{EQUIDIST-SPHERIC-10}), (\ref{EQUIDIST-SPHERIC-11})
are valid for the any point of hyperboloid including the point $\tau = 0$. The behavior of the left
side of these equalities is determined by the asymptotic of Legendre functions at $\tau \sim 0$
\be
\label{EQUIDIST-SPHERIC-14}
P^{|m|}_{-1/2 + i\rho}(\cosh\tau) \sim  \frac{\Gamma(1/2+ i\rho+|m|)}{\Gamma(1/2+i\rho-|m|)} \,
\frac{(\sinh \frac{\tau}{2})^{|m|}}{|m|!}.
\ee
Let us divide the both sides of (\ref{EQUIDIST-SPHERIC-10}) by $(\sinh\tau)^{|m|}$, and then take the limit $\tau \sim 0$. All terms with $2n + k > |m|$ go to zero. Let us consider $A_{n m}^{k}$ for $2n + k \le |m|$. We preliminary expand $(\cos\vphi)^{k}$ using the binomial formula
\bea
\label{EQUIDIST-SPHERIC-17}
(\cos\vphi)^{k} = \frac{1}{2^k}  \sum_{\ell = 0}^{k}
\frac{k!}{(k - \ell)! \ell !}  e^{i(k - 2\ell)\vphi},
\eea
and take into account  Eq. (29) from 1.5.1.\cite{BE1}
\bea
\label{EQUIDIST-SPHERIC-18}
\int\limits_{0}^{\pi} (\sin\vphi)^{\alpha} e^{i \beta \vphi} d \vphi
=
\frac{\pi}{2^{\alpha}}\,
\frac{e^{i\frac{\pi}{2} \beta} \, \Gamma(1+\alpha)}
{\Gamma\left(1+ \frac{\alpha+ \beta}{2}\right) \Gamma\left(1+ \frac{\alpha- \beta}{2}\right)},
\qquad  \Re(\alpha)> -1.
\eea
Then for the integral in each term for $A_{n m}^{k}$, we get
\be
\int\limits_{0}^{2\pi} (\sin\vphi)^{2n}  e^{i\vphi (k - 2\ell -m )}  d \vphi = (1 + (-1)^{k  - m}) \int\limits_{0}^{\pi} (\sin\vphi)^{2n}  e^{i\vphi (k - 2\ell -m )}  d \vphi, 
\ee
so $k - m = 2j$, $j \in \mathbb{Z}$ (else $A_{n m}^{k} = 0$). In such a case, $\alpha \pm \beta$ in (\ref{EQUIDIST-SPHERIC-18}) are even numbers. Therefore this integral is different from zero, if 
\bea
1+ \frac{\alpha+ \beta}{2} \ge 1 \sim 2n - m + k - 2\ell \ge 0,
\label{A_FIRST}
\\
1+ \frac{\alpha - \beta}{2} \ge 1 \sim 2n + m - k + 2\ell \ge 0.
\label{A_SECOND}
\eea
If $m \ge 0$, then (\ref{A_FIRST}) is valid, iff $k = |m| - 2n$ and $\ell = 0$. In the case $m < 0$ from (\ref{A_SECOND}) we have  $k = |m| - 2n$ and $\ell = k$. Similar conclusions can be made for integral $B_{mn}^k$. Thus we obtain
\be
\label{EQUIDIST-SPHERIC-19}
A_{n m}^{|m|-2n} = \frac{(-1)^n}{2^{|m|}},
\qquad
B_{n m}^{|m|-2n-1} = - \frac{i m}{|m|} \, A_{n m}^{|m|-2n},
\ee
and 
\bea
\label{EQUIDIST-SPHERIC-20}
{\cal U}_{\rho \nu}^{m (+)}
&=&
\frac{N_{\rho \nu}^{(+)}}{N_{\rho m}}
\frac{\Gamma(1/2+ i\rho-|m|) (-1)^{|m|} (-i \nu)_{|m|}}{\Gamma(1/2+i\rho+|m|)}\times
\nonumber
\\[2mm]
&\times&
 _4 F_3\left(
\left.\begin{array}{cccc}\frac{1}{4} + \frac{i(\nu-\rho)}{2}, & \frac{1}{4} + \frac{i(\nu + \rho)}{2},
& -\frac{|m|}{2}, &- \frac{|m|-1}{2}
\\[2mm]
\frac{1}{2}, &   \frac{1}{2} + \frac{i\nu}{2} - \frac{|m|}{2},  & 1 + \frac{i\nu}{2} - \frac{|m|}{2}
\end{array}\right| 1 \right).
\eea
\bea
\label{EQUIDIST-SPHERIC-21}
{\cal U}_{\rho \nu}^{m (-)}
&=&
 i m \frac{N_{\rho \nu}^{(-)}}{N_{\rho m}}
\frac{\Gamma(1/2+ i\rho-|m|)(-1)^{|m|} (-i\nu)_{|m| - 1}}{\Gamma(1/2+i\rho+|m|)}\times
\nonumber
\\[2mm]
&\times&
 _4 F_3\left(
\left.\begin{array}{cccc}\frac{3}{4} + \frac{i(\nu-\rho)}{2}, & \frac{3}{4} + \frac{i(\nu + \rho)}{2},
&1 - \frac{|m|}{2}, & - \frac{|m|-1}{2}
\\[2mm]
\frac{3}{2}, &   \frac{3}{2} + \frac{i\nu}{2} - \frac{|m|}{2},  & 1 + \frac{i\nu}{2} - \frac{|m|}{2}
\end{array}\right| 1 \right).
\eea
Thus, the interbasis coefficients between equidistant and pseudo-spherical bases are expressed through the generalized  balanced hypergeometric functions ${_4 F_3}(1)$ of unit argument \cite{BE1}, which are the polynomials for any integer $m$. It will be more convenient for us to work with polynomials  ${\cal U}_{\rho \nu}^{m (\pm)}$ which are further separated by even and odd values of $m$.

Using the symmetry property of ${_4 F_3}(1)$ polynomials \cite{BAILEY}
\bea
\label{EQUIDIST-SPHERIC-22}
{_4 F_3}\left(
\left.\begin{array}{cccc}
- n, \, x, \, y, \, z
\\[2mm]
u, \,   v,  \, w
\end{array}\right| 1 \right)
=
\frac{(v-z)_n (w-z)_n}{(v)_n (w)_n}
\,
{_4 F_3}\left(
\left.\begin{array}{cccc}
- n, \, u-x, \, u-y, \, z
\\[2mm]
u, \,  1-v+z-n, \, 1 - w + z - n
\end{array}\right| 1 \right)
\eea
we can rewrite the interbases coefficients ${\cal U}_{\rho \nu}^{m(\pm)}$ in form:

for even $m$
\bea
\label{EQUIDIST-SPHERIC-23}
{\cal U}_{\rho \nu}^{m (+)} 
&=&
\frac{
\left|\Gamma(\frac{1}{4} + \frac{i(\nu-\rho)}{2})
\Gamma(\frac{1}{4} + \frac{i(\nu + \rho)}{2})\right|}
{2 \pi \Gamma(\frac{1}{2}+i\rho)} 
\sqrt{\frac{\Gamma(\frac{1}{2}+i\rho-|m|)}{\Gamma(\frac{1}{2}-i\rho-|m|)}} \times
\nonumber
\\[2mm]
&\times&
{_4 F_3}\left(
\left.\begin{array}{cccc}\frac{1}{4} + \frac{i\rho}{2} - \frac{i\nu}{2},
& \frac{1}{4} + \frac{i\rho}{2} +  \frac{i\nu}{2},
& -\frac{|m|}{2}, & \frac{|m|}{2}
\\[2mm]
\frac{1}{2} , &   \frac{1}{4} + \frac{i\rho}{2},  &
\frac{3}{4}  + \frac{i\rho}{2}
\end{array}\right| 1 \right),
\label{EQUIDIST-SPHERIC-24}
\eea
\bea
{\cal U}_{\rho \nu}^{m (-)}
&=&
\frac{m \nu}{\pi}
\frac{
\left|\Gamma(\frac{3}{4} + \frac{i(\nu-\rho)}{2})
\Gamma(\frac{3}{4} + \frac{i(\nu + \rho)}{2})\right|}
{\Gamma(\frac{5}{2}+i\rho)} 
\sqrt{\frac{\Gamma(\frac{1}{2}+i\rho-|m|)}{\Gamma(\frac{1}{2}-i\rho-|m|)}}\times
\nonumber
\\[2mm]
&\times&
{_4 F_3}\left(
\left.\begin{array}{cccc}\frac{3}{4} + \frac{i\rho}{2} - \frac{i\nu}{2},
& \frac{3}{4} + \frac{i\rho}{2} +  \frac{i\nu}{2},
& - \frac{|m|}{2} +1, & \frac{|m|}{2} + 1
\\[2mm]
\frac{3}{2} , &   \frac{7}{4} + \frac{i\rho}{2},  &
\frac{5}{4}  + \frac{i\rho}{2}
\end{array}\right| 1 \right),
\eea
and for odd $m$:
\bea
\label{EQUIDIST-SPHERIC-23B}
{\cal U}_{\rho \nu}^{m (+)}
&=&
\frac{i\nu}{2\pi}
\frac{
\left|\Gamma(\frac{1}{4} + \frac{i(\nu-\rho)}{2})
\Gamma(\frac{1}{4} + \frac{i(\nu + \rho)}{2})\right|}
{\Gamma(\frac{3}{2}+i\rho)} 
\sqrt{\frac{\Gamma(\frac{1}{2}+i\rho-|m|)}{\Gamma(\frac{1}{2}-i\rho-|m|)}}\times
\nonumber
\\[2mm]
&\times&
{_4 F_3}\left(
\left.\begin{array}{cccc}\frac{1}{4} + \frac{i\rho}{2} - \frac{i\nu}{2},
& \frac{1}{4} + \frac{i\rho}{2} +  \frac{i\nu}{2},
& \frac{1}{2} - \frac{|m|}{2}, & \frac{1}{2} + \frac{|m|}{2}
\\[2mm]
\frac{1}{2} , &   \frac{3}{4} + \frac{i\rho}{2},  &
\frac{5}{4}  + \frac{i\rho}{2}
\end{array}\right| 1 \right),
\label{EQUIDIST-SPHERIC-24B}
\eea
\bea
{\cal U}_{\rho \nu}^{m (-)} 
&=&
- \frac{i m }{\pi}
\frac{
\left|\Gamma(\frac{3}{4} + \frac{i(\nu-\rho)}{2})
\Gamma(\frac{3}{4} + \frac{i(\nu + \rho)}{2})\right|}
{\Gamma(\frac{3}{2}+i\rho)} 
\sqrt{\frac{\Gamma(\frac{1}{2}+i\rho-|m|)}{\Gamma(\frac{1}{2}-i\rho-|m|)}}\times
\nonumber
\\[2mm]
&\times&
{_4 F_3}\left(
\left.\begin{array}{cccc}\frac{3}{4} + \frac{i\rho}{2} - \frac{i\nu}{2},
& \frac{3}{4} + \frac{i\rho}{2} +  \frac{i\nu}{2},
& - \frac{|m|}{2} + \frac{1}{2}, & \frac{|m|}{2} + \frac{1}{2}
\\[2mm]
\frac{3}{2} , &   \frac{3}{4} + \frac{i\rho}{2},  &
\frac{5}{4}  + \frac{i\rho}{2}
\end{array}\right| 1 \right).
\label{EQUIDIST-SPHERIC-24BB}
\eea

The obtained polynomials ${_4 F_3}(1)$ are connected to the
Wilson--Racah polynomials\cite{WILSON} (Wilson polynomials (9.1.1) \cite{KOEKOEK:2010}), and with $6j$-symbols well
known from the theory of angular momentum, or also Racah
coefficients \cite{VARSHALOVICH}. To understand the relationship we recall some facts about Wilson--Racah polynomials. They are given by the expression
\bea
\label{EQUIDIST-SPHERIC-25}
W_n (x^2)
&\equiv&
W_n(x^2; \alpha, \beta, \gamma, \delta)
=
(\alpha+\beta)_n (\alpha + \gamma)_n (\alpha+\delta)_n \times
\nonumber
\\[2mm]
&\times&
{_4 F_3}\left(
\left.\begin{array}{cccc} - n,
& \alpha+\beta+\gamma+\delta +n -1,
& \alpha - ix, & \alpha + ix
\\[2mm]
\alpha+\beta, &   \alpha+\gamma,  &
\alpha+\delta
\end{array}\right| 1 \right),
\eea
and are orthogonal with respect to the inner product (see (9.1.1)\cite{KOEKOEK:2010})
\bea
\label{EQUIDIST-SPHERIC-26}
&&\frac{1}{2\pi}
\int\limits_{0}^{\infty}  W_n (x^2) \, W_{n'} (x^2) 
\left|\frac{\Gamma(\alpha + ix)\Gamma(\beta + ix)\Gamma(\gamma + ix)
\Gamma(\delta + ix)}{\Gamma(2i x)}
\right|^2 
d x =
\nonumber
\\[2mm]
&=&
n!  (\alpha + \beta+ \gamma+ \delta +n -1)_n \Gamma(\alpha+\beta+n) 
\Gamma(\alpha+\gamma+n) \times
\nonumber
\\[2mm]
&\times&
\frac{\Gamma(\alpha+\delta+n)\Gamma(\beta+\gamma+n)
\Gamma(\beta+\delta+n)
\Gamma(\gamma+\delta+n)}
{\Gamma(\alpha+\beta+\gamma+\delta+2n)}  \delta_{n n'}.
\eea
Comparing now the hypergeometric functions in (\ref{EQUIDIST-SPHERIC-23})--\ref{EQUIDIST-SPHERIC-24BB}) with (\ref{EQUIDIST-SPHERIC-25}), we see
that for coefficients ${\cal U}_{\rho \nu}^{m(+)}$:
$\alpha=\frac{1}{4} + \frac{i\rho}{2}$, $\beta=\frac{1}{4} - \frac{i\rho}{2}$, $\gamma = \frac{1}{2}$, $\delta=0$ for even $m$, and 
$\alpha=\frac{1}{4} + \frac{i\rho}{2}$,  $\beta=\frac{1}{4} - \frac{i\rho}{2}$, $\gamma = \frac{1}{2}$, $\delta = 1$ for odd $m$;
while for coefficients ${\cal U}_{\rho \nu}^{m(-)}$: $\alpha=\frac{3}{4} + \frac{i\rho}{2}$, 
$\beta=\frac{3}{4} - \frac{i\rho}{2}$,  $\gamma = \frac{1}{2}$, $\delta=1$ for even $m$, and 
$\alpha=\frac{3}{4} + \frac{i\rho}{2}$, 
$\beta=\frac{3}{4} - \frac{i\rho}{2}$,  $\gamma = \frac{1}{2}$, $\delta = 0$ for odd $m$; $x = \nu/2$ in all cases.

Thus, the interbasis expansion coefficients ${\cal U}_{\rho \nu}^{m(+)}$ can be written in terms of Wilson-Racah  polynomials and at the same time in the form of an integral representation. For even $m$ we get
\bea
\label{EQUIDIST-SPHERIC-27}
{\cal U}_{\rho \nu}^{m(+)}
&=&
 \frac{2^{|m|}}{2\sqrt{\pi}}
\frac{\left|\Gamma\left(\frac{1}{2}+i\rho-|m|\right)
\Gamma\left(\frac{1}{4} + \frac{i(\nu-\rho)}{2}\right)
\Gamma\left(\frac{1}{4} + \frac{i(\nu + \rho)}{2}\right)\right|}{ \Gamma\left(\frac{1}{2}+ \frac{|m|}{2}\right)
\left|\Gamma\left(\frac{1}{2}+i\rho\right)\right|^2}
\nonumber
\\[2mm]
&\times&
W_{\frac{|m|}{2}} \left(\frac{\nu^2}{4}; \, \frac{1}{4} + \frac{i\rho}{2}, 
\frac{1}{4} - \frac{i\rho}{2}, \frac{1}{2},   0\right)
\nonumber
\\[2mm]
&=&
\frac{G^{(+)} (\rho, \nu) }{\pi 2^{i\rho-\frac12}}
\sqrt{\frac{\Gamma\left(\frac{1}{2}+i\rho - |m|\right)}
{\Gamma\left(\frac{1}{2}-i\rho - |m|\right)}}
\int\limits_{0}^{\infty}  \frac{\cos (|m| \arccos \tanh\mu)}{(\cosh\mu)^{i\rho + \frac12}}
\cos \nu \mu \, d\mu,
\eea
and for odd $m$
\bea
{\cal U}_{\rho \nu}^{m(+)}
&=&
 \frac{-i\nu\, 2^{|m|}}{4\sqrt{\pi}}
\frac{\left|\Gamma\left(\frac{1}{2}+i\rho-|m|\right)
\Gamma\left(\frac{1}{4} + \frac{i(\nu-\rho)}{2}\right)
\Gamma\left(\frac{1}{4} + \frac{i(\nu + \rho)}{2}\right)\right|}{\Gamma\left(\frac{|m|}{2}\right)
\left|\Gamma\left(\frac{1}{2}+i\rho\right)\right|^2}
\nonumber
\\[2mm]
&\times&
W_{\frac{|m| - 1}{2}}\left(\frac{\nu^2}{4},  \frac{1}{4} + \frac{i\rho}{2}, 
\frac{1}{4} - \frac{i\rho}{2},  \frac{1}{2},   1  \right) = 
\nonumber
\\[2mm]
\label{EQUIDIST-SPHERIC-27o}
&=&
 \frac{i G^{(+)}(\rho, \nu)}{\pi 2^{i\rho - \frac12}}
{\sqrt{\frac{\Gamma\left(\frac{1}{2} + i\rho - |m|\right)}
{\Gamma\left(\frac{1}{2} - i\rho - |m|\right)}}}
\int\limits_{0}^{\infty}  \frac{\cos (|m|\arccos \tanh\mu)}{(\cosh\mu)^{i\rho + \frac12}}
\sin \nu \mu \, d\mu,
\eea
where we denote
\bea
\label{EQUIDIST-SPHERIC-00-A27}
G^{(+)}(\rho, \nu) :=
\sqrt{\frac{\Gamma\left(\frac{1}{4} - i\frac{\rho - \nu}{2}\right)
\Gamma\left(\frac{1}{4} - i\frac{\rho + \nu}{2}\right)
}
{\Gamma\left(\frac{1}{4} + i\frac{\rho - \nu}{2}\right)\Gamma\left(\frac{1}{4} + i\frac{\rho + \nu}{2}\right)}},
\qquad
G^{(+)} G^{(+)\ast} = 1.
\eea
The representations (\ref{EQUIDIST-SPHERIC-27}) and (\ref{EQUIDIST-SPHERIC-27o}) are easy to prove, taking into account the following formulas:
\bea
\label{EQUIDIST-SPHERIC-A27}
\cos (|m| \arccos \tanh\mu)
&=&\sum\limits_{k=0}^{\frac{|m|}{2}} \frac{\left(-\frac{|m|}{2}\right)_k \left(\frac{|m|}{2}\right)_k}
{(1/2)_k \, k!} \frac{1}{(\cosh\mu)^{2k}},
\qquad
{\rm for} \  {\rm even}  \ m, 
\\[2mm]
\label{00-EQUIDIST-SPHERIC-A27}
\cos (|m| \arccos \tanh\mu)
&=& \sum\limits_{k = 0}^{\frac{|m|-1}{2}}  \frac{\left(-\frac{|m|-1}{2}\right)_k 
\left(\frac{|m|+1}{2}\right)_k}
{({3}/{2})_k  k!}  \frac{\sinh\mu}{(\cosh\mu)^{2k + 1}},
\qquad 
{\rm for}  \  {\rm odd} \ m,
\eea
 (see (11) from 2.8\cite{BE1}), their analogues for $\sin(|m| \arccos \tanh\mu)$, and (26) from 1.5.1\cite{BE1}
\bea
\label{EQUIDIST-SPHERIC-B27}
\int\limits_{0}^{\infty} \frac{\cosh 2\alpha t}{(\cosh t)^{2\beta} }
\, d t
= 4^{\beta - 1} \frac{\Gamma(\beta+\alpha)\Gamma(\beta-\alpha)}{\Gamma(2\beta)},
\qquad
\Re(\beta \pm \alpha) > 0.
\eea
Note, that using (\ref{EQUIDIST-SPHERIC-B27}) one can easily obtain the relation
\be
\label{EQUIDIST-SPHERIC-B27s}
\int\limits_{0}^{\infty} \frac{\sinh t \sinh 2\alpha t}{(\cosh t)^{2\beta} }
\, d t
= \alpha 4^{\beta - 1} \frac{\Gamma\left(\beta + \alpha -\frac12 \right) \Gamma\left(\beta-\alpha - \frac12\right)}{\Gamma(2\beta)},
\qquad
\Re (\beta \pm \alpha) > \frac12.
\ee

In the similar way, for even $m$ we get ${\cal U}_{\rho \nu}^{0(-)} = 0$, and 
\bea
{\cal U}_{\rho \nu}^{m(-)}
&=&
 \frac{m \nu\, 2^{|m|}}{8\sqrt{\pi}}
\frac{\left|\Gamma\left(\frac{1}{2}+i\rho-|m|\right)
\Gamma\left(\frac{3}{4} + \frac{i(\nu-\rho)}{2}\right)
\Gamma\left(\frac{3}{4} + \frac{i(\nu + \rho)}{2}\right)\right|}{\Gamma\left(\frac{1}{2}+ \frac{|m|}{2}\right)
\left|\Gamma\left(\frac{1}{2}+i\rho\right)\right|^2}
\nonumber
\\[2mm]
&\times&
W_{\frac{|m|}{2}-1} \left(\frac{\nu^2}{4},  \frac{3}{4} + \frac{i\rho}{2}, 
\frac{3}{4} - \frac{i\rho}{2}, \frac{1}{2},  1  \right) =
\label{EQUIDIST-SPHERIC-27A}
\\[2mm]
\label{EQUIDIST-SPHERIC-27B}
&=& \frac{m}{|m|}
\frac{G^{(-)} (\rho, \nu) }{\pi 2^{i\rho-\frac12}}
\sqrt{\frac{\Gamma\left(\frac{1}{2} + i\rho - |m|\right)}
{\Gamma\left(\frac{1}{2} - i\rho - |m|\right)}}
\int\limits_{0}^{\infty} \, \frac{\sin (|m|\arccos \tanh\mu)}{(\cosh\mu)^{i\rho + \frac12}}
\sin \nu \mu \, d\mu,  |m| = 2, 4, ...,
\nonumber
\eea
and for odd $m$
\bea
\label{EQUIDIST-SPHERIC-27Ao}
{\cal U}_{\rho \nu}^{m(-)}
&=&
 \frac{i m\, 2^{|m|}}{4\sqrt{\pi}}
\frac{\left|\Gamma\left(\frac{1}{2}+i\rho-|m|\right)
\Gamma\left(\frac{3}{4} + \frac{i(\nu-\rho)}{2}\right)
\Gamma\left(\frac{3}{4} + \frac{i(\nu + \rho)}{2}\right)\right|}{\Gamma\left(1 + \frac{|m|}{2}\right)
\left|\Gamma\left(\frac{1}{2}+i\rho\right)\right|^2}
\nonumber
\\[2mm]
&\times&
W_{\frac{|m| - 1}{2}} \left(\frac{\nu^2}{4}, \frac{3}{4} + \frac{i\rho}{2}, 
\frac{3}{4} - \frac{i\rho}{2}, \frac{1}{2},  0 \right) = 
\nonumber
\\[2mm]
&=&
\frac{- i\, m}{ |m|}
\frac{G^{(-)} (\rho, \nu) }{\pi 2^{i\rho - \frac12}}
\sqrt{\frac{\Gamma\left(\frac{1}{2} + i\rho - |m|\right)}
{\Gamma\left(\frac{1}{2} - i\rho -  |m|\right)}}
\int\limits_{0}^{\infty}  \frac{\sin(|m| \arccos \tanh\mu)}{(\cosh\mu)^{i\rho + \frac12}}
\cos \nu \mu \, d\mu,
\eea
where 
\bea
\label{EQUIDIST-SPHERIC-01-A27}
G^{(-)} (\rho, \nu) :=
\sqrt{\frac{\Gamma\left(\frac{3}{4} - i\frac{\rho - \nu}{2}\right)
\Gamma\left(\frac{3}{4} - i\frac{\rho + \nu}{2}\right)
}
{\Gamma\left(\frac{3}{4} + i\frac{\rho - \nu}{2}\right)\Gamma\left(\frac{3}{4} + i\frac{\rho + \nu}{2}\right)}},
\qquad
G^{(-)} G^{(-)*} = 1.
\eea
Let us note the general formula for integral representation of Wilson-Racah  polynomials is presented in article \onlinecite{Koornwinder:1985}. Coefficients $G^{(\pm)}$ are related to $F^{(\pm)}$ from Sec.  \ref{section:EQ_HO} as follows
\be
F^{(\pm)} G^{(\pm)} = e^{2i\phi^{(\pm)}}, \quad \phi^{(+)} =  \arg \Gamma\left(\frac{1}{4} - i\frac{\rho - \nu}{2}\right), \ \phi^{(-)} =  \arg \Gamma\left(\frac{3}{4} - i\frac{\rho - \nu}{2}\right).
\ee

It is easy to see, that in the case $\alpha^\ast = \beta$, and  $\delta, 	\gamma \in \mathbb{R}$, the Wilson-Racah polynomials are real-valued. Indeed,
\bea
W_n^\ast (x^2) &=&
(\alpha + \beta)_n (\beta + \gamma)_n (\beta+\delta)_n \times \nonumber \\
&\times&
{_4 F_3}\left(
\left.\begin{array}{cccc} - n,
& \alpha+\beta+\gamma+\delta +n -1,
& \beta + ix, & \beta - ix
\\[2mm]
\alpha+\beta, &   \beta + \gamma,  &
\beta + \delta
\end{array}\right| 1 \right) = W_n (x^2),
\eea
if  we take in (\ref{EQUIDIST-SPHERIC-22}) $u = \alpha + \beta$, $v = \alpha + \gamma$, $w = \alpha + \delta$, $y = \alpha + ix$, and use the equality $(-\beta - \delta - n + 1)_n (- \beta - \gamma - n + 1)_n = (\beta + \gamma)_n (\beta + \delta)_n$. Therefore, interbasis coefficients have the following conjugacy properties: ${\cal U}_{\rho \nu}^{m(\pm)\ast} = {\cal U}_{\rho \nu}^{m(\pm)}$, if $m$ is even; ${\cal U}_{\rho \nu}^{m(\pm)\ast} = - {\cal U}_{\rho \nu}^{m(\pm)}$, if $m$ is odd. Taking into account the simple properties of interbasis coefficients: ${\cal U}_{\rho \nu}^{-m (\pm)} = \pm  {\cal U}_{\rho \nu}^{m (\pm)}$;  ${\cal U}_{\rho, -\nu}^{m (\pm)} = \pm {\cal U}_{\rho, \nu}^{m (\pm)}$ for even $m$, and  ${\cal U}_{\rho, -\nu}^{m(\pm)} = \mp  {\cal U}_{\rho, \nu}^{m(\pm)}$ for odd one, we obtain the orthogonality relations:
\bea
\label{EQUIDIST-SPHERIC-25A}
\sum\limits_{m= -\infty}^{\infty} 
{\cal U}_{\rho \nu }^{m (\pm)}
{{\cal U}_{\rho \nu' }^{m(\mp)\ast}}  = 0,
\qquad
\int\limits_{-\infty}^{\infty}
{\cal U}_{\rho \nu }^{m (\pm)}
{{\cal U}_{\rho \nu }^{m' (\mp)\ast}}
d \nu = 0.
\eea

On the following Figures \ref{fig:U_plus_m_even} -- \ref{fig:U_minus_m_even}, one can see graphics of coefficients ${\cal U}_{\rho \nu}^{m(\pm)}$ as functions of $\nu$ for fixed value $\rho = 2$ and different values of $m$.

\begin{minipage}{0.4\textwidth}
\includegraphics[width=\textwidth]{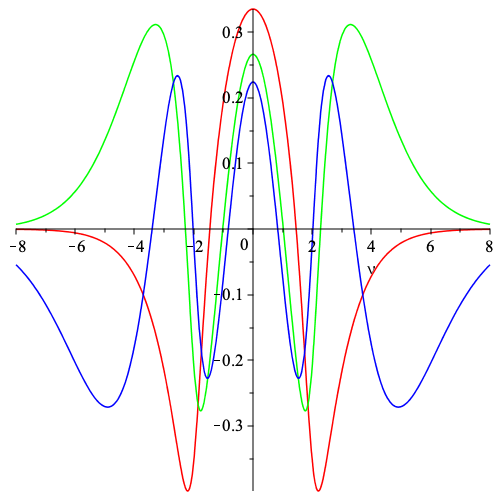}
\captionof{figure}{\small Graphics of ${\cal U}_{\rho \nu}^{m(+)}$ for $m = 2$ (red line), $m=4$ (green) and $m = 6$ (blue).}
\label{fig:U_plus_m_even}
\end{minipage}
\hfill
\begin{minipage}{0.4\textwidth}
\includegraphics[width=\textwidth]{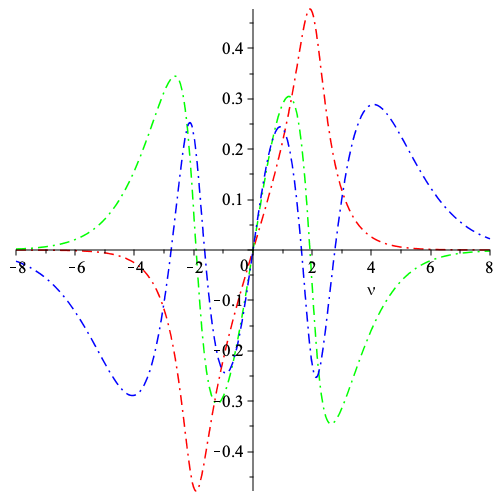}
\captionof{figure}{\small  Graphics of imaginary part of  ${\cal U}_{\rho \nu}^{m(+)}$ for $m = 1$ (red line), $m=3$ (green) and $m = 5$ (blue).}
\label{fig:U_plus_m_odd}
\end{minipage}

\begin{minipage}{0.4\textwidth}
\includegraphics[width=\textwidth]{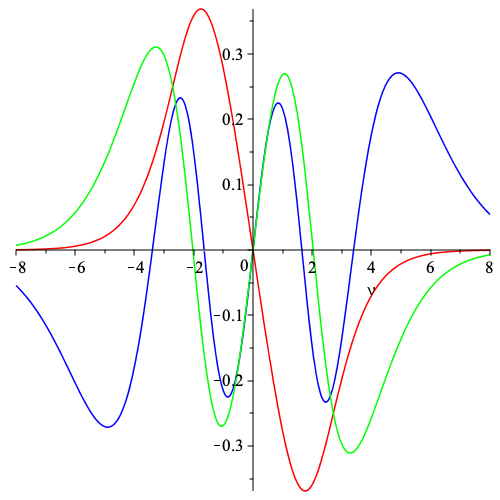}
\captionof{figure}{\small Graphics of  ${\cal U}_{\rho \nu}^{m(-)}$ for $m = - 2$ (red line), $m=4$ (green) and $m = 6$ (blue).}
\label{fig:U_minus_m_even}
\end{minipage}
\hfill
\begin{minipage}{0.4\textwidth}
\includegraphics[width=\textwidth]{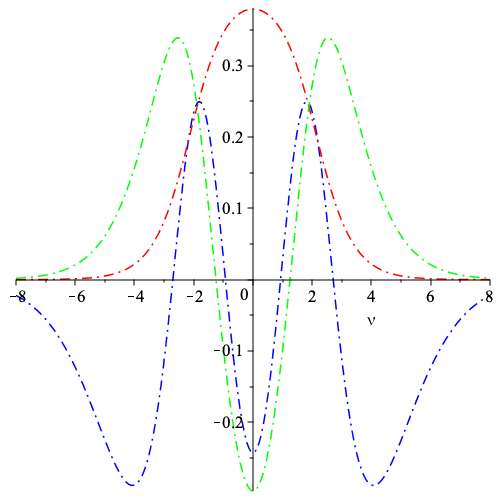}
\captionof{figure}{\small  Graphics of imaginary part of ${\cal U}_{\rho \nu}^{m(-)}$ for $m = -1$ (red line), $m=3$ (green) and $m = 5$ (blue).}
\label{fig:U_minus_m_even}
\end{minipage}

\subsubsection{Properties of the transition matrix ${\cal U}_{\rho \nu}^{m(\pm)}$
and orthogonality}
In this section, we obtain general relations that the transition matrices ${\cal U}_{\rho \nu}^{m(\pm)}$ possess, and write down the inverse expansion of the pseudo-spherical basis with respect to the equidistant one. Below we will use the integral representations obtained in the previous subsection.

{\bf 1.}  Let us consider ${\cal U}_{\rho \nu}^{m(\pm)}$ coefficients.
For even $m$ one can obtain the orthogonality relation 
\bea
\label{01-EQUIDIST-SPHERIC-0-27}
\sum\limits_{m = -\infty}^{\infty} 
{{\cal U}_{\rho \nu}^{m(+)}}
{{\cal U}_{\rho \nu'}^{m(+)\ast}}
= \frac{1}{2}\, [\delta(\nu-\nu') +  \delta(\nu+\nu')],
\eea
taking into account following identities:
\bea
\label{02-EQUIDIST-SPHERIC-01-27}
\sum\limits_{m_{\rm (even)} = -\infty}^{\infty} 
\cos( |m| \arccos \tanh\mu)\cos(|m| \arccos \tanh\mu')
= \frac{\pi \cosh\mu }{2} \delta(\mu - \mu'),
\eea
\bea
\label{02-EQUIDIST-SPHERIC-02-27}
\int\limits_{0}^{\infty} \cos\nu\mu \cos\nu'\mu\, d\mu = \frac{\pi}{2}
\left[\delta(\nu - \nu') + \delta(\nu + \nu')\right].
\eea
Similarly, one can prove for odd $m$
\bea
\label{01-EQUIDIST-SPHERIC-03-27}
\sum\limits_{m_{\rm (odd)} = -\infty}^{\infty} 
{{\cal U}_{\rho \nu}^{m(+)}}
{{\cal U}_{\rho \nu'}^{m(+)\ast}}
= \frac{1}{2} [\delta(\nu-\nu') -  \delta(\nu+\nu')].
\eea
In the same way one can calculate orthogonality conditions for coefficients ${\cal U}_{\rho \nu}^{m(-)}$. Finally,  we have
\bea
\label{01-EQUIDIST-SPHERIC-04-27}
\sum\limits_{m = -\infty}^{\infty}
{{\cal U}_{\rho \nu}^{m(\pm)}}
{{\cal U}_{\rho \nu'}^{m(\pm)\ast}}
= \delta(\nu-\nu').
\eea

{\bf 2.}
Let us now prove the following orthogonality relations for coefficients 
${\cal U}_{\rho \nu}^{m(\pm)}$:
\bea
\label{06-EQUIDIST-SPHERIC-27}
\int\limits_{-\infty}^{\infty}
{{\cal U}_{\rho \nu}^{m(+)}}
{{\cal U}_{\rho \nu}^{m' (+)\ast}}
d \nu = \frac{\delta_{m, m'} +  \delta_{m, -m'}}{2}, \quad  
\int\limits_{-\infty}^{\infty}
{{\cal U}_{\rho \nu}^{m(-)}}
{{\cal U}_{\rho \nu}^{m' (-)\ast}} 
d \nu = \frac{\delta_{m, m'} -  \delta_{m, -m'}}{2}.
\eea
Indeed, for even $m$, $m'$ and $\mu, \mu' \in [0, \infty)$ and using formula
(\ref{02-EQUIDIST-SPHERIC-02-27}) 
we get
\bea
\int\limits_{-\infty}^{\infty}
{{\cal U}_{\rho \nu}^{m(+)}}
{{\cal U}_{\rho \nu}^{m' (+)\ast}}
d \nu
&=&
\frac{2}{\pi}
\sqrt{\frac{\Gamma\left(\frac{1}{2}+i\rho - |m|\right)}
{\Gamma\left(\frac{1}{2}-i\rho - |m|\right)}}
\sqrt{\frac{\Gamma\left(\frac{1}{2}-i\rho - |m'|\right)}
{\Gamma\left(\frac{1}{2}+i\rho - |m'|\right)}} \times
\nonumber
\\[2mm]
&\times&
\int\limits_{0}^{\infty} \frac{d \mu}{\cosh\mu}
\cos (|m|\arccos \tanh\mu)
\cos (|m'| \arccos \tanh\mu).
\label{INT_ORTOGONALITY_U_+}
\nonumber
\eea
Next, making the change $\cos\phi = \tanh\mu$ and taking into account that
\bea
\frac{2}{\pi}
\int\limits_{0}^{\pi/2}
d \phi   \cos |m|\phi  \cos |m'|\phi
=
\frac{\delta_{m, m'} +  \delta_{m, -m'}}{2},
\eea
we obtain the left relation from (\ref{06-EQUIDIST-SPHERIC-27}).  By analogy, one can prove the same result for coefficients ${\cal U}_{\rho \nu}^{m(+)}$ with odd $m$, $m'$ and the right relation for ${\cal U}_{\rho \nu}^{m(-)}$ when $m$, $m'$ have the same parity.  

 Let us note, that if $m$ and $m'$ have the different parity, then in integrals  (\ref{06-EQUIDIST-SPHERIC-27}) we obtain (for example, taking $m$ even and $m'$ odd) the factor 
\be
\int\limits_{-\infty}^{\infty} 
\cos \nu \mu  \sin \nu \mu' d \nu = 0,
\ee
therefore coefficient ${\cal U}_{\rho \nu}^{m(+)}$ are ortogonal to ${\cal U}_{\rho \nu}^{m'(+)}$ if the parities of $m$ and $m'$ do not coincide. The same is true for coefficients
${\cal U}_{\rho \nu}^{m(-)}$.  Also, the relations (\ref{06-EQUIDIST-SPHERIC-27})
can be obtained from expressions (\ref{EQUIDIST-SPHERIC-27})--(\ref{EQUIDIST-SPHERIC-27Ao}) and the  orthogonality condition (\ref{EQUIDIST-SPHERIC-26}) for
Wilson-Racah polynomials. Using this formula we can rewrite the inverse expansion 
of pseudo-spherical bases over equidistant ones (\ref{HORIC-EQUIDIST-EXPAN-03}).

\subsubsection{Particular cases}

{\bf 1}. In case $\nu = 0$ we have that ${\cal U}_{\rho 0}^{m (-)} = 0$ for even $m$ and
${\cal U}_{\rho 0}^{m (+)} = 0$ for odd values of $m$. Taking the limit $\nu \sim 0$ in formulas (\ref{EQUIDIST-SPHERIC-27})
and (\ref{EQUIDIST-SPHERIC-27Ao}) and using Saalsch\"utz's  theorem (3) 4.4\cite{BE1}, 
we get
\bea
\label{BETA-100}
{\cal U}_{\rho, 0}^{m (+)} 
&=&
\frac{1}{\sqrt{2 \pi}} 
\frac{\left|\Gamma\left(\frac{1}{4} + \frac{i\rho}{2} - \frac{|m|}{2}\right)\right|}
{\left|\Gamma\left(\frac{3}{4} + \frac{i\rho}{2}- \frac{|m|}{2}\right)\right|},\quad |m| = 2n,
\\[2mm]
{\cal U}_{\rho, 0}^{m (-)} 
&=&
i \frac{\mathrm{sign}\, m}{\sqrt{2 \pi}} 
\frac{\left|\Gamma\left(\frac{1}{4} + \frac{i\rho}{2} - \frac{|m|}{2}\right)\right|}
{\left|\Gamma\left(\frac{3}{4} + \frac{i\rho}{2}- \frac{|m|}{2}\right)\right|},\quad |m| = 2n +1.
\eea
Therefore, from expansion (\ref{EQUIDIST-SPHERIC- EXPANSION-02}) we obtain as $\nu \sim 0$
\bea
\label{BETA-101}
&&
\left|\Gamma\left(\frac{1}{4} + \frac{i\rho}{2}\right)\right|^2 \,
{_2F_1} \left(\frac14- \frac{i\rho}{2}, \frac14 + \frac{i\rho}{2}; \frac12; -\sinh^2\tau\sin^2\vphi \right) = 
\nonumber\\[2mm]
&=&
2 \sum\limits_{n = 0}^{\infty} \left(1 - \frac{\delta_{n, 0}}{2}\right)
\frac{1}{4^n} \left| \Gamma\left(\frac{1}{4} + \frac{i\rho}{2} - n \right) \right|^2
P^{2n}_{- \frac12+i\rho}(\cosh\tau) \cos 2n\vphi,
\eea
and
\bea
\label{BETA-102}
&&
\left|\Gamma\left(\frac{3}{4} + \frac{i\rho}{2}\right)\right|^2 \, \sinh\tau\sin\vphi \,
{_2F_1} \left(\frac34- \frac{i\rho}{2}, \frac34 + \frac{i\rho}{2}; \frac32; -\sinh^2\tau\sin^2\vphi \right) = 
\nonumber
\\[2mm]
&=&
- \frac{1}{2} \sum\limits_{n = 0}^{\infty} \frac{1}{4^{n}} \left|\Gamma\left(- \frac{1}{4} + \frac{i\rho}{2} - n \right)\right|^2 P^{2n + 1}_{-\frac12+i\rho} (\cosh\tau) \sin (2n +1) \vphi.
\eea
Formulas (\ref{BETA-101}), (\ref{BETA-102}) can be further simplified if one chooses, for example, $\vphi = \pi/2$. 
\\
{\bf 2}. In case when $m=0$, we get that ${\cal U}_{\rho \nu}^{0 (-)} = 0$ and
\bea
\label{EQUIDIST-SPHERIC-200}
{\cal U}_{\rho \nu}^{0 (+)} 
=
\left|\Gamma\left(\frac{1}{4} + i\frac{\nu - \rho}{2}\right)
\Gamma\left(\frac{1}{4} + i\frac{\nu + \rho}{2}\right)\right|
\sqrt{\frac{\cosh\pi\rho}{4\pi^3}},
\eea
therefore expansion  (\ref{HORIC-EQUIDIST-EXPAN-03}) can be presented in the following form
\bea
\label{EQUIDIST-SPHERIC-30}
&&
P_{- \frac12 + i\rho}(\cosh\tau_1 \cosh\tau_2) =
\frac{1}{4\pi^2}
\int\limits_{-\infty}^{\infty} 
\frac{\left|\Gamma\left(\frac{1}{4} + \frac{i(\nu-\rho)}{2}\right)
\Gamma\left(\frac{1}{4} + \frac{i(\nu + \rho)}{2}\right)\right|^2}
{\left|\Gamma\left(\frac{1}{2} + i\rho\right)\right|^2 } \times
\nonumber\\[2mm]
&\times&
{_2F_1} \left(\frac14- \frac{i(\rho-\nu)}{2}, \frac14 + \frac{i(\rho+\nu)}{2};
\frac12; -\sinh^2\tau_1 \right) (\cosh\tau_1)^{i\nu}
e^{i \nu \tau_2}
d \nu.
\eea
At $\tau_1 = 0$ the expansion can be rewritten as follows
\bea
\label{EQUIDIST-SPHERIC-00-30}
P_{- \frac12 + i\rho}(\cosh\tau_2) =
\frac{1}{4\pi^2}
\int\limits_{-\infty}^{\infty} 
\frac{\left|\Gamma\left(\frac{1}{4} + \frac{i(\nu-\rho)}{2}\right)
\Gamma\left(\frac{1}{4} + \frac{i(\nu + \rho)}{2}\right)\right|^2}
{\left|\Gamma\left(\frac{1}{2} + i\rho\right)\right|^2 } e^{i \nu \tau_2}\, d \nu,
\eea
or vice versa
\bea
\label{0-EQUIDIST-SPHERIC-30}
{2\pi}
\int\limits_{-\infty}^{\infty} 
P_{- \frac12 + i\rho}(\cosh\tau_2) 
e^{- i \nu \tau_2}\,
d \tau_2
=
\frac{\left|\Gamma\left(\frac{1}{4} + \frac{i(\nu-\rho)}{2}\right)
\Gamma\left(\frac{1}{4} + \frac{i(\nu + \rho)}{2}\right)\right|^2}
{\left|\Gamma\left(\frac{1}{2} + i\rho\right)\right|^2 }.
\eea
Further simplification $\tau_2 = 0$ in (\ref{EQUIDIST-SPHERIC-00-30})  
leads to the well known formula for the Mellin-Barnes integrals (8) 1.19\cite{BE1}
\bea
\label{0-EQUIDIST-SPHERIC-30}
\int\limits_{-\infty}^{\infty} 
{\left|\Gamma\left(\frac{1}{4} + i \frac{\rho - \nu}{2}\right)
\Gamma\left(\frac{1}{4} + i \frac{\rho + \nu}{2}\right)\right|^2}
\,
d \nu
= 4\pi^2
\left|\Gamma\left(\frac{1}{2} + i\rho\right)\right|^2.
\eea

\subsection{Connection between horocyclic  and pseudo-spherical basis}
\label{subsection:HOR-SPHER}

Let us compute the interbasis coefficients ${\cal V}^m_{\rho s}$ between HO and PS basis, where 
the wave functions $\Psi^{HO}_{\rho s} (\tilde{x},\tilde{y})$ and $\Psi^S_{\rho m} (\tau, \vphi)$ are given in  (\ref{Psi_HO}) and (\ref{sol_S}). Using the orthogonality of  
$e^{i m \vphi}$ functions in interval $\vphi\in[0,2\pi)$, we obtain from the right Eq. of ({\ref{EQUIDIST-SPHERIC- EXPANSION-04})
\bea
\label{HORIC-SPHERICAL-01B}
{\cal V}_{\rho s}^m  P^{|m|}_{i\rho-1/2}(\cosh\tau)
=
\frac{1}{\sqrt{2 \pi^3}}\,
\frac{1}{|\Gamma(1/2-|m|+i\rho)|}
\int\limits_{0}^{2\pi} 
\sqrt{\tilde{y}} \, K_{i\rho}(|s|\tilde{y}) e^{i s{\tilde{x}}}
 e^{- i m\vphi} \, d \vphi.
\eea
Further calculation of ${\cal V}_{\rho s}^m $ is carried out similarly to the calculation of the coefficients ${\cal U}_{\rho \nu}^{m}$.
After the long and routine calculations we get 
\bea
\label{HORIC-SPHERICAL-05}
{\cal V}_{\rho s}^m
&=&
\frac{i (-1)^{|m|}  |m|!}{\sinh\pi\rho}
\sqrt{\frac{\pi}{2}}
\frac{|\Gamma(1/2-|m|+i\rho)|}{|\Gamma(1/2+i\rho)|^2} \times
\\[2mm]
&\times&
 \Biggl\{
\sum_{\ell=0}^{\infty} 
\frac{(|s|/2)^{2\ell + i\rho}}{\Gamma(1 + i\rho + \ell)\, \ell!} \, L_{|m|}^{1/2 + 2\ell +  i\rho} (\mp s)
-
\sum_{\ell=0}^{\infty} 
\frac{(|s|/2)^{2\ell - i\rho}}{\Gamma(1 - i\rho + \ell)\, \ell!} \, L_{|m|}^{1/2 + 2\ell -  i\rho} (\mp s)
 \Biggr\}, \nonumber
\eea
where $L_n^{\alpha} (x)$ are a Laguerre polynomials
\be
\label{HORIC-SPHERICAL-05A}
L_n^{\alpha} (x) = \frac{(\alpha+1)_n}{n!} \,
{_1F_1}(-n; \alpha+1; x),
\ee
and the sign $\mp$ corresponds to the positive and negative value of $m$. Let us note that the interbases coefficients between horocyclic and pseudo spherical bases have unexpectedly cumbersome form.

The alternative method to construct coefficients ${\cal V}_{\rho s}^m$ is based on
knowledge of transformation between horocyclic and equidistant bases, as well as between equidistant and spherical bases. This two steps transition leads to the calculation of integral
\be
\label{HORIC-SPHERICAL-06}
{\cal V}_{\rho s}^m
=
\int\limits_{-\infty}^{\infty} \left[{\cal W}_{\rho s}^{\nu (+)} 
{\cal U}_{\rho \nu}^{m (+)} + {\cal W}_{\rho s}^{\nu (-)}  {\cal U}_{\rho \nu }^{m (-)} \right]  d \nu,
\ee
which can be considered as an integral representation of coefficients
${\cal V}_{\rho s}^m$. Essentially relying on this integral representation one can obtain the following orthogonality properties for coefficients ${\cal V}_{\rho s}^m$:
\be
\label{1-HORIC-SPHERICAL-06}
\int\limits_{-\infty}^{\infty}  {\cal V}_{\rho s}^m {\cal V}_{\rho s}^{m^\prime \ast} ds
=
\delta_{m, m'},
\qquad
\sum_{m= - \infty}^{\infty} {\cal V}_{\rho s}^m {\cal V}_{\rho s'}^{m\, \ast}
=
 \delta(s-s'),
\ee
so we can write the right expansion formula (\ref{EQUIDIST-SPHERIC- EXPANSION-04}).


In particular case $m=0$ the interbasis coefficients ${\cal V}_{\rho s}^m$ are greatly simplified (it follows from (\ref{HORIC-SPHERICAL-01B}) at $\tau = 0$)
\be
\label{3-HORIC-SPHERICAL-06}
{\cal V}_{\rho s}^0  = \sqrt{\frac{2}{\pi}}
\frac{K_{i \rho} (|s|)}{|\Gamma(\frac12+i\rho)|},
\ee
therefore we obtain the decomposition
\be
\label{04-HORIC-SPHERICAL-06}
\frac{\pi^2}{4 \sqrt{\tilde{y}} \cosh\pi\rho}  
P_{-\frac12 + i\rho} \left(\frac{\tilde{x}^2 + \tilde{y}^2 + 1}{2\tilde{y}}\right)
=
\int\limits_{0}^{\infty} 
K_{i \rho} (s) 
K_{i \rho} (s \tilde{y}) 
\cos{ s \tilde{x}} \,
d s,
\ee
that coincides with the well known formula for integrals of the product of two MacDonald functions
(see 2.16.36. 2\cite{PRUDNIKOV2}).

Comparison of formula (\ref{3-HORIC-SPHERICAL-06}) with integral (\ref{HORIC-SPHERICAL-06}) gives the integral representation
\bea
K_{i\rho}(|s|) = \frac{\sqrt{\cosh\pi\rho}}{4\pi^2\sqrt{2|s|}} \int\limits_{-\infty}^{\infty} \Gamma\left(\frac{1}{4} + i\frac{\rho - \nu}{2}\right) \Gamma\left(\frac{1}{4} - i\frac{\rho + \nu}{2}\right) \left(\frac{|s|}{2}\right)^{i\nu}d\nu,
\eea
and from (\ref{HORIC-SPHERICAL-05}) we get the series representation
\bea
K_{i\rho}(|s|) = \frac{i\pi }{2 \sinh\pi\rho}
\sum_{\ell = 0}^{\infty} \frac{(|s|/2)^{2\ell}}{\ell !}   \Biggl\{
\frac{(|s|/2)^{i\rho}}{\Gamma(1 + i\rho + \ell)} 
-
\frac{(|s|/2)^{ - i\rho}}{\Gamma(1 - i\rho + \ell)}  \Biggr\}.
\eea


\section{Contractions}
\label{sec:CONT}

\subsection{Contraction of the Lie algebra $so(2,1)$ and Laplace-Beltrami operator}

To realize the contractions of Lie algebra $so(2,1)$ to $e(2)$ let us introduce the Beltrami
coordinates on the hyperboloid $H_2^{+}$ in such a way
\begin{equation}
  x_{\mu} := R\frac{u_{\mu}}{u_0} = R\frac{u_{\mu}}{\sqrt{R^2+ u_1^2 + u_2^2}},\qquad \mu = 1,2. \label{Beltrami_H2}
\end{equation}
In variables (\ref{Beltrami_H2}) generators (\ref{GEN}) look like this
\bea
\label{generators_H2}
-\frac{K_1}{R} =: \pi_2 &=& \partial_{x_2} - \frac{x_2}{R^2}(x_1\partial_{x_1} + x_2 \partial_{x_2}),
\nonumber\\[2mm]
-\frac{K_2}{R} =: \pi_1 &=& \partial_{x_1} - \frac{x_1}{R^2}(x_1\partial_{x_1} + x_2 \partial_{x_2}),
\\[2mm]
M &=&  x_1\partial_{x_2} - x_2\partial_{x_1} = x_1\pi_2 - x_2\pi_1,
\nonumber
\eea
and commutator relations of $so(2,1)$ take the form (\ref{GEN-1}):
\begin{equation}
	[\pi_1,\pi_2] = \frac{M}{R^2}, \qquad [\pi_1, M] = \pi_2, \qquad
	 [M,\pi_2] = \pi_1. 
	\label{comm_o21}
\end{equation}
Let us consider the Lie algebra $e(2) = \langle L_3, P_1, P_2\rangle$  with commutators
\begin{equation}
	[P_1, P_2] = 0, \qquad  [P_1, L_3] =  P_2 , \qquad  [L_3, P_2] =  P_1. 
	\label{comm_e2}
\end{equation}
Then in the limit $R^{-1} \sim 0$ we have $\pi_1 \sim P_1 = \partial_{x}$, $\pi_2 \sim P_2 = \partial_{y}$, $M \sim L_3 = xP_2 - yP_1$, $x,y\in\mathbb{R}$ are Cartesian coordinates on plane $E_2$. Therefore relations (\ref{comm_o21}) contract to (\ref{comm_e2}), 
so algebra $so(2,1)$ contracts  to $e(2)$.
Moreover, the $so(2,1)$ Laplace-Beltrami operator  $\Delta_{LB} = 
(K_1^2 + K_2^2 - M^2)/R^2$ contracts to the Laplace operator $\Delta$
\begin{equation}
\Delta_{LB} = \pi_1^2 + \pi_2^2 - \frac{M^2}{R^2}
\sim \Delta = P_1^2 + P_2^2.
\end{equation}
Let us now consider the analytical contractions of the subgroup coordinates, corresponding solutions and the interbasis expansions to their analogues on Euclidean plane and the Helmholtz equation  $\Delta\Psi = - k^2 \Psi$, $k > 0$.

\subsection{Contractions in pseudo-spherical bases}
\label{sec:Contractions}

In the contraction limit $R\rightarrow\infty$  the pseudo-spherical coordinates (\ref{spherical}) 
transforms \cite{POG-YAKH4} as $\tau \sim \frac{r}{R} \sim 0$, where $r$ is the radius in polar coordinates $(r, \vphi)$ on Euclidean plane $E_2$. The eigenvalues  of  operators $\Delta_{LB}$ and $L^S = M^2$ contract as $\rho\sim kR$  and $m \to m$. As a result, the Legendre differential equation 
is converted to Bessel one.  Taking into account the asymptotic formulas for gamma functions
at the large values of variable (see 1.18.\cite{BE1}  (4) and (6))
\bea
\label{Gamma_limit}
&&
\left|\Gamma(x+iy)\right| \exp\left(\frac{\pi |y|}{2}\right) |y|^{\frac{1}{2}-x} \sim
\sqrt{2\pi},
\qquad
\frac{\Gamma(z + \alpha)}{\Gamma(z + \beta)} \sim z^{\alpha - \beta},
\eea
using the explicit form of Legendre function through the hypergeometric functions (3.2.\cite{BE1} Eq. (7)), we get
in contraction limit
\bea
\label{CONTR-FUNCTION-1}
P^{|m|}_{i\rho-1/2}(\cosh\tau) &=&
\frac{\Gamma(1/2 + i\rho + |m|)}{\Gamma(1/2 + i\rho - |m|)}
 \left(\sinh{\frac{\tau}{2}}\right)^{|m|}
 \left(\cosh{\frac{\tau}{2}}\right)^{|m|}
\frac{1}{|m|!}
\nonumber\\[2mm]
&\times& {_2F_1}\left(\frac{1}{2} + |m| + i\rho,
\frac{1}{2} + |m| - i\rho; 1 + |m|; - \sinh^2\frac{\tau}{2}\right)
\nonumber\\[2mm]
&\sim& \frac{(- k^2 R r)^{|m|}}{2^{|m|}\, |m|!}
{_0F_1}\left(;1 + |m|; - \frac{k^2 r^2}{4}\right)
= (- k R)^{|m|} J_{|m|}(kr).
\eea
Thus, for the pseudo-spherical  functions (\ref{sol_S}) in the contraction limit we have
\begin{eqnarray}
\label{CONTR-FUNCTION2}
\lim_{R\rightarrow\infty}\sqrt{R}\, \Psi_{\rho m}^S(\tau, \varphi) = (-1)^{|m|}   \sqrt{k} \, J_{|m|}(kr)
\frac{e^{im\varphi}} {\sqrt{2\pi}}.
\end{eqnarray}
The pseudo-spherical basis up to the constant factor, coming from
contraction of Dirac delta-function $R \, \delta(\rho-\rho') \rightarrow \delta(k-k')$, and the phase $(-1)^{|m|}$, 
contracts into polar one \cite{ARXIV:2025}. In Figs. \ref{fig:CONTR_PS_1}, \ref{fig:CONTR_PS_2} (see also \ref{fig:SOL_PS})) one can see how the Legendre function approaches the Bessel function with increasing $R$.

\begin{minipage}{0.45\textwidth}
\includegraphics[width=\textwidth]{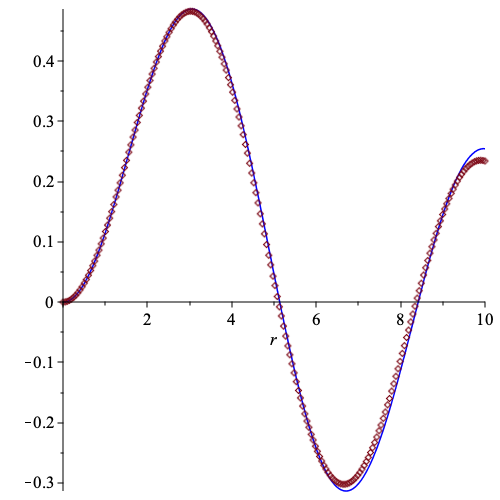}
\captionof{figure}{\small Graphics of wave function $\sqrt{R} N_{kR, m} P^{|m|}_{-1/2 + ikR}\left(\cosh\frac{r}{R}\right)$ (red points) and $(-1)^{|m|}\sqrt{k}  J_{|m|}(kr)$ (blue line), for  $R = 10$, $k=1$ and $m = 2$.}
\label{fig:CONTR_PS_1}
\end{minipage}
\hfill
\begin{minipage}{0.45\textwidth}
\includegraphics[width=\textwidth]{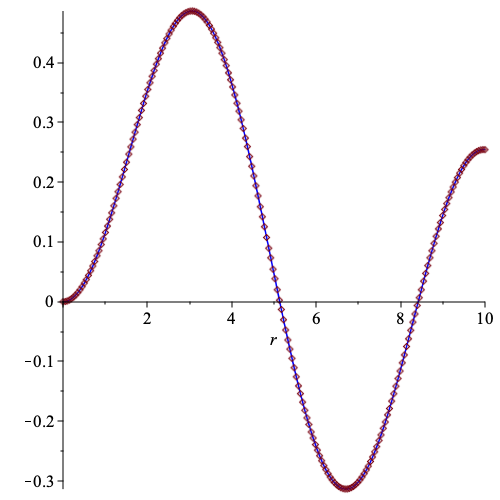}
\captionof{figure}{\small  Graphics of wave function $\sqrt{R} N_{kR, m} P^{|m|}_{-1/2 + ikR}\left(\cosh\frac{r}{R}\right)$ (red points) and $(-1)^{|m|}\sqrt{k}  J_{|m|}(kr)$  (blue line), for  $R = 100$, $k=1$ and $m = 2$.}
\label{fig:CONTR_PS_2}
\end{minipage}

Indeed taking the contraction limit  in Eqs.  
(\ref{LEGENDRE-4}) and (\ref{LEGENDRE-5}) we arrive
to the formulas 
\begin{eqnarray}
\label{BESSEL-NORM1}
\int\limits^{\infty}_{0}
 J_{|m|}(kr)  J_{|m|}(k'r)\, r dr = \frac{1}{k}  \delta(k-k'),
\end{eqnarray}
and
\begin{eqnarray}
\label{BESSEL-COMPLET1}
\int\limits^{\infty}_{0}
J_{|m|}(kr) J_{|m|}(k r')\, k dk = \frac{1}{r} 
\delta(r-r'),
\end{eqnarray}
which are the well-known orthogonality and completeness relations for Bessel functions.

From condition (\ref{LEGENDRE-ORTH}) we obtain the following equality 
\begin{eqnarray}
\label{BESSEL-COMPLET2}
\int\limits^{\infty}_{0}
J_{|m|}(kr)  J_{|m'|}(k r)  \frac{dr}{r} = \frac{2}{\pi}
\frac{\sin\left[(|m'| - |m|)\frac{\pi}{2}\right]}{|m'|^2 - |m|^2},
\end{eqnarray}
which coincides with  2.12.31.3.\cite{PRUDNIKOV2} and expresses the known result from the theory of Bessel functions, namely,  two Bessel functions with integer index are orthogonal ones if the index runs through only even or only odd values.

\subsection{Contractions in equidistant basis}
\label{sec:4.2}

To perform the contraction limit $R\rightarrow\infty$ we take:
\begin{eqnarray}
\tau_2 \sim \frac{x}{R},\quad   \tau_1 \sim \frac{y}{R}.
\label{EQ_CONTR}
\end{eqnarray}
 The eigenvalues of second invariant operator 
$L^{EQ}/R^2 \sim P_1^2 = \partial^2/\partial x^2 $ contract as  $\nu \sim k_1R$. Here we consider solution of Helmholtz  equation in Cartesian coordinates\cite{ARXIV:2025} in the form 
\be
\label{CARTESIAN_P_M}
\Psi^{(+)}_{k_1 k_2}(x,y) = \frac{e^{ik_1 x}}{\sqrt{2\pi}} \frac{\cos |k_2|y}{\sqrt{2\pi}}, \quad 
\Psi^{(-)}_{k_1 k_2}(x,y) = \frac{e^{ik_1 x}}{\sqrt{2\pi}} \frac{\sin |k_2|y}{\sqrt{2\pi}},\quad k^2 = k_1^2 + k_2^2,\ k_{1,2}\in\mathbb{R}. 
\ee
The solution of  Eq. (\ref{EQUID-EQ3}) with the parity ${\cal P}_{\tau_1}$: $\tau_1 \to - \tau_1$
are transformed correspondingly into solutions  with the parity ${\cal P}_{y}$: $y \to - y$ on Euclidean plane.
Taking into account the asymptotic relation for gamma functions (\ref{Gamma_limit}) we get that $N_{\rho \nu}^{(+)}  \sim \frac{1}{R}\sqrt{\frac{k}{\pi |k_2|}}$, $N_{\rho \nu}^{(-)}  \sim \sqrt{\frac{k |k_2|}{\pi}}$ and
\begin{eqnarray}
\label{LIMIT-HYPER-01}
\lim_{R\to\infty} 
R N_{\rho\nu}^{(+)}\psi_{\rho\nu}^{(+)}(\tau_1)
&=&
\sqrt{\frac{k}{\pi |k_2|}}\, {_0F_1}\left(;\frac12;  - \frac{k_2^2  y^2}{4}\right)
=
\sqrt{\frac{k}{\pi |k_2|}}\, \cos|k_2| y,
\\[2mm]
\lim_{R\to\infty} 
R N_{\rho\nu}^{(-)} \psi_{\rho\nu}^{(-)}(\tau_1) 
&=&
y \sqrt{\frac{k |k_2|}{\pi}} \, 
{_0F_1}\left(;\frac32;  -\frac{k_2^2 y^2}{4}\right)
=
\sqrt{\frac{k}{\pi |k_2|}} \sin |k_2| y.
\end{eqnarray}
Finally, the contraction limit for the functions $\Psi^{EQ(\pm)}_{\rho\nu}(\tau_1, \tau_2)$ takes the form
\begin{eqnarray}
\label{(EQ)_sol_contraction}
\lim_{R\to\infty}
R \Psi^{EQ(\pm)}_{\rho\nu}  =   \sqrt{\frac{2 k}{|k_2|} }\,  \Psi^{(\pm)}_{k_1 k_2}(x,y).
\end{eqnarray}
Figures \ref{fig:contr_Psi_plus_EQ} and \ref{fig:contr_Psi_plus_EQ2} show how the functions $\psi_{\rho\nu}^{(\pm)}(\tau_1)$ tend to cosine as $R$ increases.

The contraction of the left-hand side of normalization integral (\ref{COMPLET-00-005}) looks like this
\begin{eqnarray}
\int\limits^{\infty}_{-\infty}  dy  \int\limits^{\infty}_{-\infty} dx 
\frac{2k}{R^2 \sqrt{ |k_2 k_2^\prime|} }  \Psi^{(\pm)}_{k_1 k_2}(x,y)  \Psi^{(\pm)\ast}_{k_1^\prime k_2^\prime}(x,y)
=  \frac{k}{R^2 |k_2|}\delta(k_1 - k_1^\prime)\delta(|k_2| - |k_2^\prime|),
\label{CONTR_NORM_EQ}
\end{eqnarray}
where we use relations 
\be
\label{CARTESIAN_NORMAL_k1k2}
\int\limits_{-\infty}^{\infty} dx  \int\limits_{-\infty}^{\infty} dy \Psi^{(\pm)}_{k_1 k_2}(x,y) \Psi^{(\pm)\ast}_{k_1^\prime k_2^\prime}(x,y)  = \frac{1}{2}\delta(k_1 - k_1^\prime) \delta(|k_2| - |k_2^\prime|).
\ee
The result (\ref{CONTR_NORM_EQ}) coincides with the contraction of the right-hand side of (\ref{COMPLET-00-005})
\begin{eqnarray}
\delta(\rho - \rho^\prime)\delta(\nu - \nu{\,^\prime}) \sim
\frac{k}{R^2 |k_2|}\delta(k_1 - k_1^\prime)\delta(|k_2| - |k_2^\prime|).
\end{eqnarray}
For completeness condition (\ref{COMPLET-005}) we consider $\rho \sim R \sqrt{k_1^2 + k_2^2}$ and obtain
\be
\label{COMLETE_CARTESIAN_k1k2}
2 \int\limits_{-\infty}^{\infty} dk_1  \int\limits_{0}^{\infty} dk_2  \left[\Psi^{(+)}_{k_1 k_2}(x,y) \Psi^{(+)\ast}_{k_1 k_2}(x^\prime,y^\prime) +  \Psi^{(-)}_{k_1 k_2}(x,y) \Psi^{(-)\ast}_{k_1 k_2}(x^\prime,y^\prime)\right] = \delta(x - x^\prime) \delta(y - y^\prime),
\ee
which is in accordance with completeness condition for $\Psi^{(\pm)}_{k_1 k_2}(x,y)$ functions on Euclidean plane\cite{ARXIV:2025}.

The contraction limit for functions $\psi_{\rho \nu}^{(1,2)}(\tau_1)$ can be determine with the help of formulas (\ref{H2-1}) (see also Ref. \onlinecite{POG-YAKH0}). 

\begin{minipage}{0.45\textwidth}
\includegraphics[width=\textwidth]{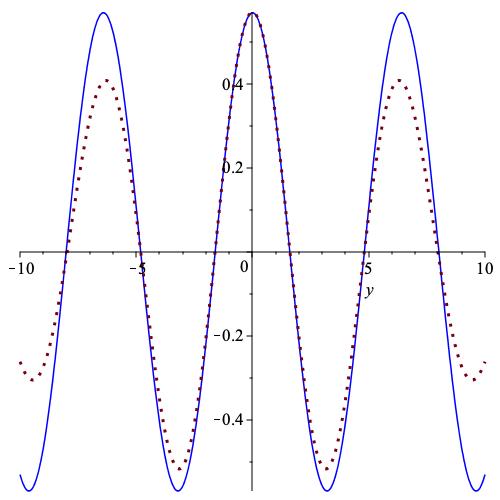}
\captionof{figure}{\small Graphics of wave function $R N_{\rho \nu}^{(+)} \psi_{\rho\nu}^{(+)}(\frac{y}{R})$ (point) and $\sqrt{\frac{k}{\pi |k_2|}} \cos |k_2| y$ (solid) for $\rho = kR$, $\nu = k_1R$, $R = 5$, $k=1$ and $k_1 = 0.2$.}
\label{fig:contr_Psi_plus_EQ}
\end{minipage}
\hfill
\begin{minipage}{0.45\textwidth}
\includegraphics[width=\textwidth]{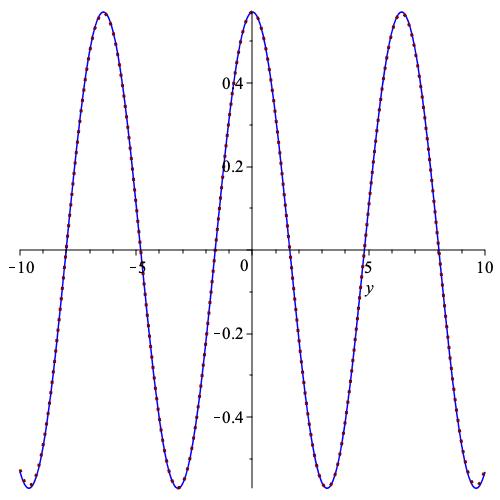}
\captionof{figure}{\small  Graphics of wave function $R N_{\rho \nu}^{(+)} \psi_{\rho\nu}^{(+)}(\frac{y}{R})$ (point) and $\sqrt{\frac{k}{\pi |k_2|}} \cos |k_2| y$ (solid) for $\rho = kR$, $\nu = k_1R$, $R = 50$, $k=1$ and $k_1 = 0.2$.}
\label{fig:contr_Psi_plus_EQ2}
\end{minipage}

\subsection{Contractions in horocyclic basis }
\label{sec:Contraction Cartesian basis}

The contraction limit for horocyclic variables (\ref{sys_horicyclic}) is determined by the following 
formulas:
\bea
\tilde{x} \sim \frac{y}{R},
\qquad
\tilde{y}
\sim  1 + \frac{x}{R},
\qquad  x \in (- R, \infty),
\qquad   y \in (-\infty, \infty).
\nonumber
\eea
The eigenvalue of the corresponding operator $L^{HO}/R^2 = (K_1/R + M/R)^2 \sim (\partial/\partial y + M/R)^2 \sim \partial^2/\partial y^2$
contracts as $s \sim k_2 R$. 
Consequently we need the asymptotic formula for MacDonald function
\begin{equation}
\label{MacDon}
K_{i\rho}(|s|\tilde{y}) \sim  K_{ikR}(|k_2|(R+x))
\end{equation}
as $R\to\infty$. Because of $kR > |k_2|R > 0$ we can use the
asymptotic formula (3.14.2)\cite{MAGNUS}
\begin{equation}
\label{K_inu}
K_{i\nu}(z) \sim \frac{\sqrt{2\pi}}{(\nu^2 - z^2)^{1/4}}
\exp\left(-\frac{\pi\nu}{2}\right)\sin\left(\frac{\pi}{4}
- \sqrt{\nu^2 - z^2} + \nu \arccosh \frac{\nu}{z} \right),
\quad
\nu > z > 0.
\end{equation}
Then we obtain that (\ref{MacDon}) takes the following form
\begin{eqnarray}
\label{MacDon-CONT-00}
K_{ikR}(|k_2|(R+x))
\sim e^{-kR\frac{\pi}{2}} \sqrt{\frac{2\pi}{|k_1|R}} \,
\sin\left(M -  |k_1| x \right),
\end{eqnarray}
where we denote
\begin{eqnarray}
\label{MacDon-CONT-04}
M := \frac{\pi}{4} + (k R) \arccosh \frac{k}{|k_2|} - |k_1| R.
\end{eqnarray}
Indeed, from (\ref{K_inu}) we get
\begin{eqnarray}
\label{MacDon-CONT-01}
K_{i kR}(|k_2|(R+x))  \sim
 \frac{\sqrt{2\pi}}{\left[k^2 R^2 -  k_2^2R^2\left(1 + \frac{x}{R}\right)^2\right]^{1/4}}
\exp\left(-\frac{\pi kR}{2}\right)
\nonumber
\\[2mm]
\times
\sin\left(\frac{\pi}{4} - \sqrt{k^2 R^2 -  k_2^2R^2\left(1 + \frac{x}{R}\right)^2} + (kR)
\arccosh \frac{k}{|k_2|(1 + x/R)} \right).
\end{eqnarray}
Taking into account that
\begin{eqnarray}
\label{MacDon-CONT-02}
R \sqrt{k^2 -  k_2^2\left(1 + \frac{x}{R}\right)^2}
&\sim&
R\left(\sqrt{k^2 - k_2^2} - \frac{k_2^2}{\sqrt{k^2 - k_2^2}} \frac{x}{R}\right),
\nonumber
\\[2mm]
\arccosh \frac{k}{|k_2|(1 + x/R)}
&\sim&
\arccosh \frac{k}{|k_2|} -  \frac{k}{\sqrt{k^2 - k_2^2}} \frac{x}{R},
\end{eqnarray}
the argument of the sine function in (\ref{MacDon-CONT-01})  contracts as follows
\begin{eqnarray}
\label{MacDon-CONT-03}
\frac{\pi}{4} - \sqrt{k^2 R^2 -  k_2^2 R^2\left(1 + \frac{x}{R}\right)^2} + (kR)\, \arccosh \frac{k}{|k_2|(1 + x/R)} \sim
\nonumber
\\[2mm]
\sim \frac{\pi}{4} - R |k_1|  + (kR) \arccosh \frac{k}{|k_2|}  -  x \frac{k^2 - k_2^2}{|k_1|}
= M -  |k_1| x.
\end{eqnarray}
Thus relation (\ref{MacDon-CONT-00}), together with the contraction limit of the normalization constant $N_{\rho s} \sim e^{iR \frac{\pi}{2}} \sqrt{k/|k_2|}/R\pi$ and $ \exp(i s \tilde{x}) \sim \exp(i k_2 y)$, give the following formula for the wave function (\ref{Psi_HO}) as $R\to \infty$
\begin{eqnarray}
\label{Sol_HO_contract}
\lim_{R \rightarrow\infty}
R \Psi^{HO}_{\rho s}(\tilde{y},\tilde{x})
= \frac{\sqrt{k}}{\pi}  \frac{\sin\left(M -  x  |k_1|  \right)}{\sqrt{|k_1|}} e^{i k_2 y}.
\end{eqnarray}
In Fig. \ref{fig:contr_HO} one can see how the values of the MacDonald function approach the values of the sine as $R$ increases for a fixed value of $x = 1$.

\begin{figure}[htbp]
\begin{center}
\includegraphics[scale=0.4]{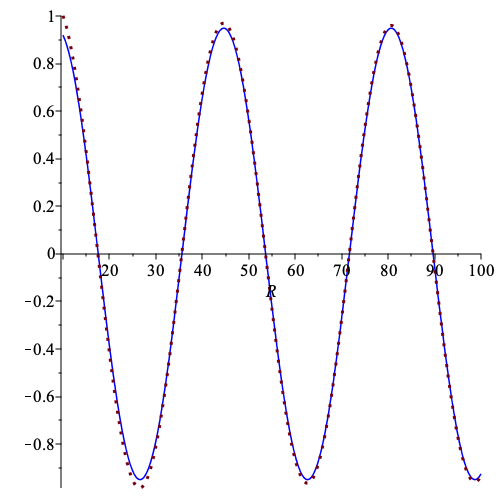}
\captionof{figure}{\small Graphics of functions $\sqrt{\frac{2k}{\pi}} \sin \frac{M - |k_1|x}{\sqrt{|k_1|}}$ (blue line) and $\frac{\sqrt{2kR\sinh\pi kR}}{\pi} \sqrt{1 + \frac{x}{R}} K_{ikR}\left(|k_2|(R + x)\right)$  (red points) for $k = 1$, $k_2 = 1/\sqrt{2}$.}
\label{fig:contr_HO}
\end{center}
\end{figure}

Using properties of Dirac delta function, in contraction limit for the right side of normalization integral (\ref{normHO})  we obtain
\begin{eqnarray}
\delta(\rho - \rho^\prime)\delta(s - s^\prime)
\sim  \frac{1}{R^2} \,  \frac{k}{|k_1|}\,
\delta(|k_1| - |k_1^\prime|) \, \delta(k_2 - k_2^\prime),
\end{eqnarray}
while its left side transforms as it follows
\begin{eqnarray}
\frac{\sqrt{kk^\prime/ |k_1 k_1^\prime|}}{R^2 \pi^2} \int\limits^{\infty}_{-\infty} e^{i(k_2 - k_2^\prime)y} d y
\int\limits^{\infty}_{- R}
\sin\left(|k_1| x - M(k_1,k_2) \right)  \sin\left(|k_1^\prime| x - M(k_1^\prime, k_2^\prime) \right) d x
= \nonumber \\
= \frac{\sqrt{kk^\prime} \delta(k_2 - k_2^\prime)}{R^2 \pi\sqrt{|k_1 k_1^\prime|}}  \int\limits^{\infty}_{- R} dx \left[ \cos\left(M - M^\prime -  (|k_1| - |k_1^\prime|)x \right) - \cos\left(M + M^\prime -  (|k_1| + |k_1^\prime|)x \right) \right]  \sim \nonumber \\
\sim \frac{k}{R^2 |k_1|} \delta(k_2 - k_2^\prime)\delta(|k_1| - |k_1^\prime|),
\label{NOR-CONTRACTION11}
\end{eqnarray}
where we use $\int\limits_{-\infty}^{\infty} \cos at dt = 2\int\limits_{0}^{\infty} \cos at dt = 2\pi \delta(a)$ and $\lim\limits_{R\to\infty}  \int\limits^{\infty}_{- R}  \sin ax\, dx = 0$. Relation (\ref{NOR-CONTRACTION11}) is in accordance with normalization of Cartesian basis with parity $x \to - x$ 
\be
\label{CARTESIAN_P_M_x}
\tilde{\Psi}^{(+)}_{k_1 k_2}(x,y) = \cos|k_1|x\, e^{i k_2 y}/2\pi,  \quad \tilde{\Psi}^{(-)}_{k_1 k_2}(x,y) = \sin|k_1|x\, e^{i k_2 y}/2\pi,
\ee
on Euclidean plane\cite{ARXIV:2025} and the right side of (\ref{Sol_HO_contract}) is a linear combination of $\tilde{\Psi}^{(\pm)}_{k_1 k_2}(x,y)$ functions.


\subsection{Contraction in interbasis coefficients ${\cal U}_{\rho \nu}^{m (\pm)} $}
\label{sec:CONT:SPH-EQUI}

Taking into account the contraction limit at $R \to \infty$ for the quantum numbers $\rho \sim k R$
and $\nu \sim k_1 R$, using asymptotic formulas for gamma functions (\ref{Gamma_limit}) and
formulas of summation for hypergeometric functions (Ref. \onlinecite{BE1} 2.8 (11), 2.9 (4), 2.8 (12)),
we obtain for (\ref{EQUIDIST-SPHERIC-20}) and (\ref{EQUIDIST-SPHERIC-21})
\bea
{_4F_3}\left(
\left.\begin{array}{cccc}
\frac{1}{4} + \frac{i(\nu-\rho)}{2}, & \frac{1}{4} + \frac{i(\nu + \rho)}{2},
& -\frac{|m|}{2}, &- \frac{|m|-1}{2}
\\[2mm]
\frac{1}{2}, &   \frac{1}{2} + \frac{i\nu}{2} - \frac{|m|}{2},  & 1 + \frac{i\nu}{2} - \frac{|m|}{2}
\end{array}\right | 1
\right)
\sim
\frac{\cos |m|\alpha}{(\cos\alpha)^{|m|}},
\eea
\bea
\label{01-U_contr}
_4F_3\left(
\left.\begin{array}{cccc}
\frac{3}{4} + \frac{i(\nu-\rho)}{2}, & \frac{3}{4} + \frac{i(\nu + \rho)}{2},
&1 - \frac{|m|}{2}, & - \frac{|m|-1}{2}
\\[2mm]
\frac{3}{2}, &   \frac{3}{2} + \frac{i\nu}{2} - \frac{|m|}{2},  & 1 + \frac{i\nu}{2} - \frac{|m|}{2}
\end{array}\right| 1
\right)
\sim
\frac{\cot\alpha}{|m|}
\,
\frac{\sin |m|\alpha}{(\cos\alpha)^{|m|}},
\eea
and
\bea
\label{U_contr}
\lim_{R \rightarrow\infty} 
\sqrt{R} \, {\cal U}_{\rho \nu}^{m(+)}
=
\frac{(-i)^{|m|}}{\sqrt{\pi |k_2|}} \cos m\alpha,
\qquad
\lim_{R \rightarrow\infty} 
\sqrt{R} \,
{\cal U}_{\rho \nu}^{m(-)}
=
- \frac{(-i)^{|m|}}{\sqrt{\pi |k_2|}}  \, \sin m\alpha,
\eea
where we introduce the angle $\alpha \in [-\pi, \pi)$, $k_1 = k \cos\alpha$, $k_2 = k\sin\alpha$.  Figs. \ref{fig:CONTR_EQ_plus} - \ref{fig:CONTR_EQ_plus2_m_3}  show graphs of the coefficients ${\cal U}_{kR, k_1R}^{m(+)}$ as functions of $k_1\in[-k, k]$ for even and odd values of $m$ and for different values of $R$. It is evident that as $R$ increases, these graphs tend to the corresponding trigonometric limit functions.

The contraction limit of expansion (\ref{EQUIDIST-SPHERIC- EXPANSION-02}) leads to the well known formula 
\bea
e^{i k_1x} e^{i k_2y} = e^{ikr\cos(\varphi - \alpha)} = \sum_{m = -\infty}^{\infty} e^{- i m \alpha}\, i^m J_m(kr) e^{i m \varphi}
\label{00-CONT-SPH_EQUI-3}
\eea
for decomposition of the flat wave through the spherical two-dimensional waves (see (27) 7.2.4 \cite{BE2} and Ref. \onlinecite{ARXIV:2025}). 

\begin{minipage}{0.45\textwidth}
\includegraphics[width=\textwidth]{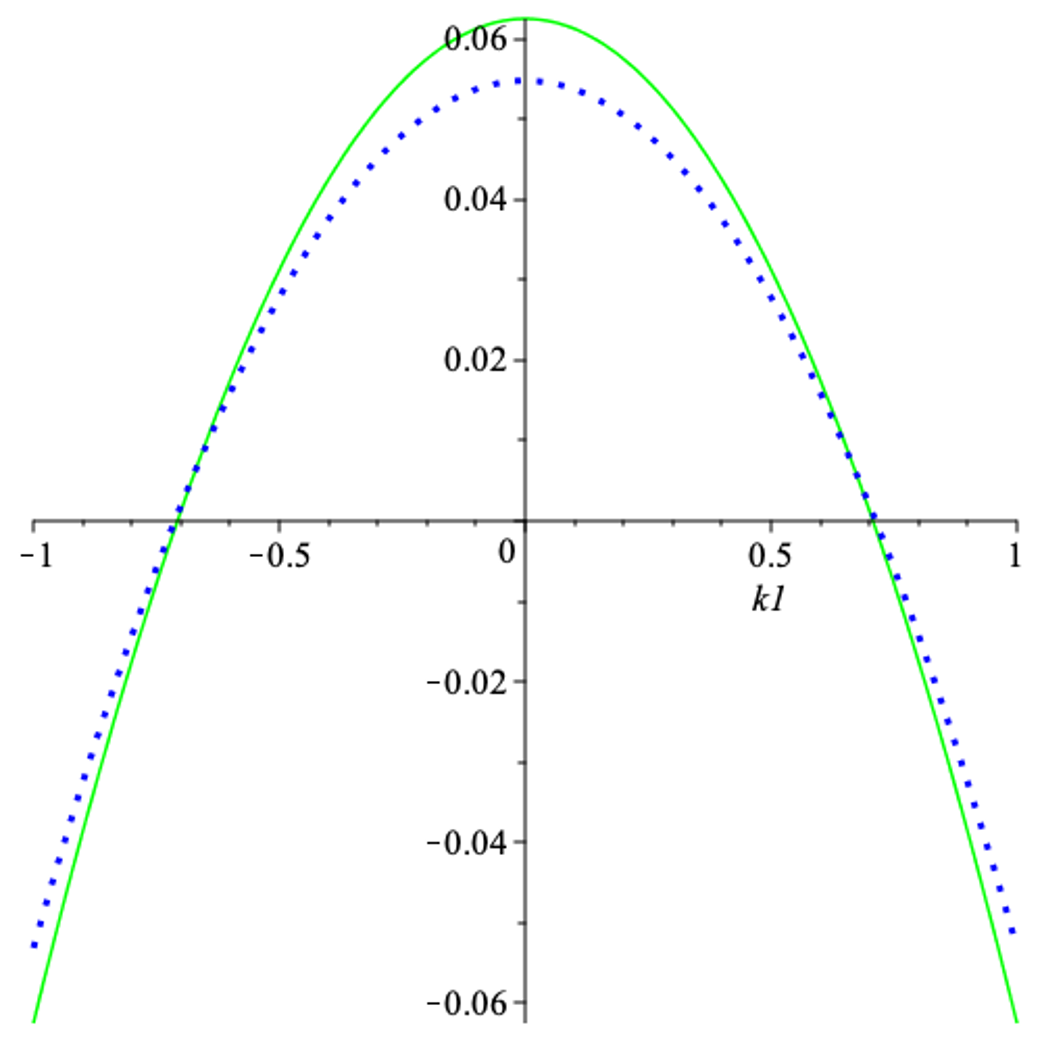}
\captionof{figure}{\small Graphics of coefficient ${\cal U}_{\rho\nu}^{m(+)}$  (blue points) and its contraction (green solid line) for $R = 4$, $k=1$ and $m = 2$.}
\label{fig:CONTR_EQ_plus}
\end{minipage}
\hfill
\begin{minipage}{0.45\textwidth}
\includegraphics[width=\textwidth]{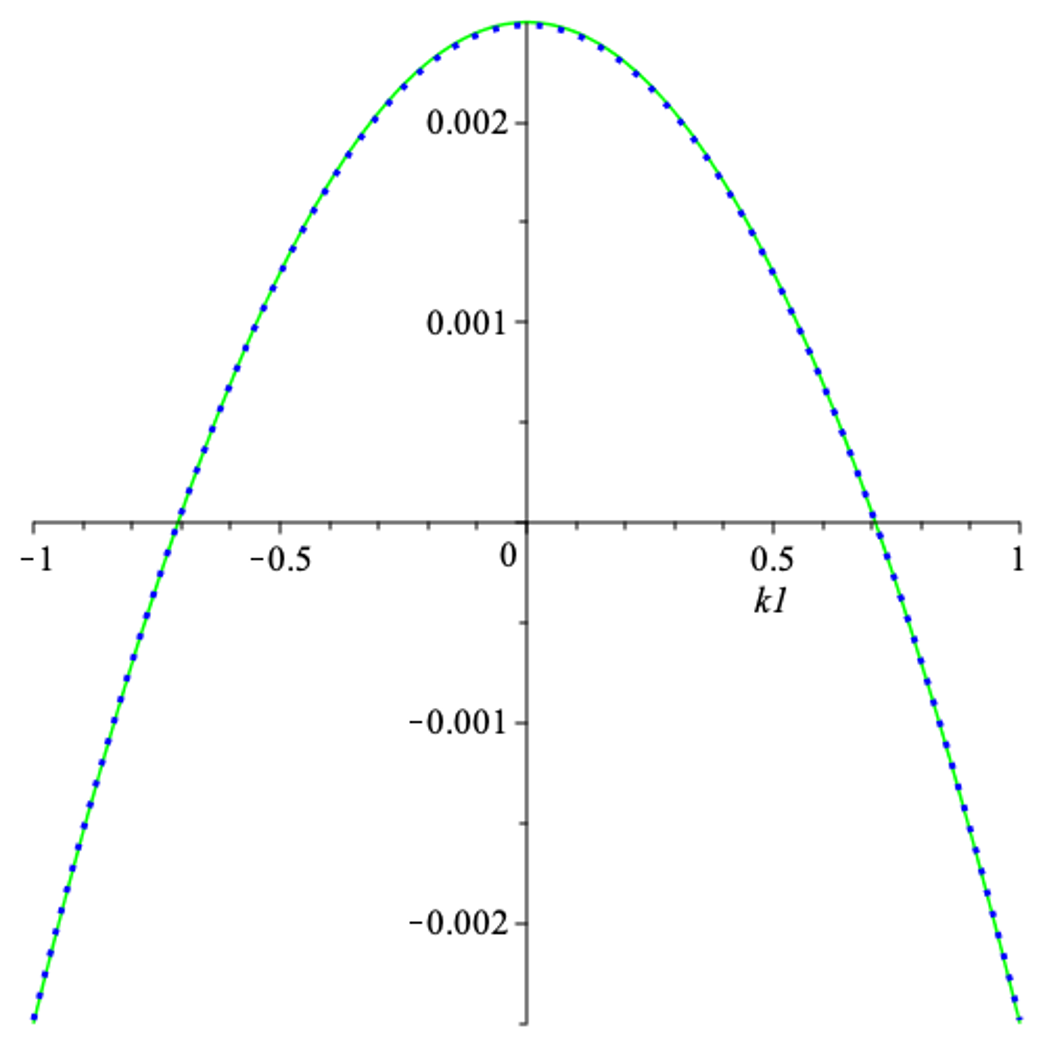}
\captionof{figure}{\small Graphics of coefficient ${\cal U}_{\rho\nu}^{m(+)}$  (blue points) and its contraction (green solid line) for $R = 20$, $k=1$ and $m = 2$.}
\label{fig:CONTR_EQ_plus2}
\end{minipage}

\begin{minipage}{0.45\textwidth}
\includegraphics[width=\textwidth]{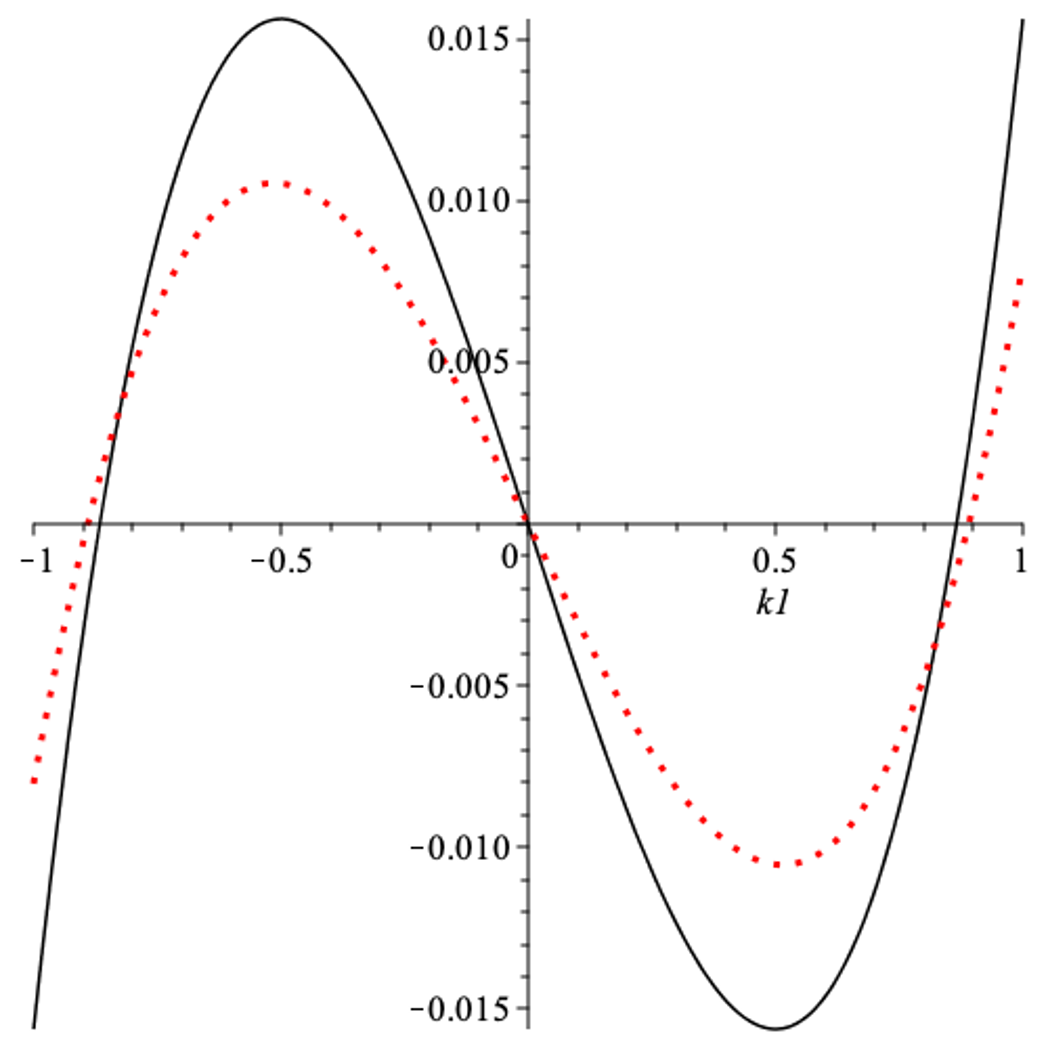}
\captionof{figure}{\small Graphics of imaginary part of coefficient ${\cal U}_{\rho\nu}^{m(+)}$  (red points) and its contraction (black solid line) for $R = 4$, $k=1$ and $m = 3$.}
\label{fig:CONTR_EQ_plus_m_3}
\end{minipage}
\hfill
\begin{minipage}{0.45\textwidth}
\includegraphics[width=\textwidth]{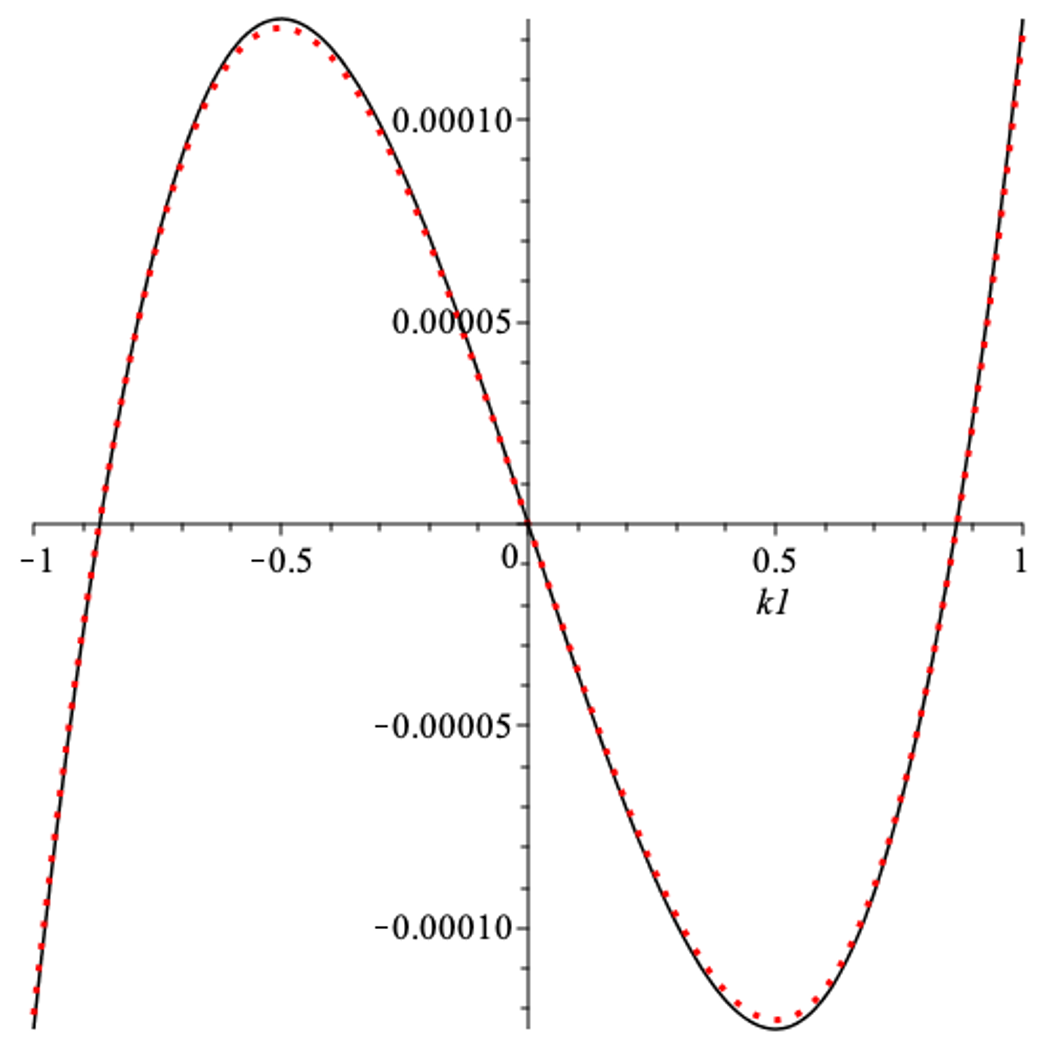}
\captionof{figure}{\small Graphics of imaginary part of coefficient ${\cal U}_{\rho\nu}^{m(+)}$ (red points) and its contraction (black solid line) for $R = 20$, $k=1$ and $m = 3$.}
\label{fig:CONTR_EQ_plus2_m_3}
\end{minipage}

To realize the limit $R \to \infty$ in the both sides of expansion (\ref{HORIC-EQUIDIST-EXPAN-03}) we consider $\nu \sim kR\cos\alpha$, with fixed $k$.  Then, after multiplication of both side by $\sqrt{R}$ one can obtain, considering (\ref{CONTR-FUNCTION2}), (\ref{(EQ)_sol_contraction})
and (\ref{U_contr})
\begin{equation}
\label{01-HORIC-EQUIDIST-EXPAN-03}
J_{|m|} (kr) e^{i m \vphi}
= \frac{(-i)^{|m|}}{\pi} 
\int\limits_{0}^{\pi} 
\cos\left(m \alpha + kr\sin\alpha\sin\vphi \right) e^{ikr \cos\alpha \cos\vphi} d \alpha,
\end{equation}
that coincides with the expansion of the polar basis through Cartesian one on the Euclidean plane\cite{ARXIV:2025} and gives (at $\vphi = - \pi/2$ and for $m \ge 0$) the well known integral representation for the Bessel function (see (2) 7.3.1\cite{BE2})
\begin{equation}
\label{01-HORIC-EQUIDIST-EXPAN-03}
J_{m} (kr) = \frac{1}{\pi} \int\limits_{0}^{\pi} \cos\left(kr \sin\alpha - m \alpha\right) d \alpha.
\end{equation}

\subsection{Contractions in coefficients ${\cal W}_{\rho s}^{\nu (\pm)} $}
\label{secti1-on:EQ_HO}

Let us trace the contraction limit of coefficients ${\cal W}_{\rho s}^{\nu(\pm)}$.  We put $s \sim k_2' R$, $\rho \sim kR$, $\nu \sim k_1 R$, where $k$ is fixed and $k = \sqrt{k_1^2+k_2^2} = \sqrt{k_1^{\prime2} + k_2^{\prime 2}}$. Using the asymptotic expansions for gamma-functions (2) 1.18.\cite{BE1}
$\Gamma(z) \sim \sqrt{2\pi} e^{-z + \left(z - {1}/{2}\right)\ln z}$, $z\sim\infty$, 
one can obtain from (\ref{HORIC-EQUIDIST-106})
\bea
\label{W_contracted}
{\cal W}_{\rho s}^{\nu(+)} \sim \frac{e^{iRk_1}}{2\sqrt{\pi|k_2'|R}} \left|\frac{k_2'}{k_2} \right|^{iRk_1} \left|\frac{k_1-k}{k_1+k}\right|^{iRk/2}.
\eea
For ${\cal W}_{\rho s}^{\nu(-)} $ one can see that
$
(({\cosh\pi\rho - i\sinh\pi\nu})/({\cosh\pi\rho + i\sinh\pi\nu}))^{\frac{1}{2}} \sim 1
$
 due to inequality $k > |k_1|$. Therefore, from relations (\ref{HORIC-EQUIDIST-08}), (\ref{HORIC-EQUIDIST-0-8})  we have ${\cal W}_{\rho s}^{\nu(-)} \sim {i k_2'}/{|k_2'|} \, {\cal W}_{\rho s}^{\nu(+)}$. Graphics of real and imaginary parts of coefficient ${\cal W}_{kR, k'_2R}^{k_1R (+)}$ and their asymptotic as functions of $R$ are shown in Figs. \ref{fig:CONTR_W} and \ref{fig:CONTR_W_1}.

\begin{minipage}{0.45\textwidth}
\includegraphics[width=\textwidth]{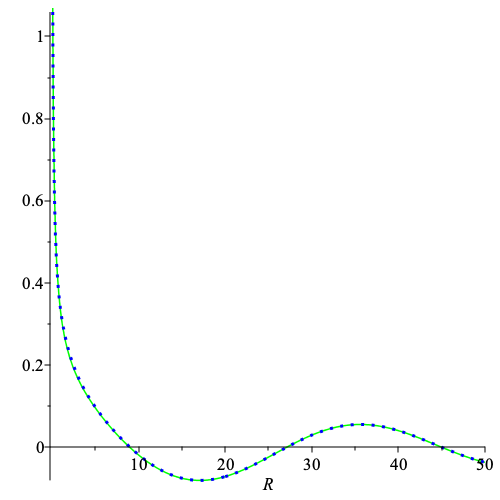}
\captionof{figure}{\small Graphics of real part of coefficient ${\cal W}_{kR, k'_2R}^{k_1R (+)}$  (blue dots) and its asymptotic (green  line) for $k = 1$, $k_1 = k'_2 = 1/\sqrt{2}$.}
\label{fig:CONTR_W}
\end{minipage}
\hfill
\begin{minipage}{0.45\textwidth}
\includegraphics[width=\textwidth]{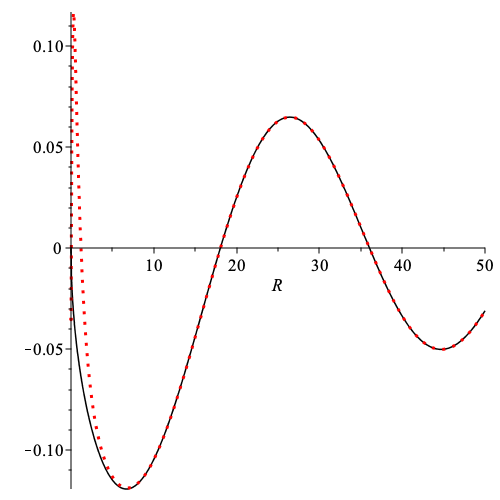}
\captionof{figure}{\small Graphics of imaginary part of coefficient ${\cal W}_{kR, k'_2R}^{k_1R (+)}$ (red dots) and its asymptotic (black  line) for $k = 1$, $k_1 = k'_2 = 1/\sqrt{2}$.}
\label{fig:CONTR_W_1}
\end{minipage}

Now we can realize the contraction limit in the both sides of (\ref{HORIC-EQUIDIST-02}) and (\ref{HORIC-EQUIDIST-03}). At the fixed parameter $k$ we get:
\bea
\label{LHS}
\sqrt{\frac{2\pi}{|k_1'| }} \sin\left(M' -  |k'_1| x \right) \cos k_2' y
&=&
\frac{1}{2} \sqrt{\frac{R}{ |k_2'| }} \int\limits_{-\infty}^{\infty} \frac{\cos |k_2| y}{\sqrt{|k_2|}} e^{i k_1 x} e^{iR p(k_1)} d k_1,
\\[2mm]
\label{0-LHS}
\sqrt{\frac{2\pi}{|k_1'| }} \sin\left(M' -  |k'_1| x \right) \sin k_2' y
&=&
\frac{\mathrm{sign}(k'_2)}{2} \sqrt{\frac{R}{ |k_2'| }} \int\limits_{-\infty}^{\infty} \frac{\sin |k_2| y}{\sqrt{|k_2|}} e^{i k_1 x} e^{iR p(k_1)} d k_1,
\eea
where we use the notation
\bea
p := k_1 (1 + \ln |k_2'|) - k_1\ln |k_2| + \frac{k}{2} \ln{ \frac{k-k_1}{k+k_1}}.
\eea
The integrals are equal to the sum of two integrals of the form
\bea
J_1 = \int\limits_{-\infty}^{\infty} h(k_1) \cos(k_1 x)\, e^{iRp(k_1)} dk_1,\quad J_2 = i \int\limits_{-\infty}^{\infty} h(k_1) \sin(k_1 x)\, e^{iRp(k_1)} dk_1,
\eea
with $h(k_1) :=  \left(k^2 - k_1^{2}\right)^{- \frac{1}{4}} \cos y\sqrt{k^2 - k_1^2}$ for (\ref{LHS}) and $h(k_1) :=  \left(k^2 - k_1^{2}\right)^{- \frac{1}{4}} \sin y\sqrt{k^2 - k_1^2}$ for (\ref{0-LHS}).  The approximation of each of them we can calculate using method of stationary phase (see Ref. \onlinecite{Olver:74}, Chapter II, par. 11)
 \bea
\int\limits^\infty _{-\infty} e^{iR p(t)} q(t) dt \sim \sqrt{2\pi}  \sum\limits_{j=1}^2 \frac{q(a_j)}{\sqrt{\left|p^{\prime\prime}(a_j)\right|R}}\exp\left\{iRp(a_j) + i\frac{\pi}{4}\, \mathrm{sign}\left[p''(a_j)\right]\right\},
\eea
where $a_{1,2} =  \pm k_1'$ are two stationary points of function $p(k_1)$. Then $p''(a_{1,2}) = \pm {k_1'}/{(k_2')^2}$ and we obtain for integral in (\ref{LHS}):
\bea
J_1 = 2 \sqrt{2\pi |k_2'|} \frac{\cos|k_2'| y}{\sqrt{R|k_1'|}}\cos{x |k_1'|} \cos \left(R p(a_1) + \frac{\pi}{4}\right), \\[2mm]
J_2 = - 2 \sqrt{2\pi |k_2'|} \frac{\cos|k_2'| y}{\sqrt{R|k_1'|}}\sin{x |k_1'|} \sin \left(R p(a_1) + \frac{\pi}{4}\right).
\eea
Taking into account that $R p(a_1) = - M' + \frac{\pi}{4}$ and returning to the right-hand side of (\ref{LHS}), we obtain an expression that coincides with the left-hand side. The relation (\ref{0-LHS}) can be proved in the same way.

\subsection{Contractions in interbasis coefficients ${\cal V}^m_{\rho s}$}
\label{1-subsection:HOR-SPHER}

Taking the contraction limit $s \sim k_2R$, $\rho \sim kR$ at the both sides of 
formula (\ref{HORIC-SPHERICAL-01B}), using (\ref{CONTR-FUNCTION-1}), (\ref{MacDon-CONT-00}) and (\ref{MacDon-CONT-04}),  we have 
\bea
\label{HORIC-SPHERICAL-01B1}
{\cal V}_{\rho s}^m 
J_{|m|}(kr)
&\sim&
\frac{(-1)^{|m|}}{2i \sqrt{2 \pi^3 |k_1| R}}
\Biggl\{e^{i M} 
\int\limits_{0}^{2\pi}
e^{-i |k_1| r \cos\vphi}
 e^{i k_2 r \sin\vphi}
 e^{- i m\vphi} d \vphi
\nonumber\\[2mm]
&-&
e^{-i M} 
\int\limits_{0}^{2\pi} 
e^{i |k_1| r \cos\vphi}
 e^{i k_2 r \sin\vphi}
 e^{- i m\vphi}  d \vphi
\Biggr\}.
\eea
Using now that $k_1 = k \cos\alpha$, $k_2 = k \sin\alpha$ and expansion of function 
$e^{i z \sin\beta}$ over the Bessel functions (see (\ref{00-CONT-SPH_EQUI-3})), we obtain 
\bea
\label{00-HORIC-SPHERICAL-01B}
\int\limits_{0}^{2\pi} 
e^{-i |k_1| r \cos\vphi + i k_2 r \sin\vphi}
e^{- i m\vphi}  d \vphi 
= \left\{
\begin{array}{c}
  2\pi (-i)^{|m|} J_{|m|}(kr) e^{i m\alpha},\ \cos\alpha > 0, \\[2mm]
2\pi i^{|m|} J_{|m|}(kr) e^{- i m\alpha},\ \cos\alpha < 0,
\end{array}
\right.
\eea
and finally ($l\in\mathbb{Z}$)
\bea
\label{01-HORIC-SPHERICAL-01B}
{\cal V}_{\rho s}^{2l} 
\sim
(-1)^l\sqrt{\frac{2}{\pi |k_1|R}}
\left\{
\begin{array}{c}
  \sin(M + 2l\alpha),\ \cos\alpha > 0, \\[2mm]
  \sin(M - 2l\alpha),\ \cos\alpha < 0;
\end{array}
\right.
\eea
\bea
\label{01-HORIC-SPHERICAL-01B_odd}
{\cal V}_{\rho s}^{2l+ 1} 
\sim
(-1)^l\, \sign(2l + 1) \sqrt{\frac{2}{\pi |k_1|R}}
\left\{
\begin{array}{c}
  \cos(M + (2l + 1)\alpha),\ \cos\alpha > 0, \\[2mm]
  - \cos(M - (2l + 1)\alpha),\ \cos\alpha < 0.
\end{array}
\right.
\eea

In Figs. \ref{fig:CONTR_V} and \ref{fig:CONTR_V_1} one can observe the graphics of coefficients  ${\cal V}_{kR, k_2R}^{m}$  with 50 first terms in the sum (\ref{HORIC-SPHERICAL-05}) and its asymptotic  (\ref{01-HORIC-SPHERICAL-01B}) and (\ref{01-HORIC-SPHERICAL-01B_odd}) as functions of $R$. It can be seen that as the value of $R$ increases, the graphs become closer.

\begin{minipage}{0.45\textwidth}
\includegraphics[width=\textwidth]{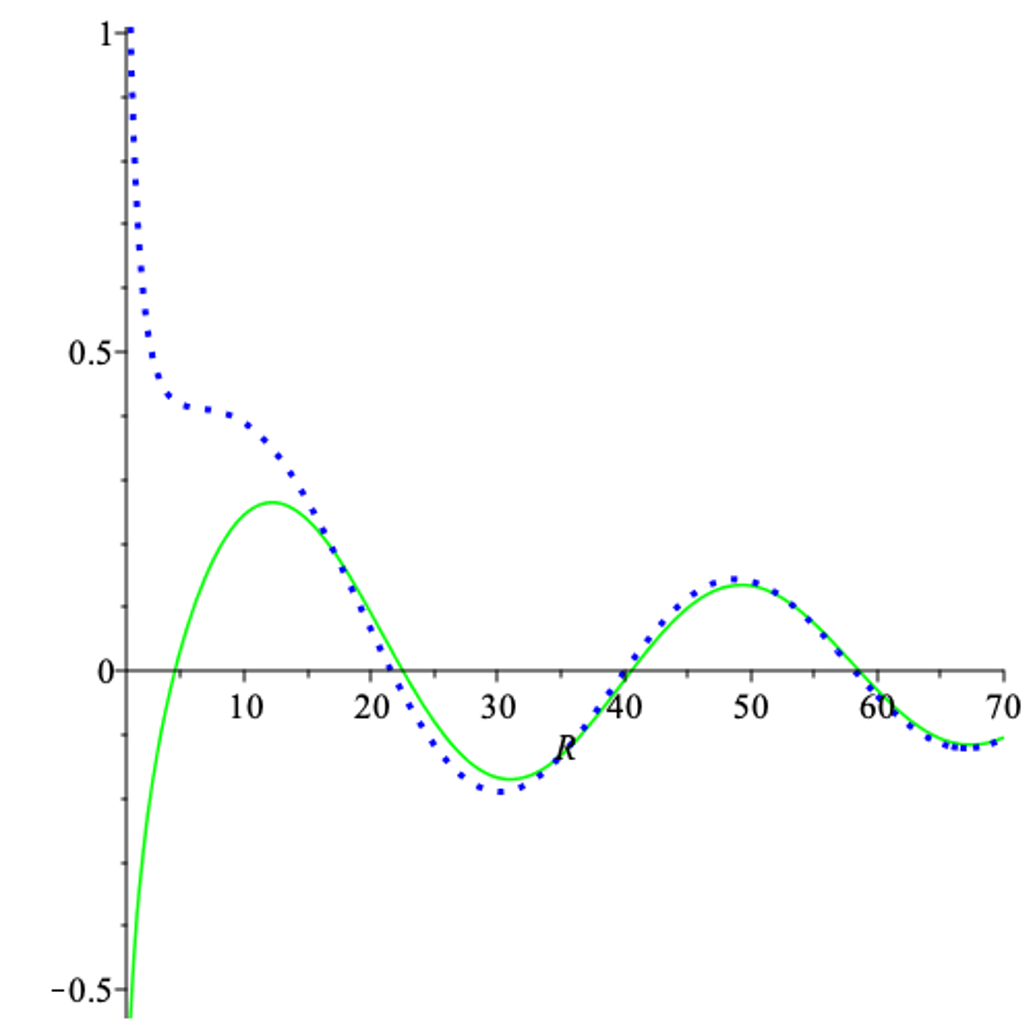}
\captionof{figure}{\small Graphics of coefficient ${\cal V}_{kR, k_2R}^{m}$  (blue points) and its asymptotic (green solid line) for $k=1$, $k_2 = 1/\sqrt{2}$ and $m = 2$.}
\label{fig:CONTR_V}
\end{minipage}
\hfill
\begin{minipage}{0.45\textwidth}
\includegraphics[width=\textwidth]{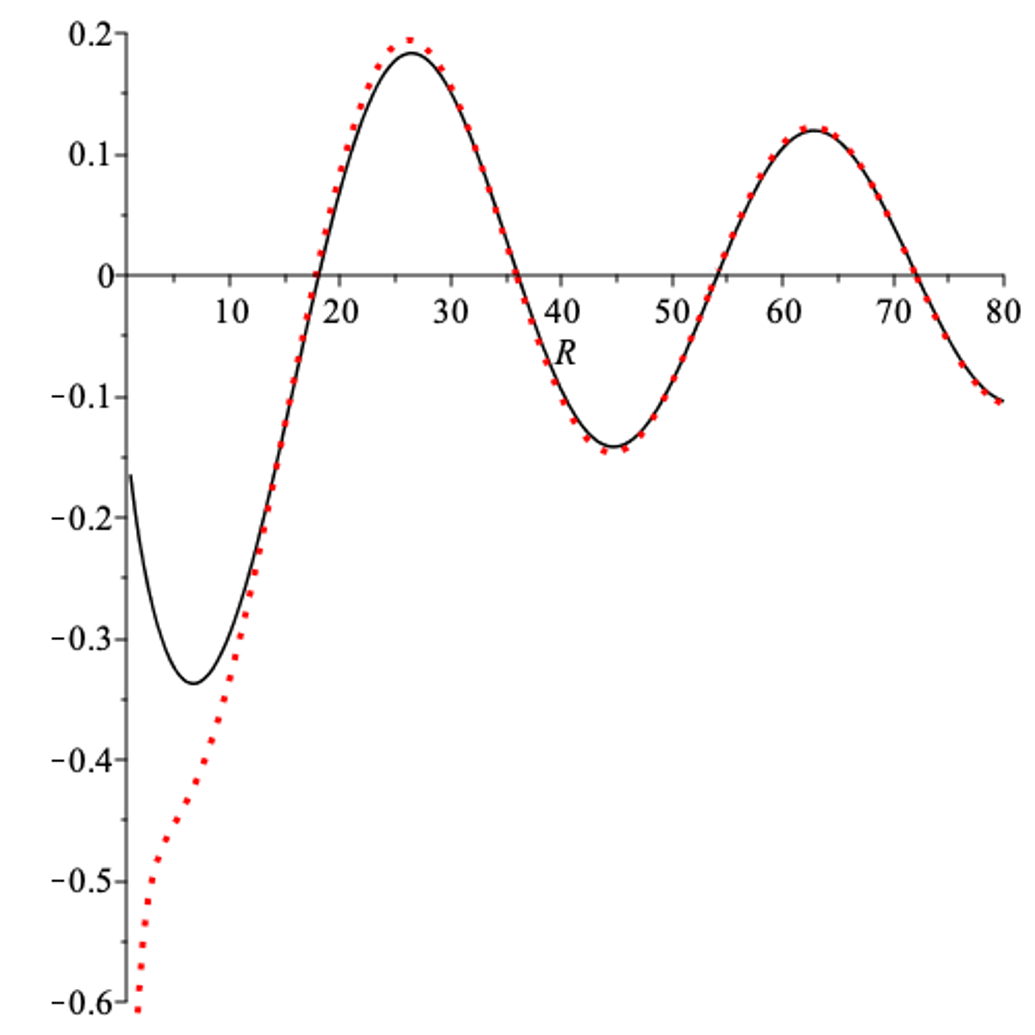}
\captionof{figure}{\small Graphics of coefficient ${\cal V}_{kR, k_2R}^{m}$ (red points) and its asymptotic (black solid line) for $k=1$, $k_2 = 1/\sqrt{2}$ and $m = 1$.}
\label{fig:CONTR_V_1}
\end{minipage}

\section{Conclusions}

In this paper, we first examined in detail three types of subgroup eigenfunctions of the Laplace-Beltrami operator on a two-sheeted hyperboloid, and secondly, we calculated all the interbasis expansions between them.  
The simplest form is acquired by the interbasis transitions between the EQ and HO 
solutions.  They are expressed through the gamma function.  Apparently, this is due to the simple fact that both bases are transformed into a Cartesian base upon contraction. We have shown that the decomposition between the SP and EQ bases is realized by means of Wilson-Racah polynomials.  The most complex form, which is quite unexpected, is that of the transition coefficients between the HO and SP bases, which are written through infinite sums of Laguerre polynomials.
 
Knowledge of interbasis decompositions allows one to prove the orthogonality and completeness conditions for subgroup bases in a simple way. The obtained interbasis expansions generalize some known relations between special functions. We have traced how the coefficients defining the expansions and the expansions themselves between subgroup bases contract from the two-sheeted hyperboloid to the Euclidean plane.

In a future paper we are planning to construct the interbasis expansions between non subgroup bases.

\section*{APPENDIX: Completeness and orthonormality conditions for subgroup bases}
\label{bsection: Appendix A}

\subsection{Completeness and orthonormality conditions of horocyclic bases}
\label{bsection: Completeness of HO}


The MacDonald functions $K_{i\rho} (|s| {\tilde y})$ form  a  complete  orthogonal set, due to relations:
\begin{eqnarray}
\label{MACDONALD-7}
\frac{2x}{\pi^2} \sinh \pi x
\int\limits^{\infty}_{0} K_{i x}(y)
K_{- ix'}(y)
 \frac{d y}{y} = \delta(x - x'),
\\[2mm]
\label{COM-HOR-2}
\frac{2}{\pi^2}  \int_{0}^{\infty} x \sinh\pi x \,   K_{ix}(y) K_{-ix} (y')  d x  = y \delta(y-y').
\eea
To prove the orthogonality relation (\ref{MACDONALD-7}), one can use the integral representation for MacDonald function (see 3.7\cite{MAGNUS})
\begin{eqnarray}
\label{MACDONALD--ORT-7}
z^{\alpha} K_{i\rho}(z) = \sqrt{\frac\pi2} \int\limits^{\infty}_{1}(t^2-1)^{-\frac\alpha2  - \frac14} P_{- \frac12+ i\rho}^{\alpha+\frac12}(t) \, e^{- z t} d t,
\qquad \Re(\alpha) < \frac12,
\
\Re(z) > 0.
\end{eqnarray}
Substituting (\ref{MACDONALD--ORT-7}) in the left expression of  (\ref{MACDONALD-7}) and taking into 
account formula (6) from 3.14\cite{BE1} 
\begin{eqnarray}
\label{MACDONALD--ORT-8}
\pi
P_{- \frac12+ i\rho}(-x) 
 = 
 \cosh \pi \rho \int\limits^{\infty}_{1} 
P_{- \frac12+ i\rho}(v) \frac{d v}{v-x},
\qquad  x < 1,
\end{eqnarray}
we get  
\begin{eqnarray}
\label{MACDONALD--ORT-9}
&&\frac{2}{\pi} \int\limits^{\infty}_{0} 
K_{i \rho}(y)  K_{i \rho'}(y) \frac{d y}{y}
= 
\int\limits^{\infty}_{0} d y
\int\limits^{\infty}_{1} dt \int\limits^{\infty}_{1} d t'
P_{- \frac12+i\rho}(t) P_{- \frac12+i\rho'}(t') 
 e^{- y (t+t')}
\nonumber\\[3mm]
&=&
\int\limits^{\infty}_{1} dt \int\limits^{\infty}_{1}
P_{- \frac12+i\rho}(t) \,P_{- \frac12+i\rho'}(t') 
\frac{ dt'}{t+t'}
=
\frac{\pi}{\cosh\pi \rho} \,  
\int\limits^{\infty}_{1}\, P_{- \frac12+i\rho}(t) \,P_{- \frac12+i\rho'}(t) \,
dt.
\nonumber
\end{eqnarray}
Finally, the orthogonality relation for Legendre functions (\ref{LEGENDRE-4}) gives (\ref{MACDONALD-7}).   

Condition of completeness (\ref{COM-HOR-2}) follows from Lebedev formula  (75) from 7.10.5\cite{BE2} 
\bea
\label{00-MACDONALD-007}
x f(x) = \frac{2}{\pi^2} 
\int\limits^{\infty}_{0}
K_{i\rho}(x)  {\rho \sinh \pi\rho} \, d \rho \int\limits^{\infty}_{0} K_{i\rho}(y)f(y)  d y,
\eea
if one takes $y = x = |s| \tilde{y}$, $f(y) = \delta(y - |s|\tilde{y}')$.


\subsection{Completeness and orthonormality conditions of the equidistant  basis}
\label{bsection:Completeness_of_EQ}

Let us consider functions $u^{(\pm)}_{\rho\nu}(\tau_1) = \psi^{(\pm)}_{\rho\nu}(\tau_1) \sqrt{\cosh\tau_1}$, then 
Eq. (\ref{00-EQUID-EQ1}) takes the following form
\be
u^{(\pm)\prime\prime}_{\rho\nu} + \left(\rho^2  - \frac{\nu^2 + 1/4}{\cosh^2\tau_1}\right)u^{(\pm)}_{\rho\nu} = 0.
\ee
From the above equation one can obtain (considering conjugated equation and integrating by parts the difference)
\be
\label{INT_EQ_U}
\int\limits_{-\infty}^{\infty}u^{(\pm)}_{\rho\nu}(\tau_1) u^{(\pm)\ast}_{\rho' \nu}(\tau_1) d\tau_1 
= \frac{1}{{\rho'}^2 - \rho^2}\left.\left(u^{(\pm)\ast}_{\rho' \nu} \frac{d u^{(\pm)}_{\rho \nu}}{d\tau_1} 
- u^{(\pm)}_{\rho \nu} \frac{d u^{(\pm)\ast}_{\rho' \nu}}{d\tau_1}\right)\right|^{\infty}_{-\infty}.
\ee
To analyze the asymptotic behavior of the wave functions $u_{\rho\nu}^{(\pm)}(\tau_1)$ we use
the relation  connecting the hypergeometric functions with the variables 
$z$ and $(z-1)/z$ (see (4) from 2.10\cite{BE1})
\begin{eqnarray}
\label{H2-000}
_2F_1 (\alpha, \beta; \gamma; z) =
\frac{\Gamma(\gamma)\Gamma(\gamma-\alpha-\beta)}{\Gamma(\gamma-\beta)
\Gamma(\gamma-\alpha)} \, z^{-\alpha}
\,
_2F_1 \left(\alpha, \alpha+1-\gamma; \alpha+\beta+1-\gamma;
\frac{z-1}{z}\right)
\nonumber\\[3mm]
+ 
\frac{\Gamma(\gamma)\Gamma(\alpha+\beta-\gamma)}{\Gamma(\beta)
\Gamma(\alpha)}
\, z^{\alpha - \gamma} (1-z)^{\gamma-\alpha-\beta}
\,
_2F_1 \left(1-\alpha,  \gamma-\alpha;  \gamma + 1 - \alpha-\beta;
\frac{z-1}{z}\right).
\end{eqnarray}
Considering  (\ref{H2}) and  (\ref{H3}) we get for $\tau_1 \sim \pm \infty$
\bea
u^{(+)}_{\rho\nu} \sim \sqrt{\pi}
\left\{
\frac{\Gamma(-i\rho) \left(\cosh \tau_1\right)^{-i\rho}}{\Gamma\left(\frac14 - i\frac{\rho + \nu}{2}\right) \Gamma\left(\frac14 - i\frac{\rho - \nu}{2}\right)}
+
\frac{\Gamma(i\rho) \left(\cosh \tau_1\right)^{i\rho}}{\Gamma\left(\frac14 + i\frac{\rho + \nu}{2}\right) \Gamma\left(\frac14 + i\frac{\rho - \nu}{2}\right)}
\right\},
\eea
\bea
u^{(-)}_{\rho\nu} \sim \pm \frac{\sqrt{\pi}}{2}
\left\{
\frac{\Gamma(-i\rho) \left(\cosh \tau_1\right)^{-i\rho}}{\Gamma\left(\frac34 - i\frac{\rho + \nu}{2}\right) \Gamma\left(\frac34 - i\frac{\rho - \nu}{2}\right)}
+
\frac{\Gamma(i\rho) \left(\cosh \tau_1\right)^{i\rho}}{\Gamma\left(\frac34 + i\frac{\rho + \nu}{2}\right) \Gamma\left(\frac34 + i\frac{\rho - \nu}{2}\right)}
\right\}.
\eea
Substituting these asymptotics into the right-hand side of the expression (\ref{INT_EQ_U}) yields:
\be
\int\limits_{-\infty}^{\infty}u_{\rho\nu}^{(+)}(\tau_1) u^{(+)\ast}_{\rho' \nu}(\tau_1) d\tau_1 = 
\frac{ 4\pi^3 \delta(\rho - \rho')}{\rho\sinh\pi\rho 
\left|\Gamma\left(\frac14 + i\frac{\rho + \nu}{2}\right)\right|^2  \left|\Gamma\left(\frac14 + i\frac{\rho - \nu}{2}\right)\right|^2},
\ee
\be
\int\limits_{-\infty}^{\infty}u_{\rho\nu}^{(-)}(\tau_1) u^{(-)\ast}_{\rho' \nu}(\tau_1) d\tau_1 = \frac{\pi^3 \delta(\rho - \rho')}{\rho\sinh\pi\rho \left|\Gamma\left(\frac34 + i\frac{\rho + \nu}{2}\right)\right|^2  \left|\Gamma\left(\frac34 + i\frac{\rho - \nu}{2}\right)\right|^2}.
\ee
Comparison of the above relations with (\ref{EQUIDIS-NORM-00}) leads to constants  
(\ref{EQUIDIS-NORM-01}).

To prove the completeness of the equidistant wave functions $\Psi_{\rho \nu}^{EQ(\pm)}(\tau_1,\tau_2)$, 
we exploit the interbasis expansions (\ref{HORIC-EQUIDIST-EXPAN-01}). 
Using the completeness condition of the horocyclic basis (\ref{COMPLET-01}) and relations
(\ref{HORIC-EQUIDIST-1-01}),  we obtain 
\bea
\label{HORIC-EQUIDIST-02-01}
&& \int\limits_{0}^{\infty}
d\rho
\int\limits_{-\infty}^{\infty}
\left[\Psi_{\rho \nu}^{EQ(+)}(\tau_1,\tau_2)
\Psi_{\rho \nu}^{EQ(+)\ast}(\tau_1',\tau_2')
+
\Psi_{\rho \nu}^{EQ(-)}(\tau_1,\tau_2)
\Psi_{\rho \nu}^{EQ(-)\ast}(\tau_1',\tau_2')
\right]
d\nu
\nonumber\\
[3mm]
&&
=
\int\limits_{0}^{\infty}
d \rho
\int\limits_{-\infty}^{\infty}
d s
\int\limits_{-\infty}^{\infty} ds' \Psi^{HO}_{\rho s}(\tilde{x},\tilde{y}) 
\Psi_{\rho s'}^{HO\ast} (\tilde{x}',\tilde{y}')  \int\limits_{-\infty}^{\infty}
\left[
{{\cal W}_{\rho s}^{\nu (+)}} {{\cal W}_{\rho s'}^{\nu (+)\ast}}
+
{{\cal W}_{\rho s}^{\nu (-)}} {{\cal W}_{\rho s'}^{\nu (-)\ast}}
\right]
d \nu
\nonumber\\
[3mm]
&&
= \int\limits_{0}^{\infty}
d \rho
\int\limits_{-\infty}^{\infty}
\Psi_{\rho s}^{HO}(\tilde{x},\tilde{y}) 
\Psi_{\rho s}^{HO*} (\tilde{x}',\tilde{y}') 
d s
=
\frac{\tilde{y}^2}{R^2}
\delta(\tilde{y}- \tilde{y}')
\delta(\tilde{x}- \tilde{x}').
\eea
Taking into account (\ref{COOR-01}) and the equality
\bea
\label{HORIC-EQUIDIST-03-01}
\tilde{y}^2
\delta(\tilde{y}- \tilde{y}')
\delta(\tilde{x}- \tilde{x}')
=
\tilde{y}
\delta(\tilde{y}- \tilde{y}')
\delta(\sinh\tau_1 - \sinh\tau_1') = \nonumber \\
= \tilde{y} \delta\left(\frac{e^{\tau_2}}{\cosh\tau_1} - \frac{e^{\tau_2^\prime}}{\cosh\tau_1}\right) \frac{\delta(\tau_1 - \tau_1')}{\cosh\tau_1} =
\frac{\delta(\tau_1-\tau_1')\, \delta(\tau_2-\tau_2')}{\cosh\tau_1},
\eea
finally we come to relation (\ref{COMPLET-005}).

\subsection{Completeness and orthonormality of the pseudo-spherical basis}
\label{bsection:Completeness_of_SH}

As in previous case, we can use interbasis expansion (\ref{HORIC-EQUIDIST-EXPAN-03}), conditions (\ref{COMPLET-005}) with the help of orthogonality properties (\ref{COMPLET-00-005}),  
(\ref{EQUIDIST-SPHERIC-25A}) and (\ref{01-EQUIDIST-SPHERIC-04-27}), (\ref{06-EQUIDIST-SPHERIC-27})  to prove the orthogonality and completeness of the pseudo-spherical basis  $\Psi^{S}_{\rho m} (\tau, \vphi)$.  
Indeed for the orthogonality condition we get
\bea
\label{COM-EQUIDIST-SPHERIC-00}
&& R^2 \int\limits_{0}^{\infty} \sinh\tau d \tau \int\limits_{0}^{2\pi}
\Psi^S_{\rho m}(\tau, \vphi)  \Psi^{S\ast}_{\rho^\prime m^\prime}(\tau, \vphi) d \vphi =
\nonumber\\[2mm]
&=&
\delta(\rho-\rho')
\int\limits_{- \infty}^{\infty}
\left[{\cal U}_{\rho \nu}^{m(+)} {\cal U}_{\rho \nu}^{m'(+)*}
+ {\cal U}_{\rho \nu}^{m(-)} {\cal U}_{\rho \nu}^{m(-)*}\right] d \nu
=
\delta(\rho-\rho')  \delta_{m m'}.
\eea

Completeness follows from
\bea
\label{COM-EQUIDIST-SPHERIC}
&& R^2 \, \sum\limits_{m = - \infty}^{\infty} 
\int\limits_{0}^{\infty}
\Psi^S_{\rho m} (\tau, \vphi)  \Psi_{\rho m}^{S \ast}(\tau', \vphi') d \rho =
\nonumber\\[2mm]
&=&
R^2  \int\limits_{0}^{\infty} d \rho \int\limits_{- \infty}^{\infty}\,
\left[\Psi_{\rho \nu}^{EQ(+)}(\tau_1, \tau_2)\Psi_{\rho \nu}^{EQ(+)*}(\tau_1^\prime, \tau_2^\prime)
+ \Psi_{\rho \nu}^{EQ(-)}(\tau_1, \tau_2)\Psi_{\rho \nu}^{EQ(-)*}(\tau_1^\prime, \tau_2^\prime)
\right]  d \nu =
\nonumber\\[2mm]
&=&
\frac{1}{\cosh \tau_1}  \delta(\tau_2 -  \tau_2^\prime) \delta(\tau_1 - \tau_1^\prime) = \frac{1}{\sinh\tau} \delta(\tau - \tau^\prime) \delta(\vphi-\vphi^\prime),
\eea
where we use (\ref{EQUIDIST-SPHERIC-1-03}) and the relation
\bea
\label{0-COM-EQUIDIST-SPHERIC}
\delta(\tau_1 - \tau_1^\prime)  \delta(\tau_2 -  \tau_2^\prime) = \cosh\tau_2 \delta(\tau_1 - \tau_1^\prime) \delta(\sinh\tau_2 - \sinh\tau_2^\prime) = \nonumber \\  
 = \frac{\cosh\tau_2}{|\tanh\tau_1|} \delta(\tau_1 - \tau_1^\prime) \delta(\cot\vphi - \cot\vphi^\prime) \nonumber = \\
=  \frac{\cosh\tau \sin^2 \vphi \cosh\tau_1}{|\sinh\tau_1|}\delta(\vphi - \vphi^\prime) \delta(\sinh\tau_1 - \sinh\tau_1^\prime) = \nonumber\\
 = \frac{ |\sin \vphi| \cosh\tau_1}{|\sinh\tau_1|}\delta(\vphi - \vphi^\prime) \delta(\tau  - \tau^\prime). \nonumber
\eea

\bibliography{bib_Georgy}

\end{document}